\documentclass{jfm}

\usepackage{amssymb}
\usepackage{graphicx}
\usepackage{epstopdf,epsfig}
\usepackage{newtxtext}
\usepackage{newtxmath}
\usepackage{natbib}
\usepackage{hyperref}
\hypersetup{
    colorlinks = true,
    urlcolor   = blue,
    citecolor  = black,
}

\newcommand{\RomanNumeralCaps}[1]
\linenumbers

\title{A closure mechanism for screech coupling in rectangular twin jets}

\author{Jinah Jeun\aff{1}
  \corresp{\email{jinahjeun@gmail.com}},
  Gao Jun Wu\aff{2}
 \and Sanjiva K. Lele\aff{3}}

\affiliation{\aff{1}Center for Turbulence Research, Stanford University, Stanford, CA 94305, USA
\aff{2}Department of Aeronautics and Astronautics, Stanford University, Stanford, CA 94305, USA
\aff{3}Department of Mechanical Engineering and Department of Aeronautics and Astronautics, Stanford University, Stanford, CA 94305, USA}

\begin{document}
\maketitle

\begin{abstract}
Twin-jet configuration allows two different scenarios to close the screech feedback. For each jet, there is one loop involving disturbances which originate in that jet and arrive at its own receptivity point in-phase (self-excitation). The other loop is associated with free-stream acoustic waves that radiate from the other jet, reinforcing the self-excited screech (cross-excitation). In this work, the role of the free-stream acoustic mode and the guided jet mode as a closure mechanism for twin rectangular jet screech is explored by identifying eligible points of return for each path, where upstream waves propagating from such a point arrive at the receptivity location with an appropriate phase relation. Screech tones generated by these jets are found to be intermittent with an out-of-phase coupling as a dominant coupling mode. Instantaneous phase difference between the twin jets computed by the Hilbert transform suggests that a competition between out-of-phase and in-phase coupling is responsible for the intermittency. To model wave components of the screech feedback while ensuring perfect phase-locking, an ensemble average of leading spectral proper orthogonal decomposition modes is obtained from several segments of large-eddy simulations data that correspond to periods of invariant phase difference between the two jets. Each mode is then extracted by retaining relevant wavenumber components produced via a streamwise Fourier transform. Spatial cross-correlation analysis of the resulting modes shows that most of the identified points of return for the cross-excitation are synchronised with the guided jet mode self-excitation, supporting that it is preferred in closing rectangular twin-jet screech coupling.
\end{abstract}

\begin{keywords}
\end{keywords}


\section{Introduction}
\label{sec:intro}
Supersonic aircraft during take-off and landing from an aircraft carrier deck operate at off-design conditions, producing deafening sound from its engine exhausts characterised by three distinctive noise components. Turbulent mixing noise, which is attributed to large-scale coherent structures contained in jet turbulence, dominantly radiates at low aft angles (about 30 to 60 degrees)~\citep{Tam1995,Jordan2013}. In addition, the shock train formed due to the non-ideal expansion generates the broadband shock-associated noise and, sometimes, even screech, via the interaction with the Kelvin-Helmholtz (KH) instability waves. Screech is associated with drastic amplification in sound pressure level within a very narrow-banded frequency bin and radiates mostly upstream, causing significant potential structural damage to the airframe~\citep{Hay1970,Berndt1984,Seiner1988}. 

\citet{Powell1953} first discovered jet screech, and since then, it has drawn a continuous interest from the aeroacoustic community. He first described that screech is an aeroacoustic resonance, involving interaction between the shock and the KH instability waves, which produces upstream-travelling sound waves. The receptivity at the nozzle lip then excites new downstream-travelling disturbances, and once they sufficiently develop, they generate upstream-travelling sound, sustaining the feedback cycle. The feedback cycle involving upstream- and downstream-travelling waves is now a generally accepted scenario, but there are still many unknowns that need to be addressed. 

While the detailed mechanisms in each process of screech generation mentioned above are not fully understood, the nature of the upstream-propagating waves has received considerable attention in recent research. In Powell's early work, the feedback due to free-stream acoustic waves propagating outside the jet was stressed. \citet{Shen2002} were the first who proposed that waves closing the screech feedback of the A2 (axisymmetric) and C (helical) modes of round jets could in fact be an intrinsic neutral mode identified by~\citet{Tam1989}. However, they still suggested the A1 (axisymmetric) and B (flapping) modes favour the free-stream acoustic mode as a closure mechanism. In more recent years, there have been several studies using experimental and numerical evidence~\citep{Edgington-Mitchell2018,Gojon2018,Li2020} that support the guided jet mode as a unified closure mechanism for both axisymmetric modes. \citet{Mancinelli2019} demonstrated that screech frequency prediction based on the guided jet mode showed enhanced accuracy, compared to the result obtained by the free-stream acoustic mode. \citet{Nogueira2022b} confirmed that axisymmetric screech modes can be regarded as an absolute instability associated with the interaction between the downstream-propagating KH waves and upstream-propagating guided jet mode. Furthermore, \citet{Nogueira2022a} found that transition from the A1 to A2 modes is closed by the interaction of the KH mode and the shock system with variations in the spatial wavenumber using absolute instability analysis. Later, this work was extended to rectangular and elliptical jets, showing that their modal staging behavior can also be driven by such triadic interaction~\citep{Edgington-Mitchell2022}. For rectangular jets, \citet{Gojon2019} and \citet{Wu2020,Wu2023} revealed that the screech feedback was closed by the guided jet mode more effectively than by the free-stream acoustic mode.

The addition of an extra jet adds complexity in studying the screech closure mechanism. Whereas a single jet admits self-excited resonance only, twin systems introduce external acoustic waves originating from one jet, reinforcing the self-excitation screech feedback of its pair. Arriving at the receptivity point of its twin with a phase difference consistent with a natural coupling mode between the two jets at a given screech frequency can reinforce the other jet's own screech feedback.  

Another difficulty arising in twin-jet systems lies in modelling their coupling mode. It can also be crucial for developing effective control strategies~\citep{Samimy2023}. In the literature, while efforts are mostly given to identifying the coupling modes of twin jets from various nozzle geometries by utilizing experimental~\citep{Raman1999,Alkislar2005,Kuo2017,Bell2018,Knast2018,Esfahani2021} and numerical data~\citep{Gao2018,Jeun2022}, prediction models for them are scarce. An empirical model proposed by \citet{Webb2023} successfully predicted the preferred coupling modes of twin rectangular jets at a range of operating conditions, but their model still lacked the distinction between the guided jet mode and the free-stream acoustic mode as a closure mechanism for screech. Rather, thanks to the comparable propagation velocities of the two modes, the free-stream acoustic mode was able to be assumed without further discussion. Moreover, in many cases screech tones produced by twin jets are reported to be intermittent, constantly modifying their coupling mode in time for both axisymmetric and rectangular jets~\citep{Bell2021,Karnam2021,Jeun2022}. In this sense, developing realistic models for the twin-jet screech coupling becomes extremely challenging.

In rectangular jets there exists a preferential flapping mode along the minor axis, which somewhat simplifies the modelling work for coupling modes in them. In this paper, we consider the twin version of such rectangular jets, which presents the out-of-phase coupling about the center axis at the fundamental screech frequency. Screech tones in this twin system are also intermittent, but the near-field noise data can be divided into several segments in time that manifest steady out-of-phase coupling for a sufficiently long time to ensure the frequency resolution required for detecting sharp screech tones under the assumption of the flow stationarity. An ensemble average of spectral proper orthogonal decomposition (SPOD) modes~\citep{Towne2018} from the resulting segments is then decomposed by a streamwise Fourier transform, to isolate each wave component of the screech feedback loop. By computing spatial cross-correlation of the decomposed waves, we aim to discuss which feedback paths dominate the closure mechanism for the rectangular twin-jet screech coupling. To the best of our knowledge, the present study is the first to investigate the nature of the upstream-propagating waves that close the rectangular twin-jet screech coupling, hoping that it can aid to develop a unifying explanation for the jet screech in complex interacting jets.  

The remainder of the paper is organised as follows. High-fidelity large-eddy simulation database and SPOD of it are briefly introduced in \S~\ref{sec:les}. Screech feedback scenarios in rectangular twin jets are described in \S~\ref{sec:twinjet_feedback_scenarios}. Intermittency of screech tones in our twin jets is investigated in \S~\ref{sec:intermittency}. Ensemble averaged SPOD modes are computed accordingly in \S~\ref{sec:spod_modes}, followed by the extraction of wave components active in the screech feedback from these modes in the same section. Based on the spatial cross-correlation analysis, the preferred closure mechanism of the twin-jet screech coupling is discussed in \S~\ref{sec:feedback}. Lastly, \S~\ref{sec:conclusion} summarizes the main conclusions of the paper.

\section{Large-eddy simulation database}
\label{sec:les}
\subsection{Large-eddy simulation (LES)}
In this work we utilize high-fidelity LES data for jets issuing from twin rectangular nozzles with an aspect ratio of 2~\citep{Jeun2022}, which were computed by a fully compressible unstructured flow solver, charLES, developed by Cascade Technologies~\citep{Bres2017}. The twin nozzle had a sharp converging-diverging throat, from which internal oblique shocks formed, and a design Mach number $M_d$ = 1.5. The two nozzles were placed closely to each other with the nozzle center-to-center spacing of $3.5h$, mimicking military-style aircraft. Aeroacoustic coupling between the twin jets thus becomes quite important for this flow configuration. The system was scaled by the nozzle exit height $h$ with respect to the origin chosen at the middle of the nozzle exits. The coordinate system was chosen so that the $+x$ axis was defined along the streamwise direction, while $y$ and $z$ axes were defined along the minor and major axis directions of the nozzle at the exit, respectively. The total simulation duration was 1,400 acoustic times ($h/c_\infty$ where $c_\infty$ is the ambient speed of sound), which correspond to approximately 400 screech cycles.

\begin{figure}
  \centering
  \begin{tabular}{cc}
    \includegraphics[width=0.48\textwidth]{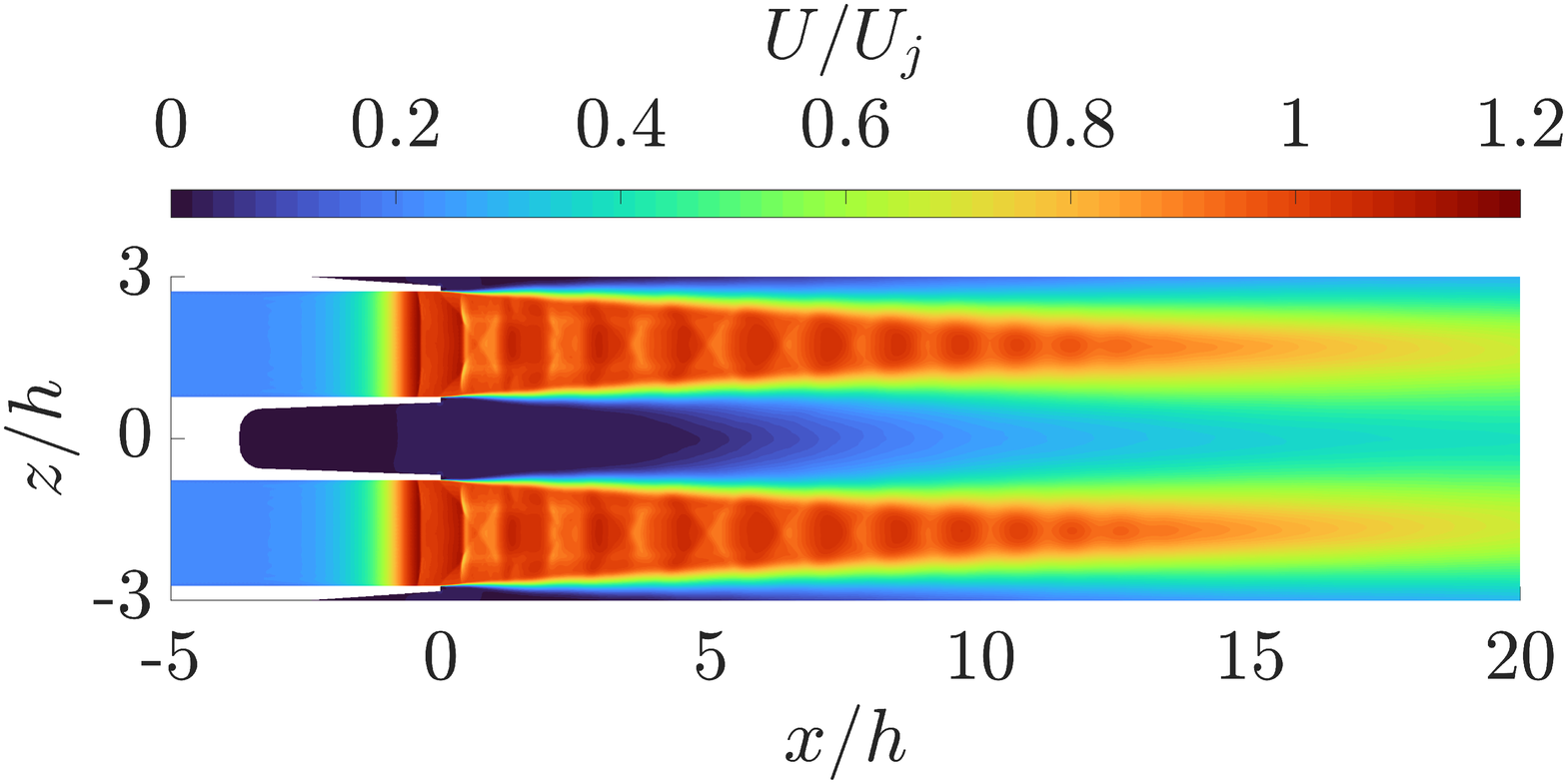} &
    \includegraphics[width=0.48\textwidth]{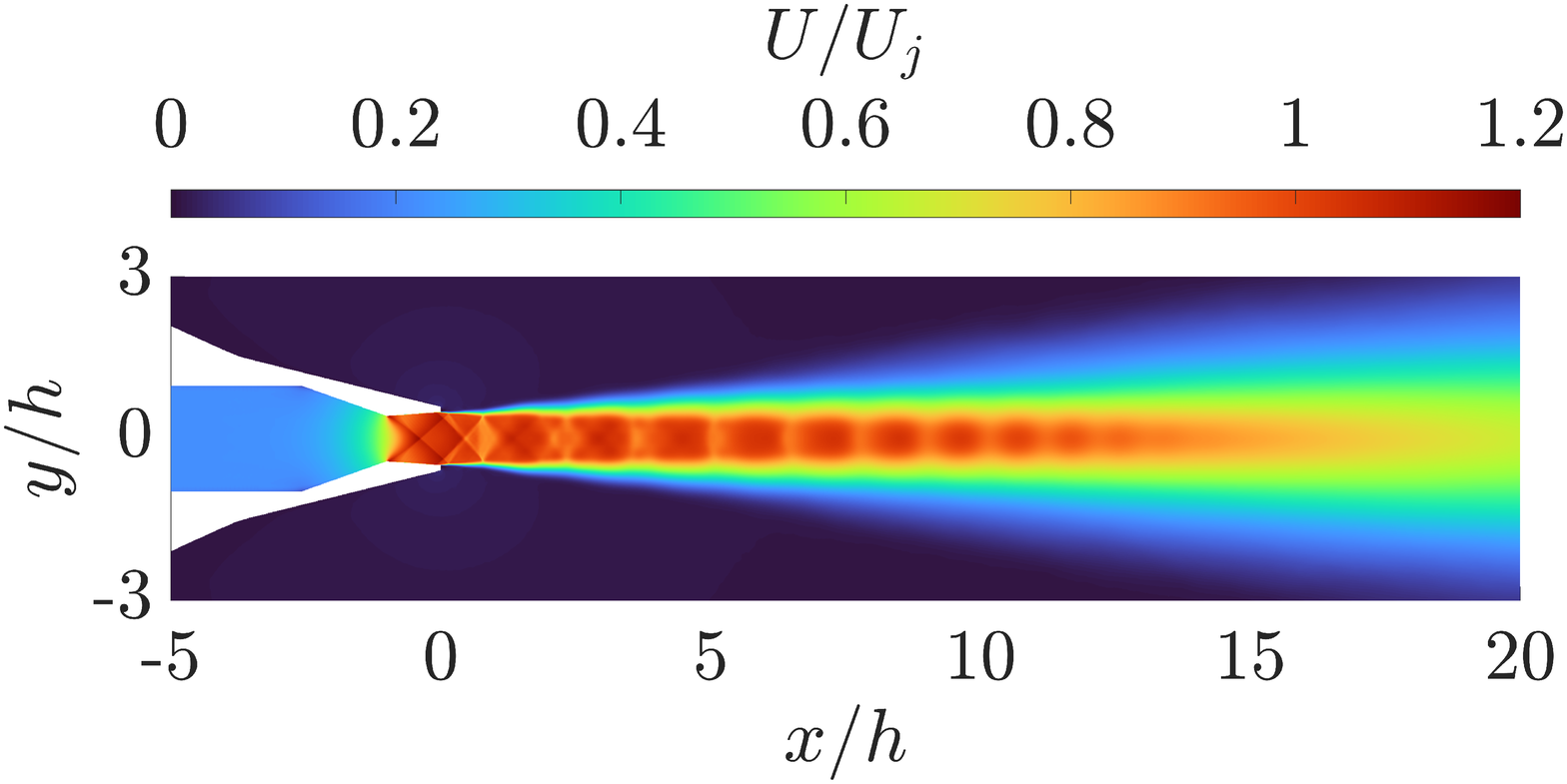} \\
    (a) & (b) \\
  \end{tabular}
  \caption{Contours of the time-averaged streamwise velocity normalised by the fully expanded jet velocity in the major axis (a) and in the minor axis (b).}
\label{fig:les}
\end{figure}

The LES database was systematically validated under various operating conditions against the experiments conducted at the University of Cincinnati. Among the three cases simulated, the present work considers overexpanded twin jets at NPR = 3 (where NPR is the nozzle pressure ratio as computed by total pressure over ambient static pressure), which registered the maximum screech. Figure~\ref{fig:les} shows the mean streamwise velocity contours in the major and minor axes. The near-field data used for the present analysis were measured in the minor axis planes extending from 0 to 20 in the streamwise direction and from -5 to 5 in the vertical direction, respectively, at the center of each nozzle ($z/h$ = $\pm$1.75). In these planes probe points were uniformly distributed in both directions with $\Delta x$ = $\Delta y$ = 0.05$h$. The LES data were collected at every 0.1$h/c_\infty$. More details on the flow solver and the LES database can be found in~\citet{Bres2017} and \citet{Jeun2022}, respectively.

\subsection{Identification of the coupling mode via SPOD}
\label{subsec:spod}
As an extension of proper orthogonal decomposition in the frequency domain, SPOD extracts coherent structures varying in both space and time. At a frequency, SPOD yields a ranked set of modes that optimally represent flow energy~\citep{Towne2018}. SPOD is therefore a suitable and effective tool to extract coherent structures associated with twin-jet screech and identify the coupling mode between them at the screech frequency. 

For a state vector $q(\boldsymbol{x},t)$ from a zero-mean stochastic process, a data matrix $Q$ can be formed by a series of flow observations sampled at a uniform rate,
\begin{equation}
    Q = \left[ q^{(1)} \ q^{(2)} \ \cdots \ q^{(N)} \right], \quad Q \in\mathbb{R}^{M\times N},
\end{equation}
where $q^{(k)}$ represents $k$-th instance of the ensemble of realizations, $M$ is the number of flow variables multiplied by the number of points in space, and $N$ is the number of sampled realizations. After taking the Fourier transform in time, the transformed data matrix $\hat{Q}$ can be constructed as
\begin{equation}
    \hat{Q} = \left[ \hat{q}^{(1)} \ \hat{q}^{(2)} \ \cdots \ \hat{q}^{(N)} \right], \quad \hat{Q} \in\mathbb{C}^{M\times N}.
\end{equation}
Then, SPOD computes the eigen-decomposition of the cross-spectral density (CSD) tensor 
\begin{equation}
    \hat{C} = \frac{1}{N-1} \hat{Q}^{H}\hat{Q}
\end{equation}
such that
\begin{equation}
    \hat{C}W\hat{\Phi} = \hat{\Phi}\hat{\Lambda}.
\end{equation}
Here, the superscript $H$ denotes the Hermitian, and $W$ is a positive definite matrix that represents numerical quadrature weights. The matrix $\hat{\Phi}$ contains the eigenvectors of $\hat{C}$, and $\Lambda$ is a matrix whose diagonals are the eigenvalues sorted in descending order. These eigenvectors are SPOD modes, and the eigenvalues represent modal energy contributed by each mode. Note that in this work SPOD modes and eigenvalues are computed via the so-called method of snapshots. Besides, the CSD matrix is estimated by the Welch's method~\citep{Welch1967} with 75\% overlap to exploit the stationarity.

The optimality and orthogonality properties of the SPOD modes depend on the choice of norm, which is incorporated through the weight matrix $W$. In this work we compute modes that are orthogonal in the compressible energy norm
\begin{equation}
\langle\boldsymbol{q}_1,\boldsymbol{q}_2\rangle_E = \iiint\boldsymbol{q}^*_1 \mathrm{diag} \left( \frac{\overline{T}}{\gamma\overline{\rho}M_j^2}, \overline{\rho}, \overline{\rho}, \overline{\rho}, \frac{\overline{\rho}}{\gamma\left(\gamma-1\right)\overline{T}M_j^2} \right)\boldsymbol{q}_2 dxdydz
\end{equation}
for the state vector $\boldsymbol{q} = [\rho, u, v, w, T]^T$, as derived by~\cite{Chu1965}. Here, $\rho$ denotes density, $u$, $v$, and $w$ are Cartesian velocity components, and $T$ represents temperature. The overlines denote the mean quantities of the corresponding flow variables. 

\begin{figure}
  \centering
  \includegraphics[width=0.55\textwidth]{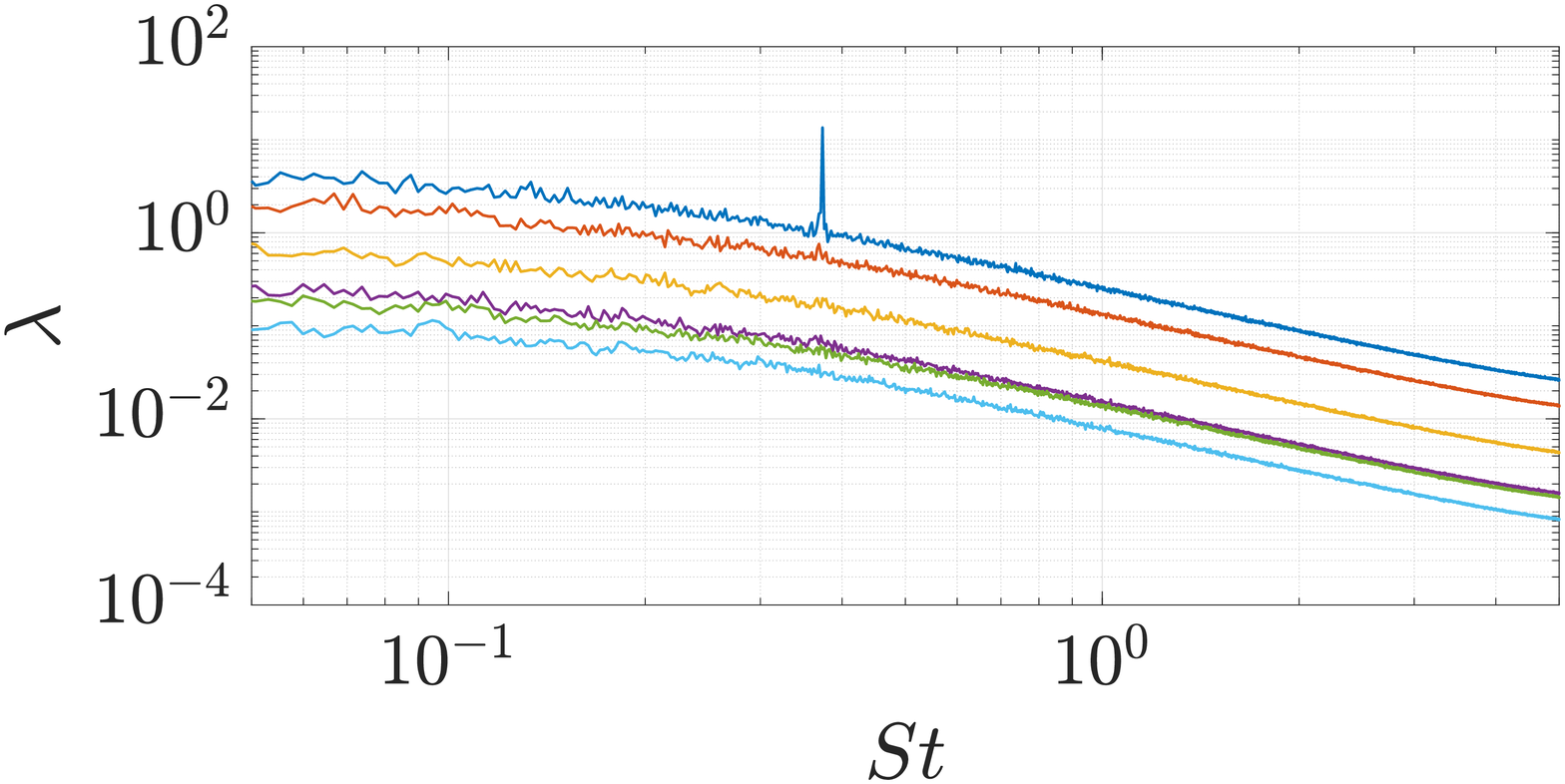}
  \caption{SPOD energy spectra obtained from the flow fluctuations extracted along the minor axis plane at the center of each nozzle ($z/h$ = $\pm$1.75). The data from both jets are combined into a single matrix, on which SPOD is subsequently performed.}
\label{fig:blind_spod_xy_spectra}
\end{figure}

\begin{figure}
  \centering
  \begin{tabular}{cc}
    \centering
    \includegraphics[width=0.48\textwidth]{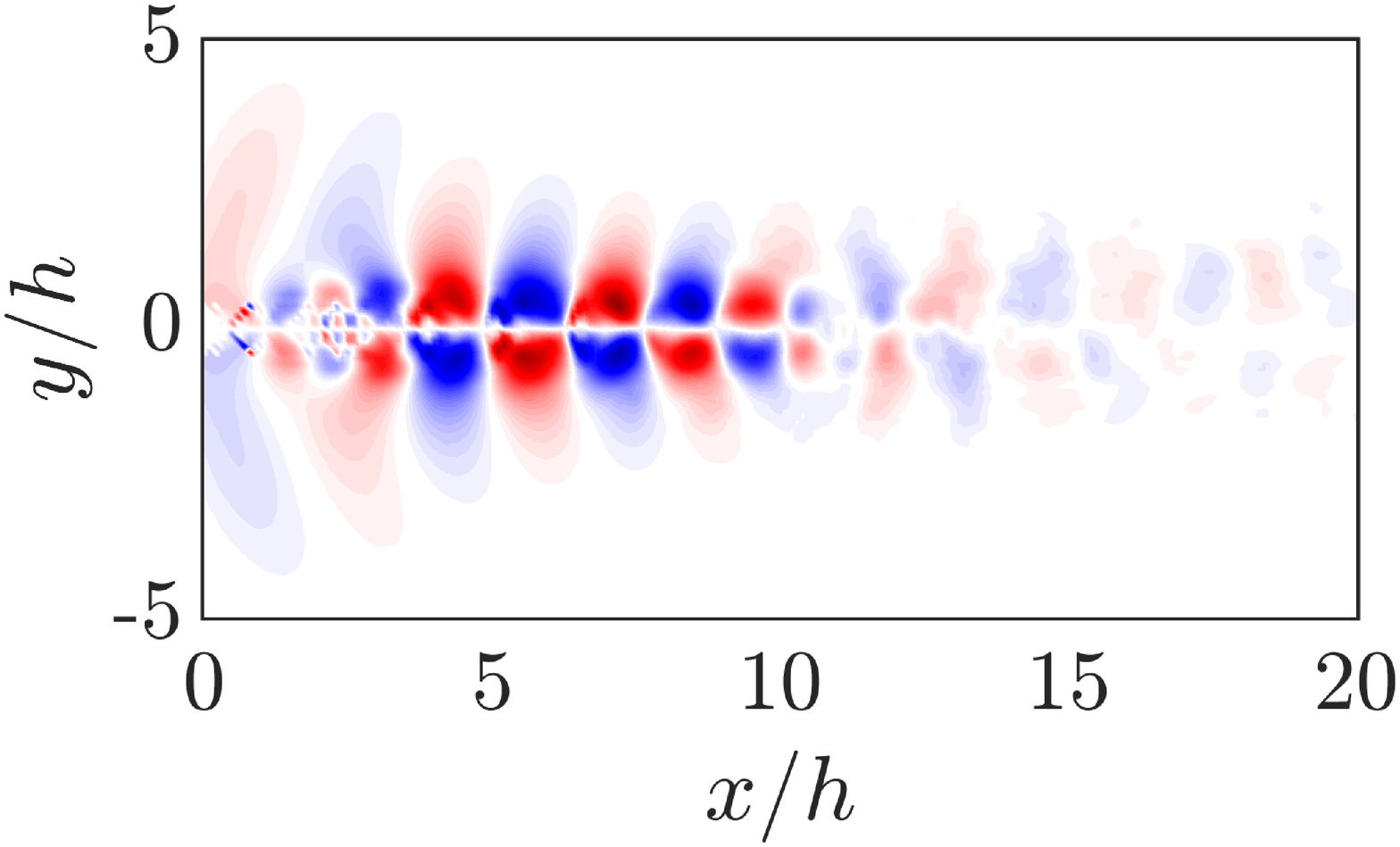} &
    \includegraphics[width=0.48\textwidth]{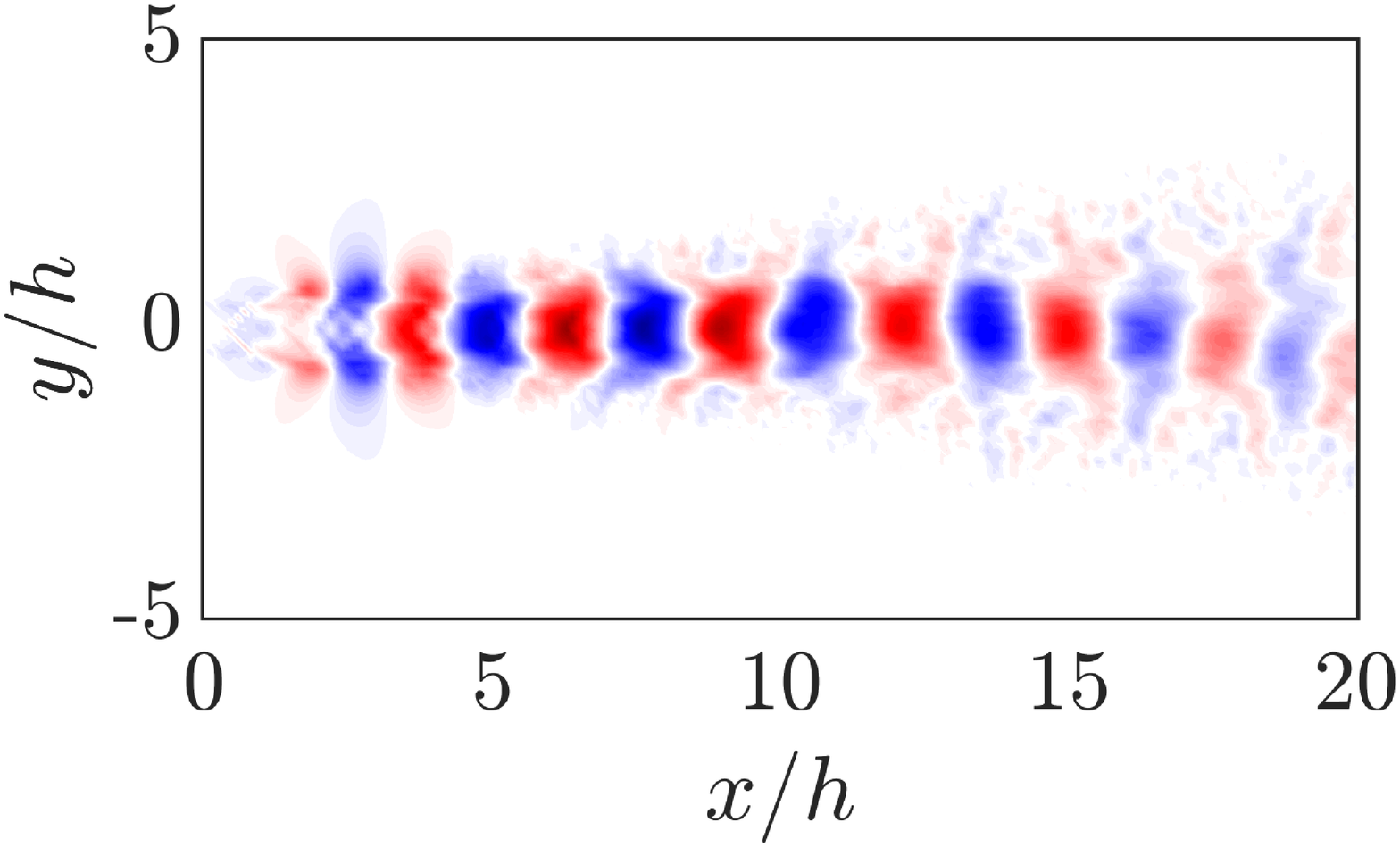} \\
    (a) & (b) \\
    \includegraphics[width=0.48\textwidth]{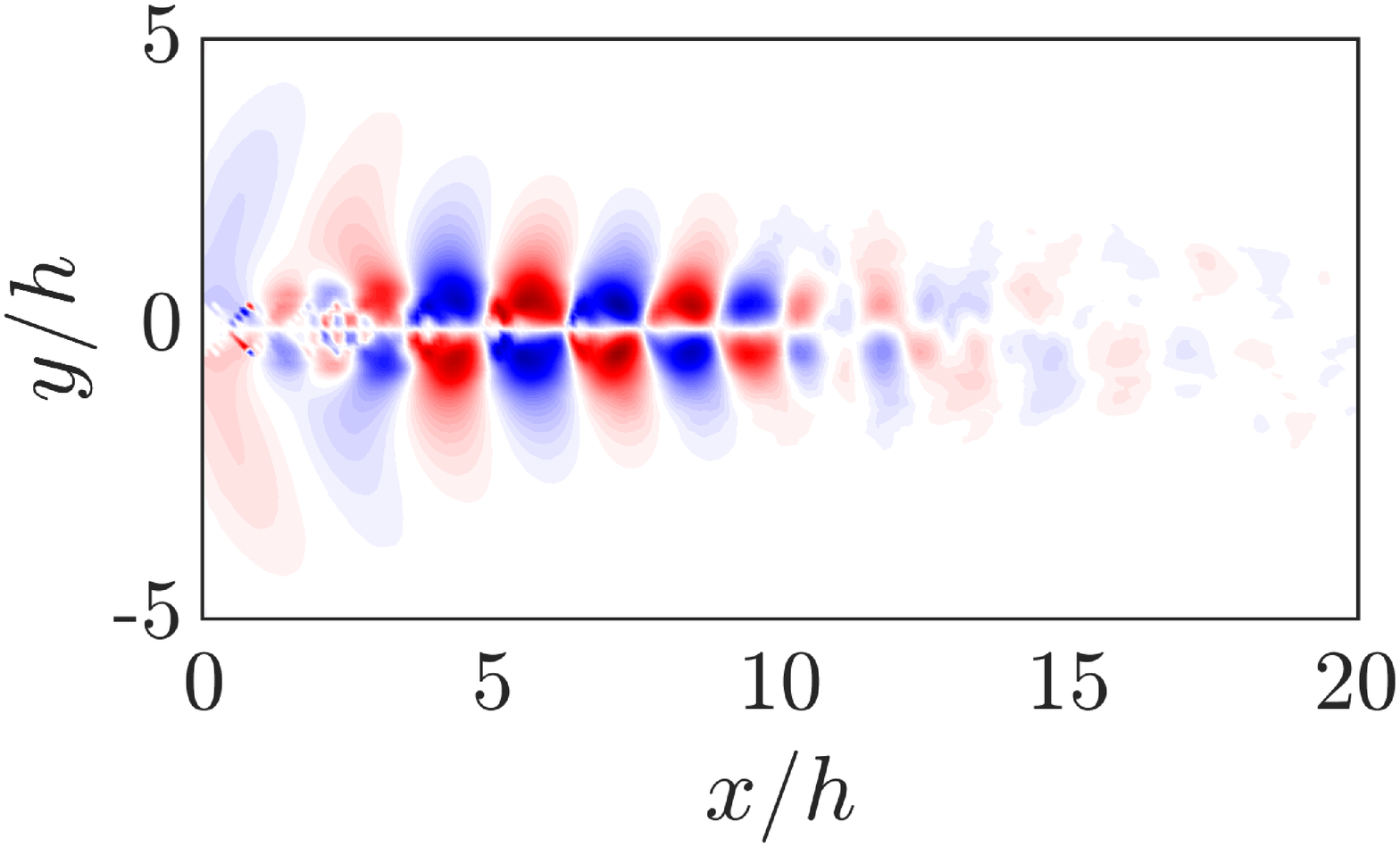} &
    \includegraphics[width=0.48\textwidth]{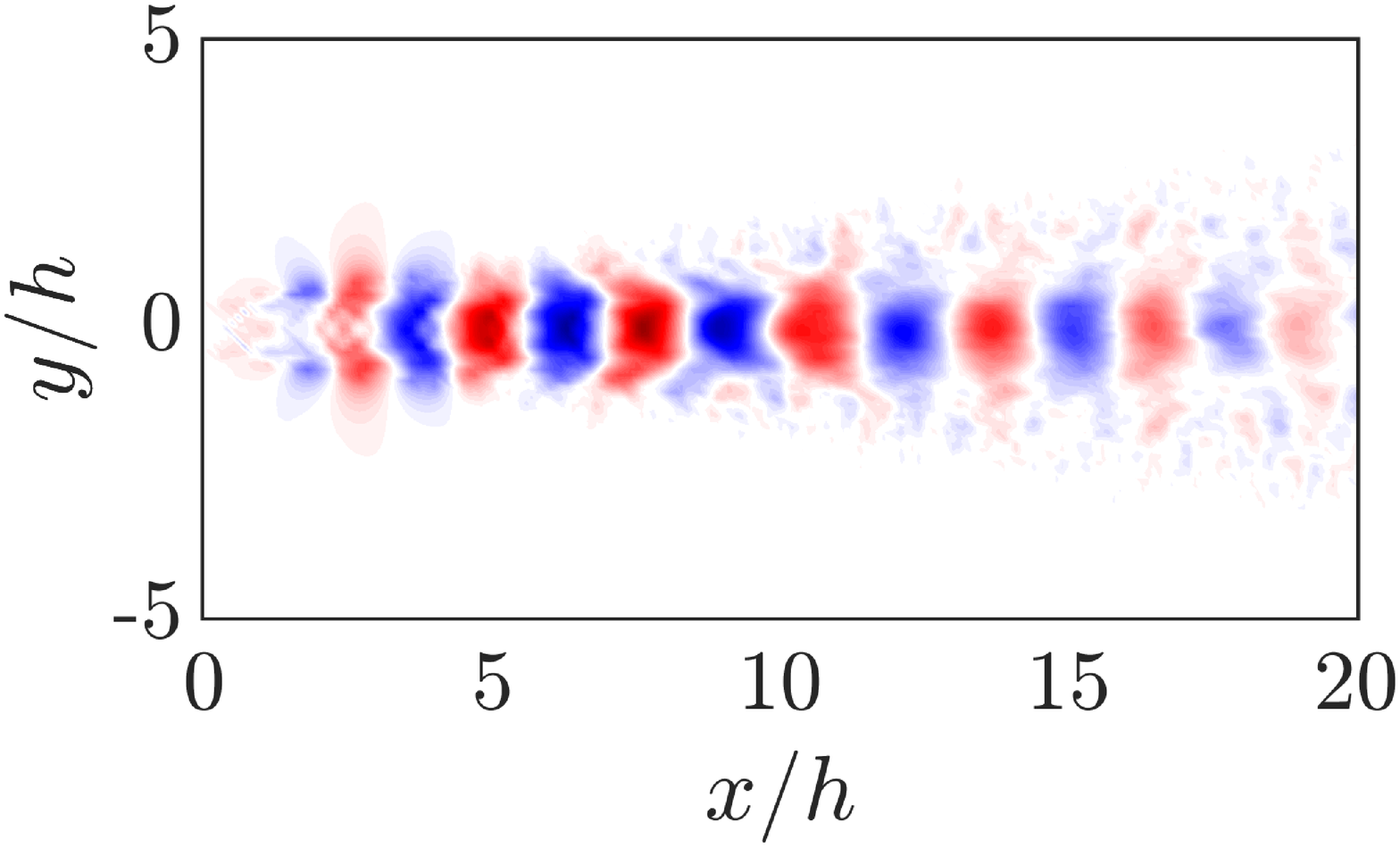} \\
    (c) & (d) \\
  \end{tabular}  
  \caption{Comparisons are made between the SPOD mode shapes for Jet 1 (a,b) and Jet 2 (c,d): (a,c) real part of the leading SPOD mode for the fluctuating pressure component; (b,d) real part of the leading SPOD mode for the fluctuating transverse velocity component. Each contour is normalised by its maximum value. The colour ranges from -1 to 1.}
\label{fig:blind_spod_xy_modes}
\end{figure}

\begin{figure}
  \centering
  \begin{tabular}{cc}
    \includegraphics[width=0.48\textwidth]{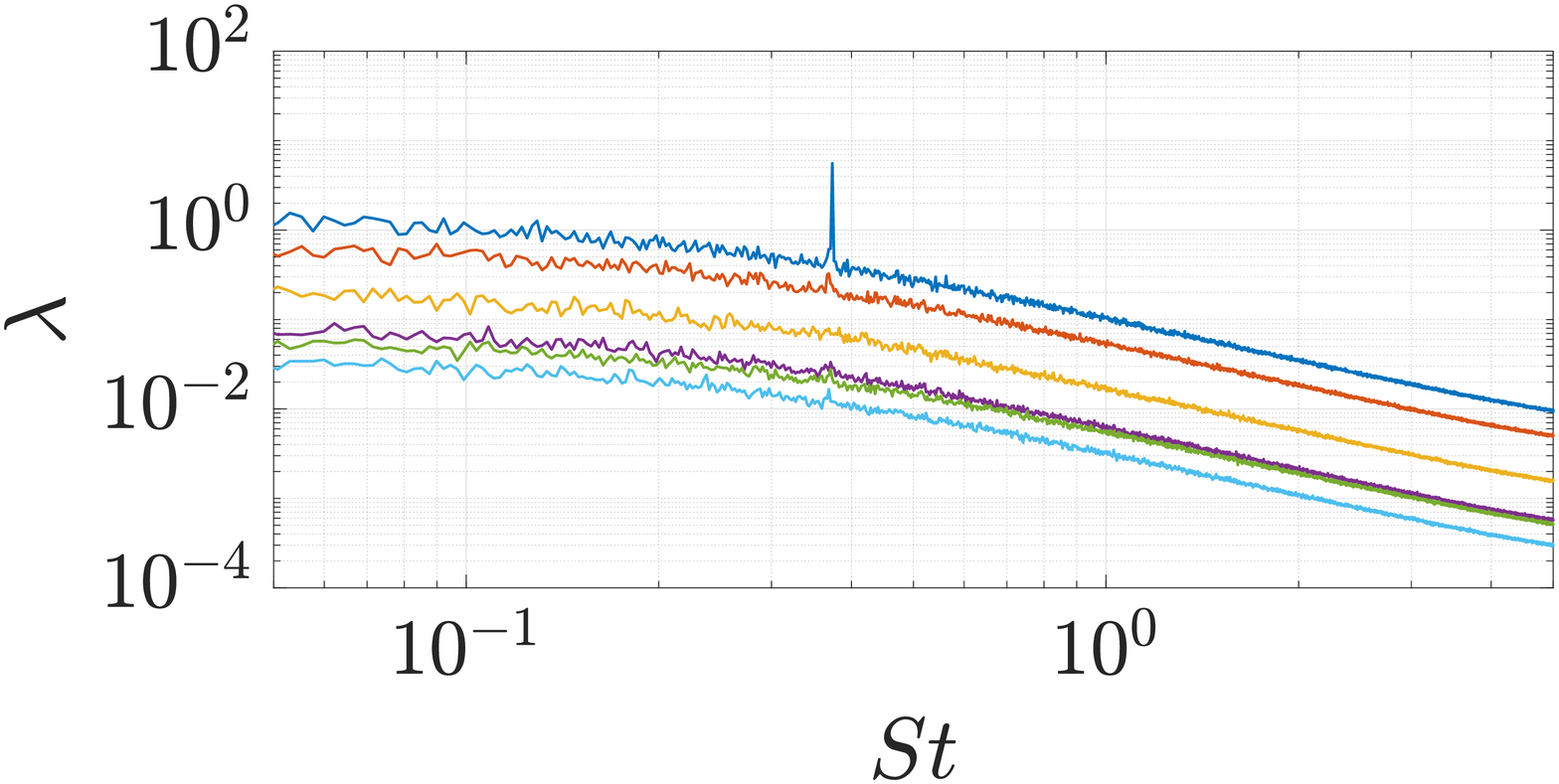} &
    \includegraphics[width=0.48\textwidth]{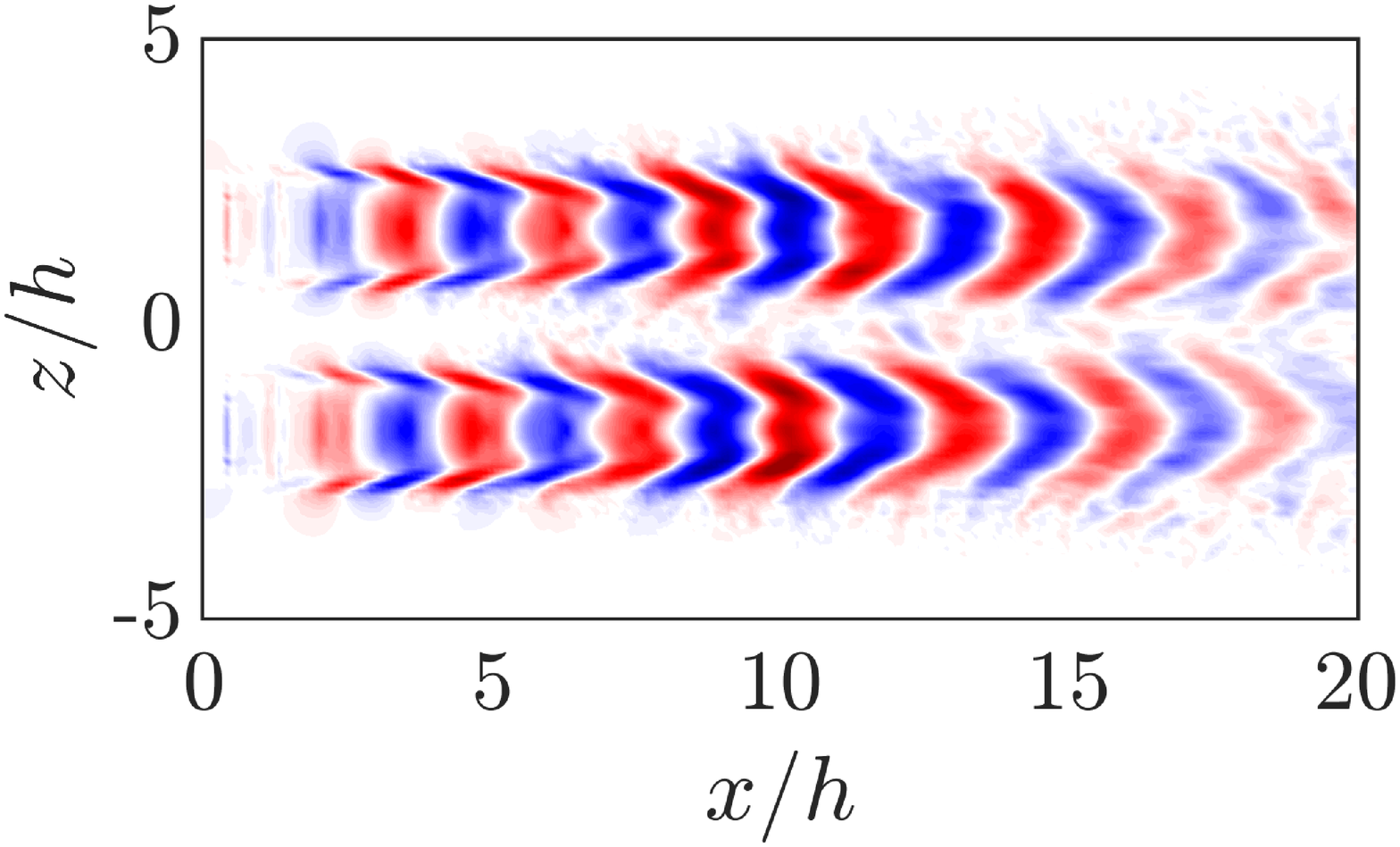} \\
    (a) & (b) \\
  \end{tabular}
  \caption{$v'$-SPOD computed by the flow fluctuations extracted along the major axis plane ($y/h$ = 0): (a) SPOD energy spectra; (b) real part of the leading SPOD mode.}
\label{fig:blind_spod_xz}
\end{figure}

Figure~\ref{fig:blind_spod_xy_spectra} shows the resulting SPOD energy spectra obtained using the flow fluctuations measured in the minor axis planes at the center of each nozzle. To consider the spatial correlation between the two jets, SPOD is applied onto a single matrix of flow data, consisting of a pair of 2D slices extracted from the mid-plane cross-section of each jet at $z/h$ = $\pm 1.75$. For our twin jets, the leading modes exhibit high amplitude tonal peaks at the fundamental screech frequency ($St_{sc} = 0.37$ where the Strouhal number $St$ is defined based on the equivalent jet diameter $D_e$ and the fully expanded jet velocity $U_j$)~\citep{Jeun2022}. A significant energy separation between the leading and higher order modes is observed at this frequency. Figure~\ref{fig:blind_spod_xy_modes} illustrates the fluctuating pressure ($p'$-SPOD) and the fluctuating transverse velocity ($v'$-SPOD) components of the leading SPOD mode shapes for each jet computed at the screech frequency. The SPOD modes indeed encompass a pair of coherent flow structures associated with each jet. In this figure, the modes are subsequently separated for visualization purposes, revealing an out-of-phase coupling of the two jets about the center axis ($z/h$ = 0). The fluctuating pressure component  is derived from the temperature and density components using the linearised equation of state. On the other hand, SPOD can be performed using only transverse velocity fluctuations in the major axis plane ($y/h$ = 0) that contains both jets. SPOD modes are now orthogonal in the sense of the $L_2$-norm. As shown in figure~\ref{fig:blind_spod_xz}, the out-of-phase coupling between the two jets can be more directly seen through the corresponding leading SPOD mode shape extracted in this plane. By examining both views, we present a comprehensive evidence for the out-of-phase coupling behavior of the twin jets about the center axis.

\section{Twin-jet screech feedback scenarios}
\label{sec:twinjet_feedback_scenarios}
\begin{figure}
  \centerline{\includegraphics[width=.65\textwidth]{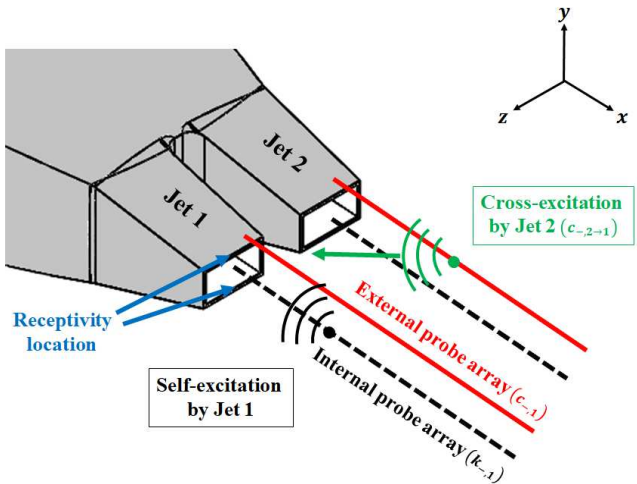}}
  \caption{Schematic representation of feedback processes in rectangular twin jets.}
\label{fig:excitation_scenarios}
\end{figure}

As shown in the previous section, the jets in the present study predominantly exhibit out-of-phase synchronisation to each other. By leveraging the preferred coupling mode, the present analysis views the twin-jet system as assembly of two isolated jets with the corresponding phase difference imposed. As such, we propose that the twin-jet screech feedback loop can be divided into two different processes as schematically described in figure~\ref{fig:excitation_scenarios}. In addition to the upstream waves originating in each jet (self-excitation), external acoustic waves radiating from the other jet (cross-excitation) can also play a role in reinforcing one jet's screech feedback loop. For simplicity, each self-excitation path is isolated from the other by extracting flow structures in the mid-plane cross-section along the minor axis at the center of each nozzle ($z/h$ = $\pm1.75$).

Note that the gain provided by the cross-excitation may be miniscule but still strong enough for the jets to prefer coupling over remaining uncoupled~\citep{Wong2023} or synchronizing to other types of coupling mode. The interplay between self-excitation and cross-excitation may contribute to intermittency in the phase relationship. However, it should be emphasized that the focus of this paper does not extend to studying the underlying mechanisms behind intermittency in coupling. Specifically, our objective is to examine whether it is the free-stream acoustic mode or the guided-jet mode that closes the screech coupling, given the twin jets exhibiting a preferred coupling mode.

The dominant flow structures associated with the screech are modelled by the leading SPOD mode at the fundamental screech frequency. As shown in the previous section, the SPOD energy spectra exhibited sharp peaks at the fundamental screech frequency, with clear energy separation between the leading mode and the higher-order modes~\citep{Jeun2022}. The relative contribution by the leading mode was measured to be almost 96\% of the total energy, justifying the use of the leading SPOD mode. The free-stream acoustic waves $c_{-}$ and the guided jet mode $k_{-}$ are then educed by exploiting a streamwise Fourier decomposition onto the leading SPOD mode, complemented by a filtering based on the phase velocity of waves. The extraction process will be further detailed later in \S\ref{sec:feedback}. The guided jet mode is evanescent in the transverse direction, decaying exponentially outside the jet plume. The free-stream acoustic mode has extended support in the transverse direction by comparison. Therefore, the free-stream acoustic mode $c_{-}$ can be traced along the external probe arrays, while the KH $k_{+}$ and guided jet mode $k_{-}$ can be done along the probe arrays inside the jet plume, as shown in figure~\ref{fig:excitation_scenarios}.

The idea of isolating the self-excitation for each jet is an attempt to explain the overall behavior using the simplest physical models. Linear stability analysis can compute modes that are naturally coupled to each other, so the resulting modes from it seem to be more physically comprehensive in analysing the screech coupling mechanism. Relying only on the periodicity in the base flow~\citep{Tam1982}, it was demonstrated that the screech is an absolute instability involving with the KH mode and the guided jet mode~\citep{Nogueira2021}. Even the presence of nozzles was deemed secondary in this theory, although many studies have shown the importance of nozzle and upstream reflecting surfaces in screech dynamics. In contrast, this paper provides an explanation as to why the twin-jet coupling is realized in the way that is observed in the experimentally validated numerical data. Recent work by~\citet{Stahl2022} and \citet{Webb2023} have shown that modelling each jet and their coupling can describe the flow resonance by twin-jet systems.

Throughout the paper, we use the terms in-phase/out-of-phase to describe coupling between the two jets (cross-excitation), marked by their relative phase. To avoid any confusion, the terms antisymmetric/symmetric are reserved for discussing each jet's response. Also note that we will denote the upstream-propagating guided jet mode by $k_{-}$, the upstream-propagating free-stream acoustic mode by $c_{-}$, and the downstream-propagating KH mode by $k_{+}$, following the same definition in~\citet{Wu2020,Wu2023}.

\section{Intermittent screech tones}
\label{sec:intermittency}
Time-frequency analysis demonstrated that the screech tones we herein consider are indeed intermittent~\citep{Jeun2022}. To investigate why screech tones appear to be irregular in time, the instantaneous phase difference between the two jets is extracted using the Hilbert transform. For a given signal $x(t)$, the Hilbert transform $\widetilde{x}(t) = \mathcal{H}[x(t)]$ is computed as
\begin{equation}
    \widetilde{x}(t) = \mathcal{H}[x(t)] = x(t) \otimes \frac{1}{\upi t},
\end{equation}
where $\otimes$ denotes the convolution operator. From this, an analytic function of the original signal $\widetilde{z}(t)$ can be defined as
\begin{equation}
    \widetilde{z}(t) = x(t) + \mathrm{i}\widetilde{x}(t),
    \label{eq:analytic_fn}
\end{equation}
where $\mathrm{i} = \sqrt{-1}$. In polar form, \eqref{eq:analytic_fn} can be rewritten as
\begin{equation}
    \widetilde{z}(t) = a(t) \mathrm{exp}\left[\mathrm{i}\theta(t)\right],
\end{equation}
where 
\begin{equation}
    \theta(t) = \arctan{ \left[ \frac{\widetilde{x}(t)}{x(t)} \right] }
\end{equation}
represents the instantaneous phase. Finally, the phase difference between the two jet signals is expressed as
\begin{equation}
    \Delta \theta(t) = \theta_{1}(t) - \theta_{2}(t), 
\end{equation}
where the subscripts 1 and 2 represent the jets centered at $z/h$ = 1.75 and -1.75, respectively.

\begin{figure}
  \centering
  \begin{tabular}{c}
    \includegraphics[width=0.7\textwidth]{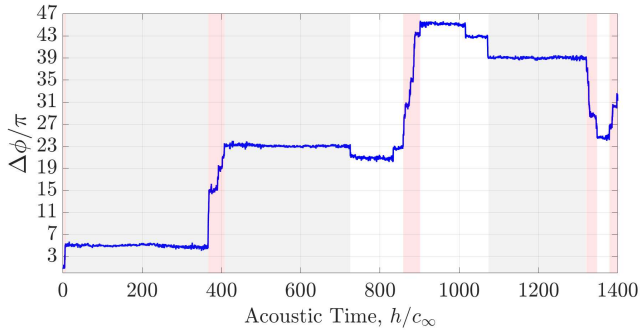} \\
  \end{tabular}
\caption{Phase differences between the two jet signals, recovered by the Hilbert transform.}
\label{fig:inst_phase_diff}
\end{figure}

\begin{figure}
  \centering
  \begin{tabular}{cc}
    \includegraphics[width=0.48\textwidth]{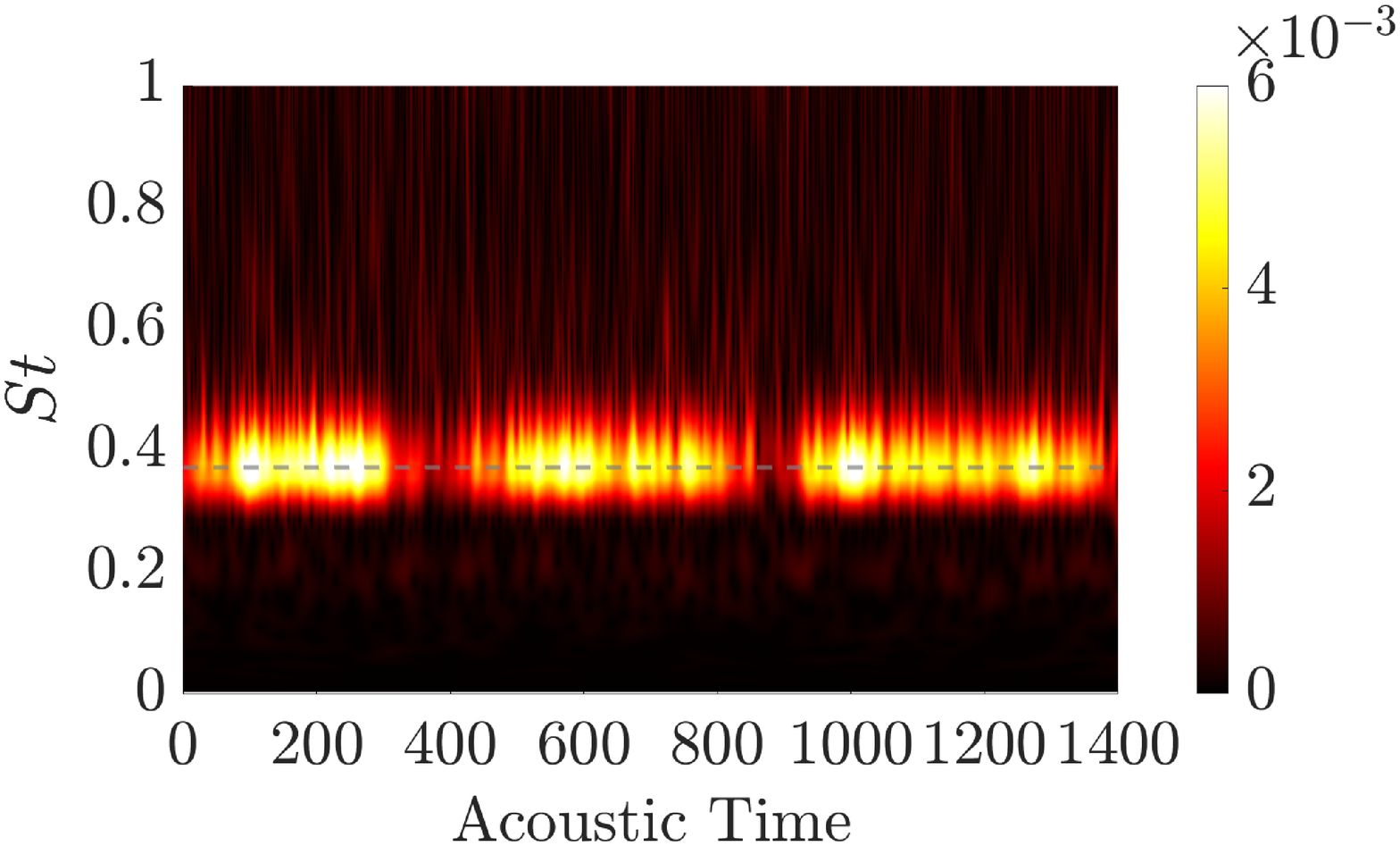} &
    \includegraphics[width=0.48\textwidth]{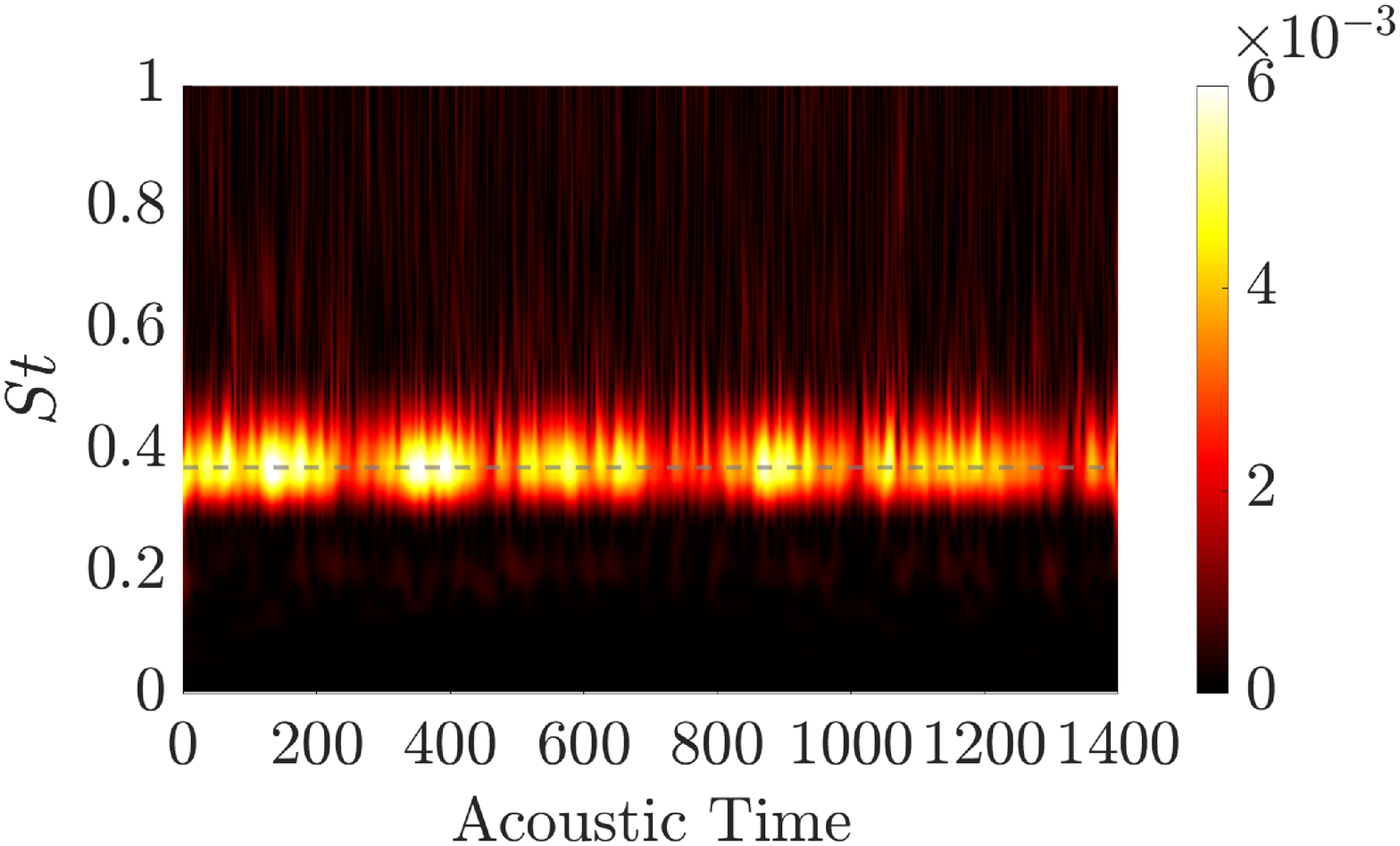} \\
    (a) & (b) \\
  \end{tabular}
\caption{Scalograms of the acoustic signals for (a) Jet 1 and (b) Jet 2. Grey dashed lines represent the screech frequency at $St$ = 0.37. Reproduced from~\citet{Jeun2022}.}
\label{fig:scalograms}
\end{figure}

To characterize screech coupling, pressure disturbances in each jet are measured just above the corresponding nozzle exit ($(x/h,y/h)$ = $(0,1)$). The resulting phase difference between the two signals is shown in figure~\ref{fig:inst_phase_diff} as a function of time. By recalling the scalograms of the corresponding signals as shown in figure~\ref{fig:scalograms}, one may notice that the phase difference varies rapidly when screech amplitudes are observed to change in time. Overall, the phase difference represents predominant out-of-phase coupling (odd multiples of $\upi$), which appears as multiple wide plateaus that are bridged by irregular switches between odd and even multiples of $\upi$.

To explain the intermittency in coupling, the pressure field $p'(x,y,z,t)$ is decomposed into antisymmetric and symmetric parts using the $D_2$ decomposition proposed by \citet{Yeung2022}. In the group theory, the dihedral group $D_2$ is a set of the symmetries of a rectangle, including reflections across two axes. Given the two-way symmetry along the minor and major axes, our jets also belong to $D_2$. Through the $D_2$ decomposition, four symmetry components are permitted in the rectangular twin jets as
\begin{equation}
  p'_{SS}(x,y,z,t) = \frac{1}{4} [p'(x,y,z,t) + p'(x,-y,z,t) + p'(x,y,-z,t) + p'(x,-y,-z,t)],   
  \label{eq:p_ss}
\end{equation}
\begin{equation}
  p'_{SA}(x,y,z,t) = \frac{1}{4}[p'(x,y,z,t) + p'(x,-y,z,t) - p'(x,y,-z,t) - p'(x,-y,-z,t)],   
\end{equation}
\begin{equation}
  p'_{AS}(x,y,z,t) = \frac{1}{4}[p'(x,y,z,t) - p'(x,-y,z,t) + p'(x,y,-z,t) - p'(x,-y,-z,t)],   
\end{equation}
and
\begin{equation}
  p'_{AA}(x,y,z,t) = \frac{1}{4}[p'(x,y,z,t) - p'(x,-y,z,t) - p'(x,y,-z,t) + p'(x,-y,-z,t)].   
  \label{eq:p_aa}
\end{equation}
Note that $p'(x,y,z,t)$ = $p'_{SS}(x,y,z,t) + p'_{SA}(x,y,z,t) + p'_{AS}(x,y,z,t) + p'_{AA}(x,y,z,t)$. In \eqref{eq:p_ss}-\eqref{eq:p_aa} the first subscript denotes the antisymmetry (A) or symmetry (S) about the major axis ($y/h$ = 0), while the second subscript pertains to the center axis ($z/h$ = 0). This notation aligns with the nomenclature suggested by~\citep{Rodriguez2018}. In this way each symmetry component has the same information across all four quadrants of the $yz$-plane. Thus, it is sufficient to consider only the symmetry components of the Jet 1 signals, provided the stationarity of the LES data.

\begin{figure}
  \centering
  \begin{tabular}{c}   
  \includegraphics[width=0.7\textwidth]{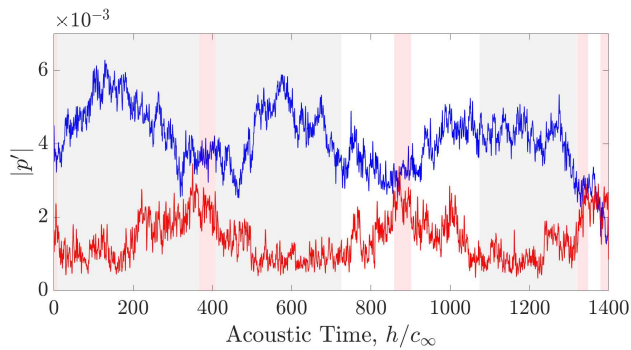} \\
  \end{tabular}
\caption{Instantaneous amplitudes of the antisymmetric components: blue, $p'_{AA}$; red, $p'_{AS}$. The symmetric components $p'_{SA}$ and $p'_{SS}$ exhibit much weaker amplitudes compared to the antisymmetric components and are omitted.}
\label{fig:inst_amp}
\end{figure}

Figure~\ref{fig:inst_amp} shows a comparison of the instantaneous amplitudes of the four symmetry modes measured at $(x/h,y/h,z/h) = (0,1,1.75)$. It is noteworthy that the two symmetric components about the major axis $p'_{SA}$ and $p'_{SS}$ exhibit weaker amplitudes compared to the antisymmetric components $p'_{AA}$ and $p'_{AS}$ and are omitted in this figure. This observation aligns with the fact that the rectangular jet screech is typically associated with the intense flapping (antisymmetric) motion along the minor axis. Large amplitudes of the two antisymmetric components $p'_{AA}$ and $p'_{AS}$ also agree well with the SPOD conducted by~\citet{Yeung2022}. They demonstrated that the eigenspectra of these two components have tonal peaks at the screech frequency, while the two symmetric components about $y$ = 0 are damped. That is, in our jets $p'_{AA}$ and $p'_{AS}$ can be used as representatives of the out-of-phase and in-phase coupling modes, respectively. Throughout most of the time, the dominance by the out-of-phase coupling is evident (grey shaded region); however, when the phase-locking between the two jets disrupted (red shaded region), the two modes appear to be nearly equally strong. In other words, the interruption of screech tones in our twin rectangular jets seems to be linked to a competition between the out-of-phase and in-phase coupling of the two jets, analogous to the behavior of intermittent tones in underexpanded round twin jets~\citep{Bell2021}. 

In this regard, a complete model for this type of screech feedback loop should be able to include the unsteady phase relationship between the two jets. Needless to say that constructing such a model is not obvious. As a first step forward, the present work considers a simplified model problem based on the assumption of perfect phase-locking of the twin jets. A way to neglect the effects of intermittency is to select the LES data over periods where the jet-to-jet coupling is perfectly phase-locked only.

\section{Dominant coherent structures in screech generation}
\label{sec:spod_modes}
\subsection{Ensemble averaged SPOD modes}
In the previous section we show that screech tones in the twin rectangular jets are intermittent including long duration when the jets are preferably coupled out-of-phase with each other. To clarify the role of coupling modes in modelling the screech feedback loop, dominant coherent structures in screech generation are extracted by retaining data windows where the jets are synchronised to out-of-phase coupling only.

With the requirement of perfect phase-locking, three different portions of the original LES data, which respectively correspond to the data collected over acoustic time = [5.5,366], [407,725], and [1074,1320] (grey shaded area in figure~\ref{fig:inst_phase_diff}), are selected. SPOD is subsequently performed for each partitioned data, resulting in three estimates of the leading SPOD modes for the two jets. Here, to explicitly include their spatial correlation, flow snapshots of the two jets are incorporated into a single data matrix. For each jet, an ensemble average of the three estimates of the leading SPOD modes is computed. Despite the truncation, all three mode estimates computed from each partition and their average exhibit flow structures highly similar to those observed in the modes based on the entire simulation data, with perfect out-of-phase coupling between the twin jets~\citep{Jeun2022}. The original SPOD modes may appear to be more rigorous, albeit at the cost of including the influence of the unsteady parts. However, in line with the assumption of perfect phase-locking (as will be addressed in~\eqref{eq:cross_excitation_phase_criteria}), we introduce the ensemble averaged SPOD modes to minimize the influence of the unsteady part as much as possible. By considering partitioned data, the convergence of the resulting SPOD modes becomes questionable. To assess the convergence of the resulting SPOD modes for each partition, we repeated SPOD by varying the number of snapshots per block. It was found that a choice of 1,888 snapshots per block with taking a 75\% overlap with the next block was deemed to result in sufficiently converged modes. Therefore, the use of ensemble averaged modes is considered acceptable, ensuring the stationarity of the partitioned data while minimizing the unsteady effects.

Due to the truncation, the frequency resolution for each segment is reduced to $\Delta St$ = 0.007. Nevertheless, the length of each partition is still long enough to recover sufficiently narrow screech tones. Depending its length, each partition is split into 2-4 blocks that are windowed by a Hann function with an overlap of 75\% to each other such that the length of each realization is given by one-third of the desired number of snapshots that needs to match the experimental frequency resolution. In this way, despite the reduced frequency resolution, SPOD modes can be computed exactly at the (measured) fundamental screech frequency, while still utilizing averaging several realizations for a more accurate estimate.

\begin{figure}
  \centering
  \begin{tabular}{cc}
    \includegraphics[width=0.48\textwidth]{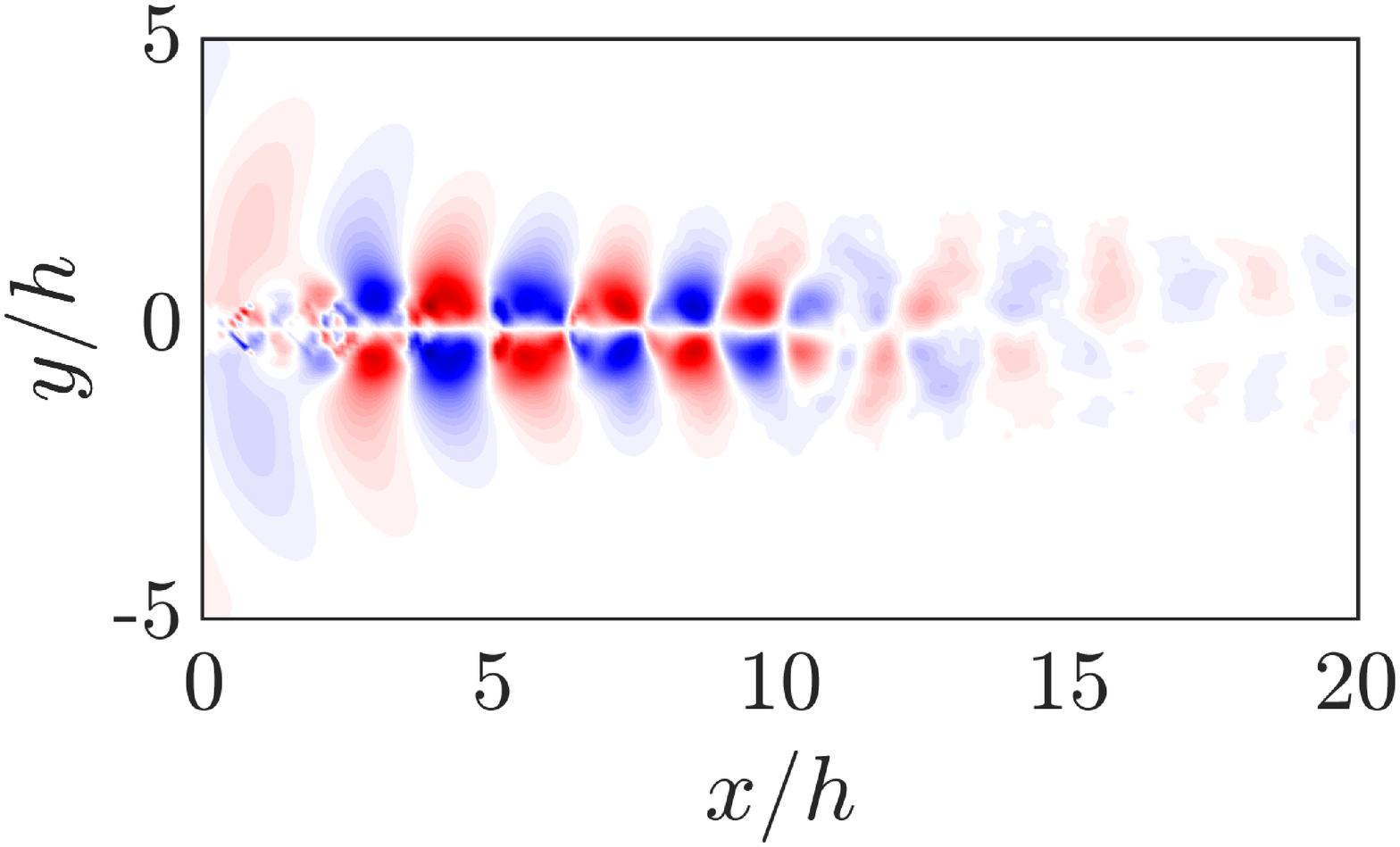} & 
    \includegraphics[width=0.48\textwidth]{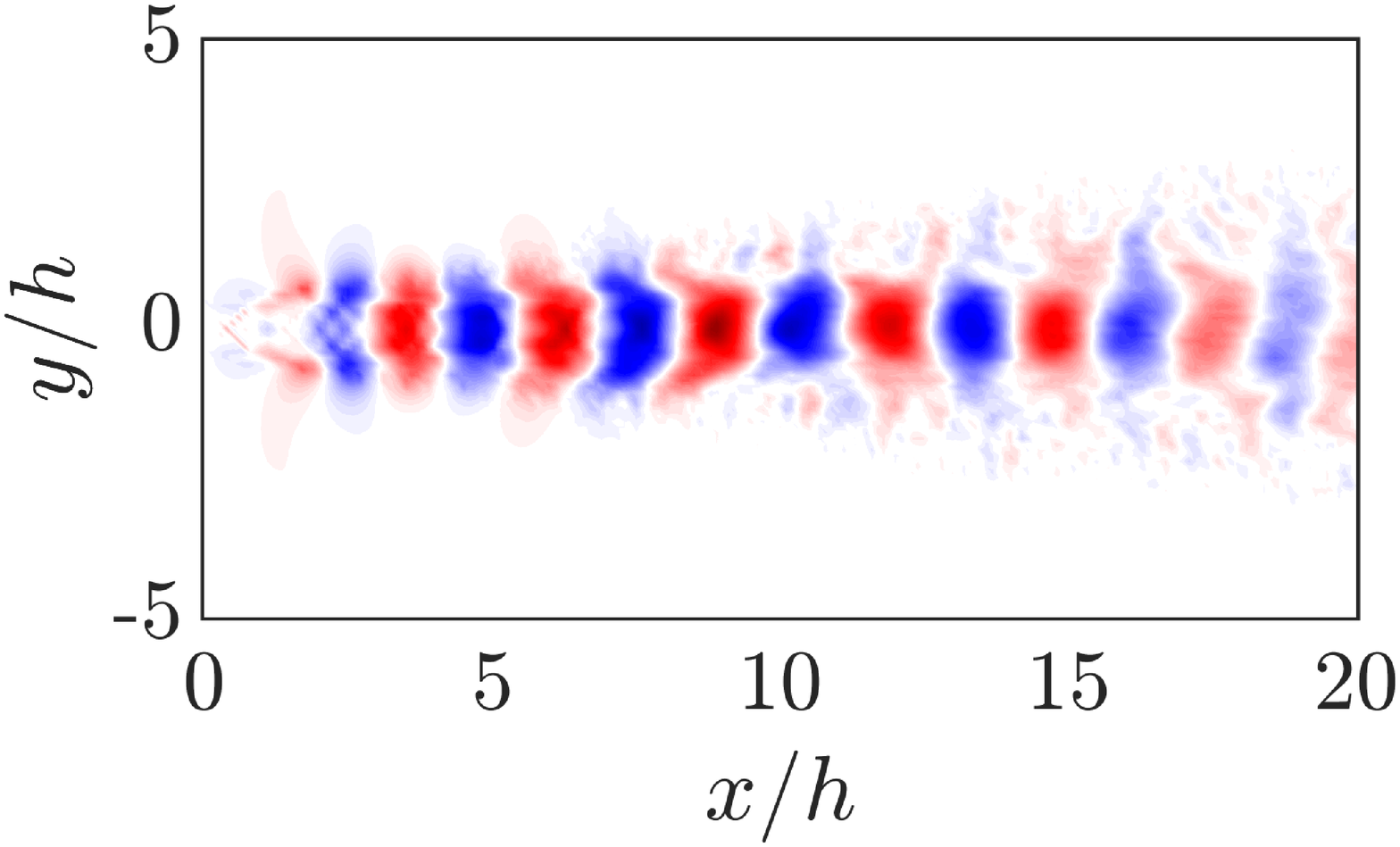} \\
    (a) & (b) \\
    \includegraphics[width=0.48\textwidth]{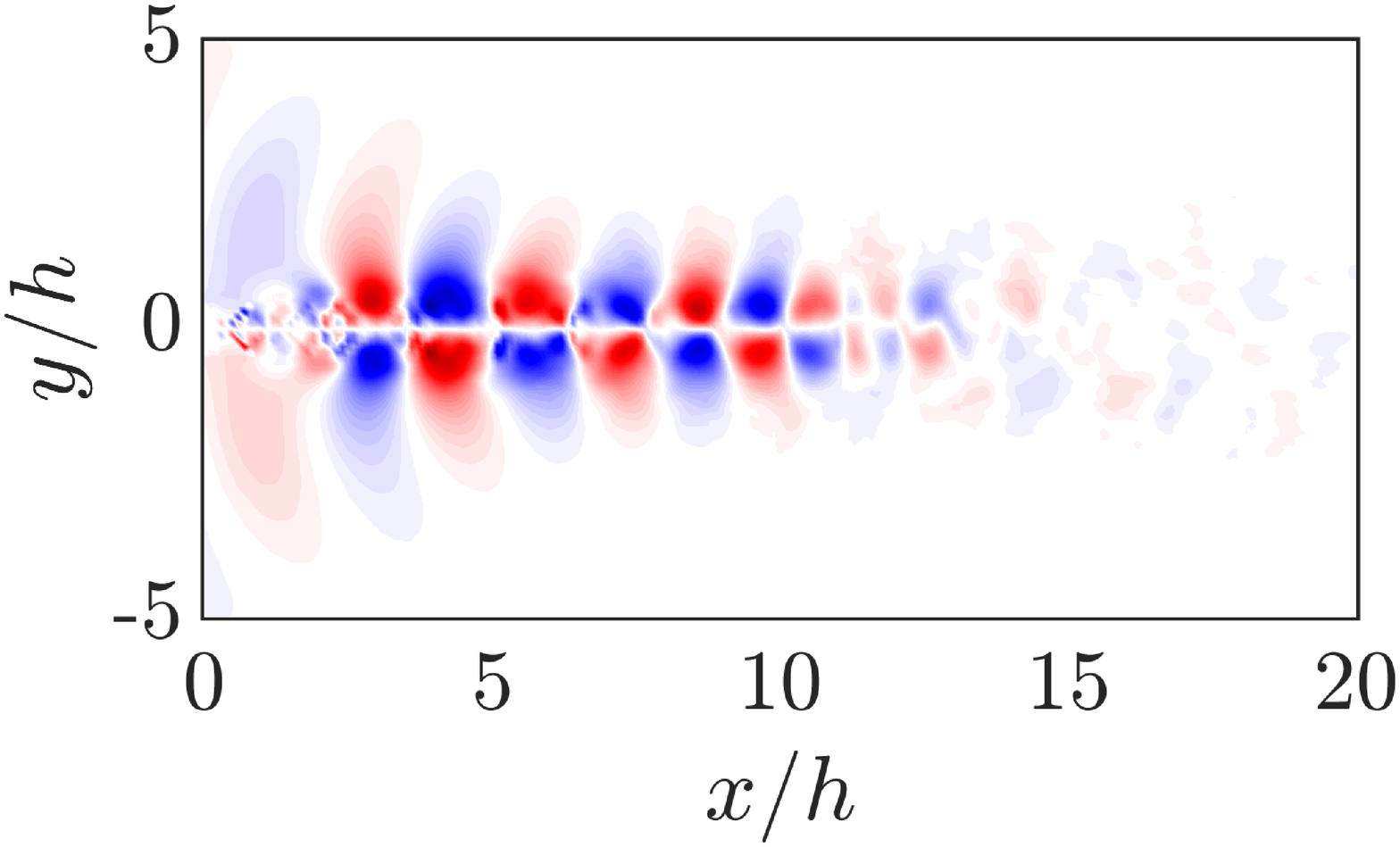} &
    \includegraphics[width=0.48\textwidth]{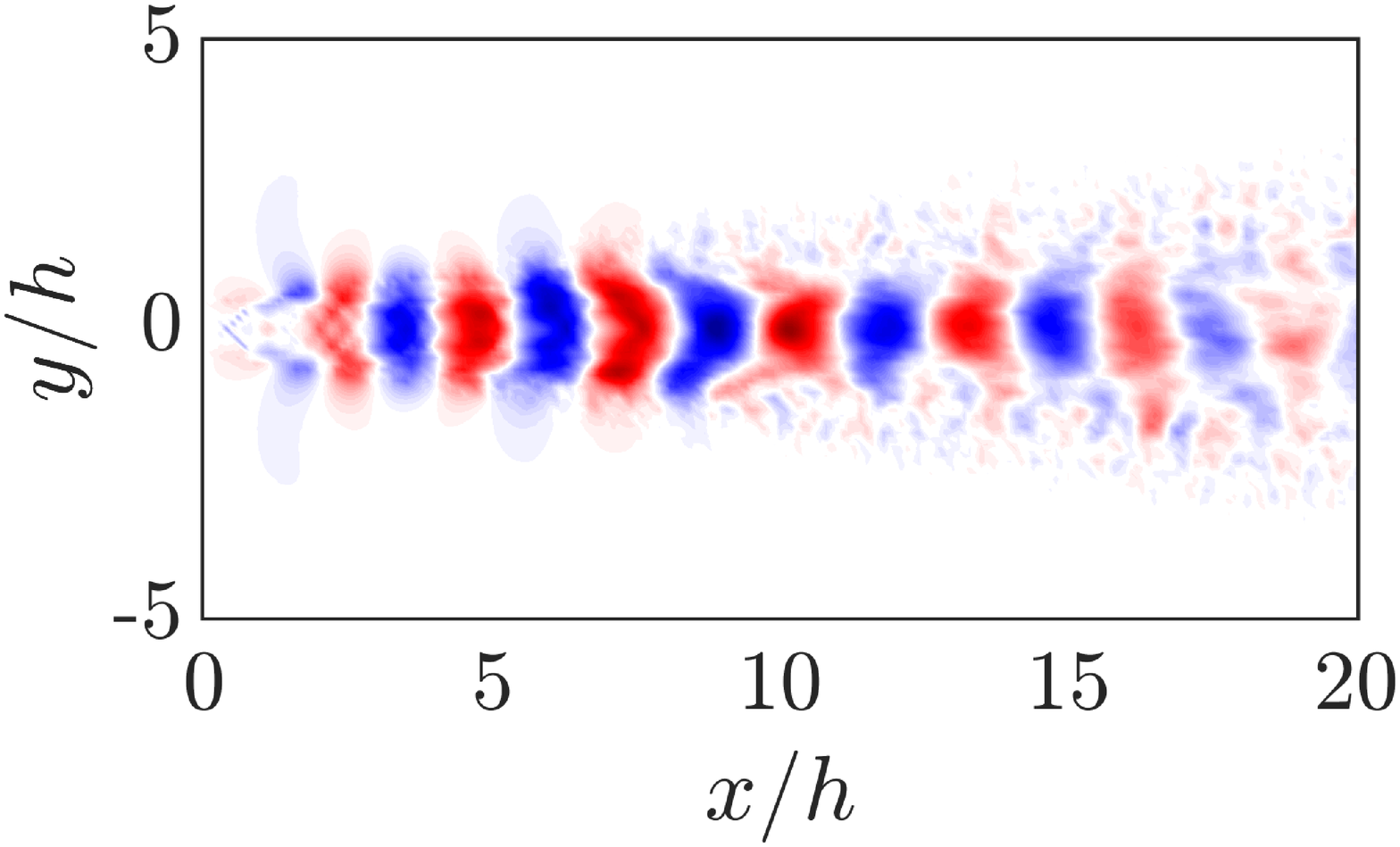} \\
    (c) & (d) \\
  \end{tabular}  
  \caption{Ensemble averaged leading SPOD modes for Jet 1 (a,b) and Jet 2 (c,d). Mode shapes are visualized by real part of the leading SPOD mode for the pressure field (a,c) and for the transverse velocity field (b,d). Each contour is normalised by its maximum value. The colour ranges from -1 to 1.}
\label{fig:ensemble_averaged_modes}
\end{figure}

Figure~\ref{fig:ensemble_averaged_modes} shows the ensemble averaged leading SPOD modes for both the pressure and transverse velocity components. Due to the exclusion of unsynchronised segments, these modes have undergone a slight phase shift from the modes depicted in figure~\ref{fig:blind_spod_xy_modes}, which were derived from the flow data spanning the entire simulation. Notwithstanding this shift, the coherent flow structures associated with the screech remain intact in the ensemble averaged modes.

\subsection{Identification of the guided jet mode, free-stream acoustic mode, and KH mode via a streamwise Fourier decomposition of the SPOD modes}
To extract the wave components at play in the screech feedback, the ensemble averaged SPOD modes are further decomposed into upstream- and downstream-propagating components based on their wavenumber in $x$~\citep{Edgington-Mitchell2018,Wu2020,Wu2023}. Here, the direction of the group velocity of a wave is determined by examining the direction of the phase velocity ($u_p$ = $\omega / k_x$ at a certain frequency $\omega$) as a proxy for it. More specifically, this involves considering the sign of the streamwise wavenumber $k_x$ at the screech frequency $\omega$ = $\omega_{sc}$.

\begin{figure}
  \centering
  \begin{tabular}{cc}
    \includegraphics[width=0.48\textwidth]{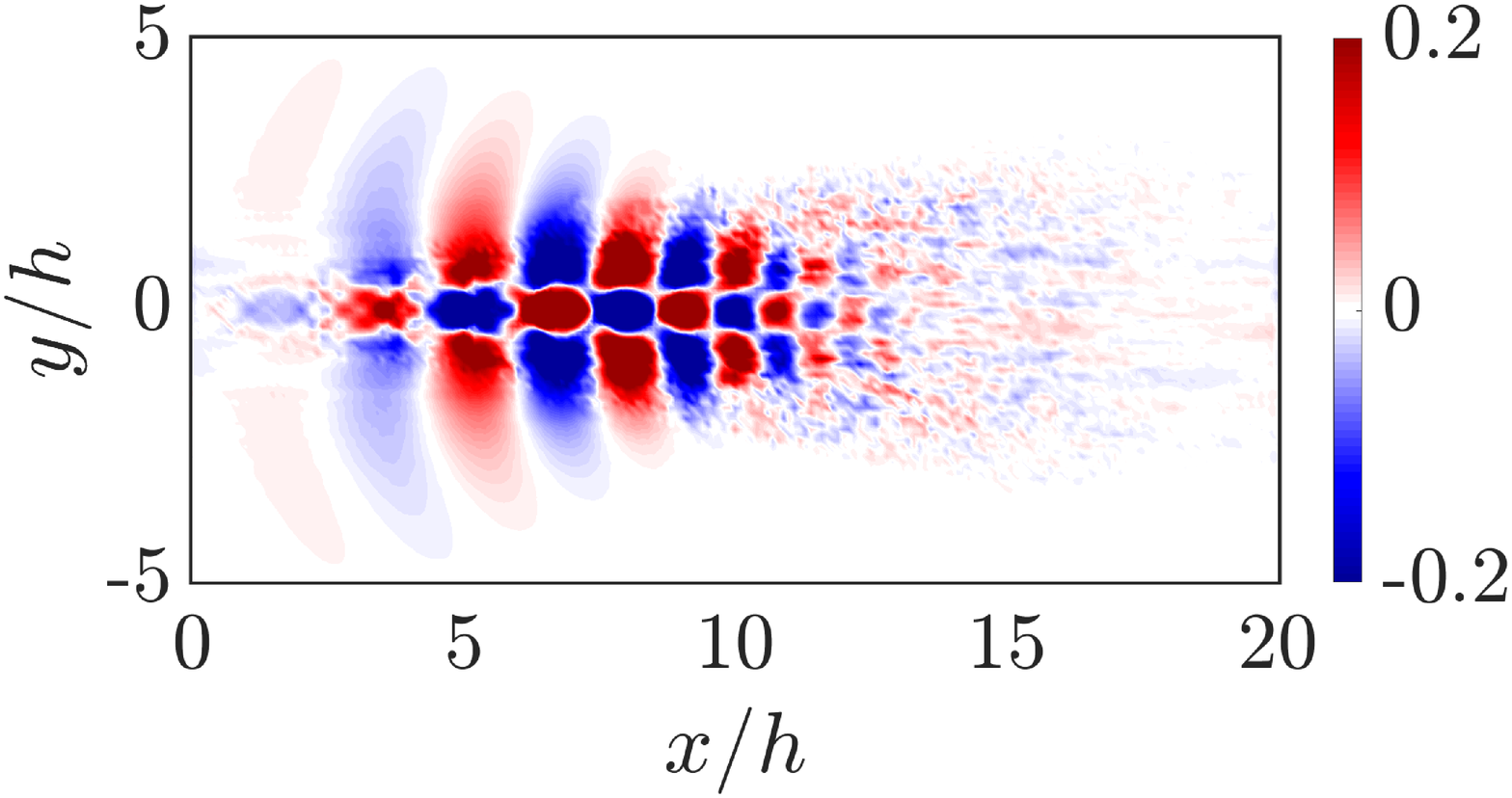} &
    \includegraphics[width=0.48\textwidth]{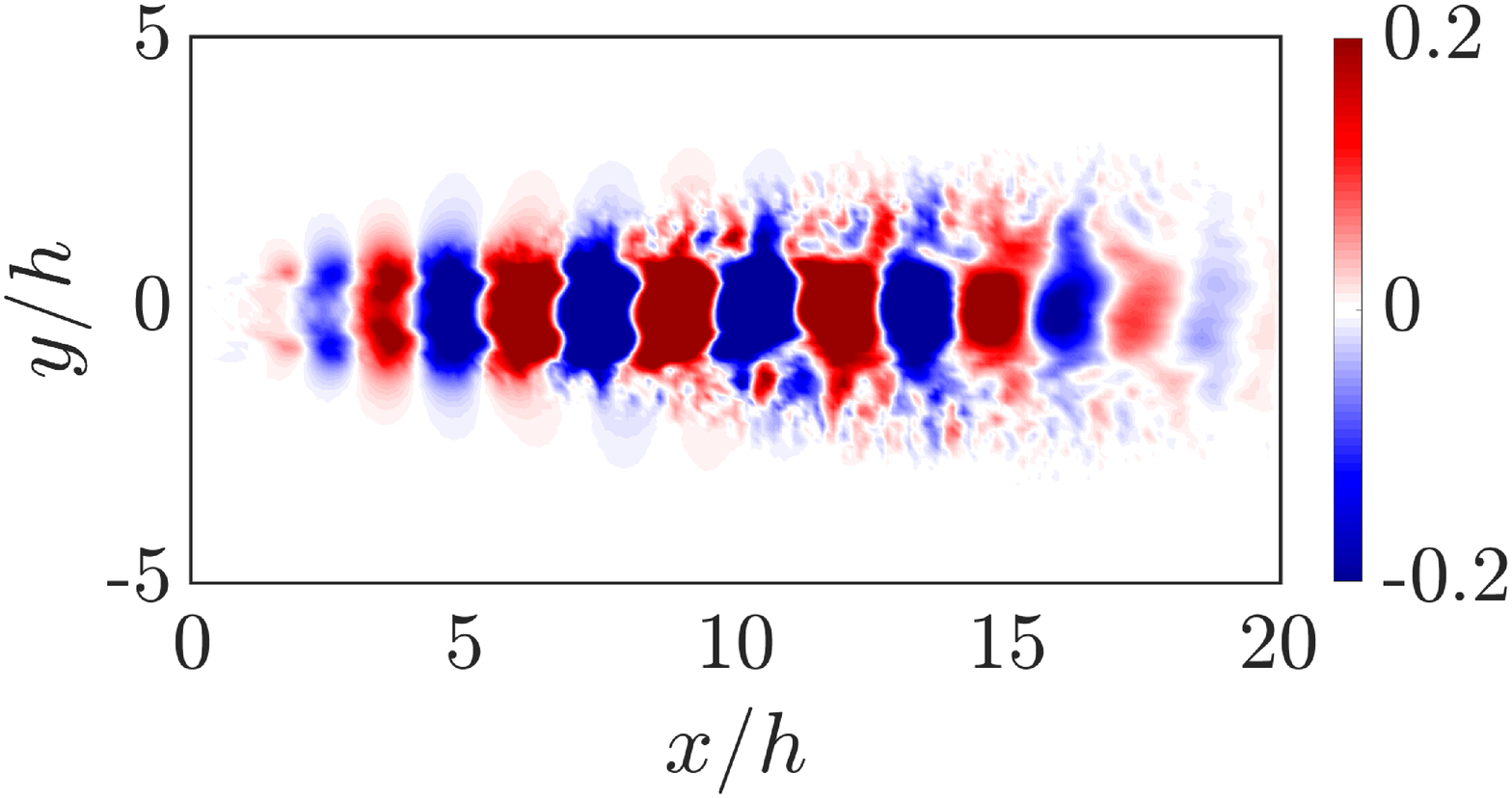} \\
    (a) & (b) \\
    \includegraphics[width=0.48\textwidth]{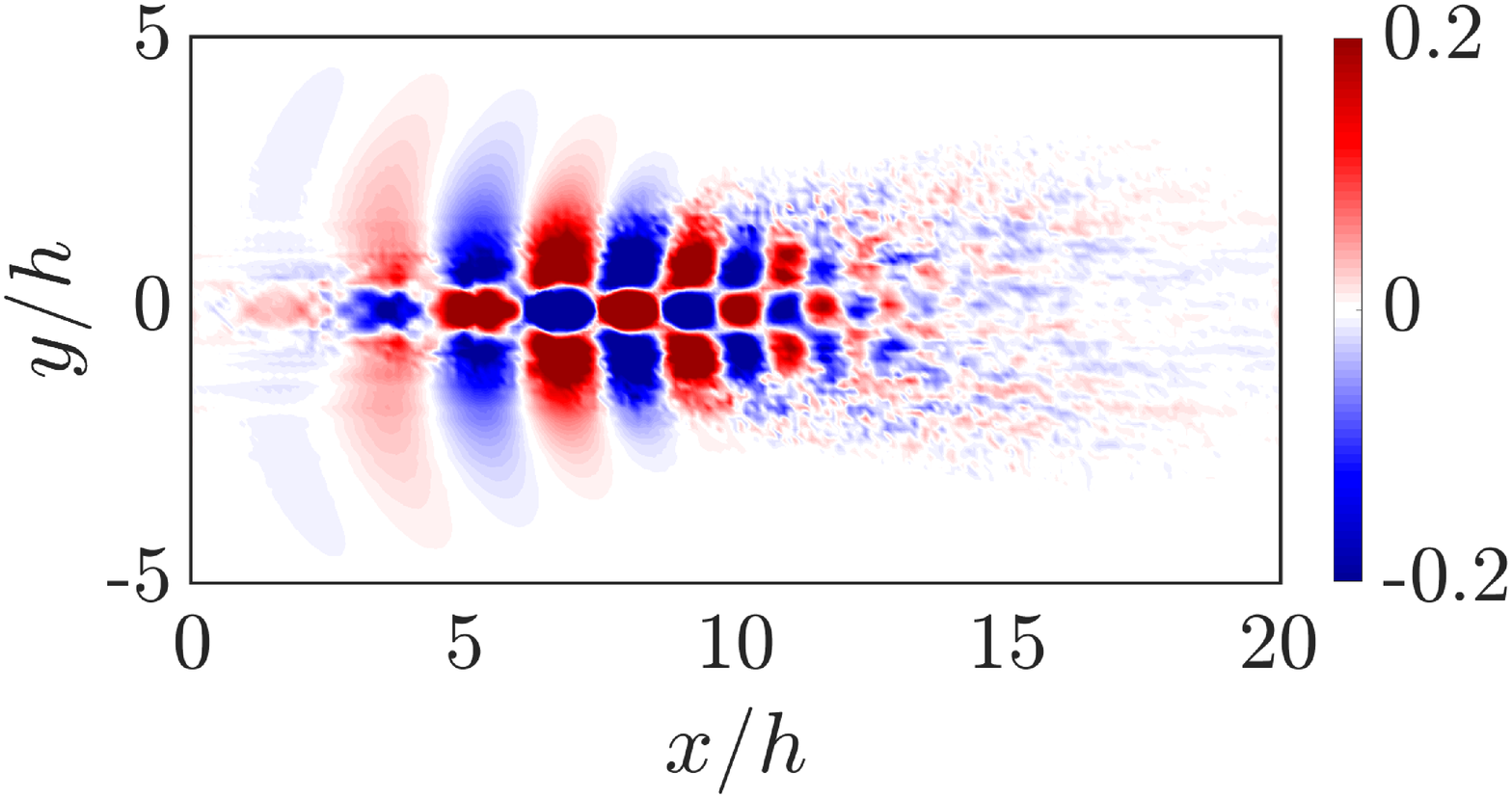} &
    \includegraphics[width=0.48\textwidth]{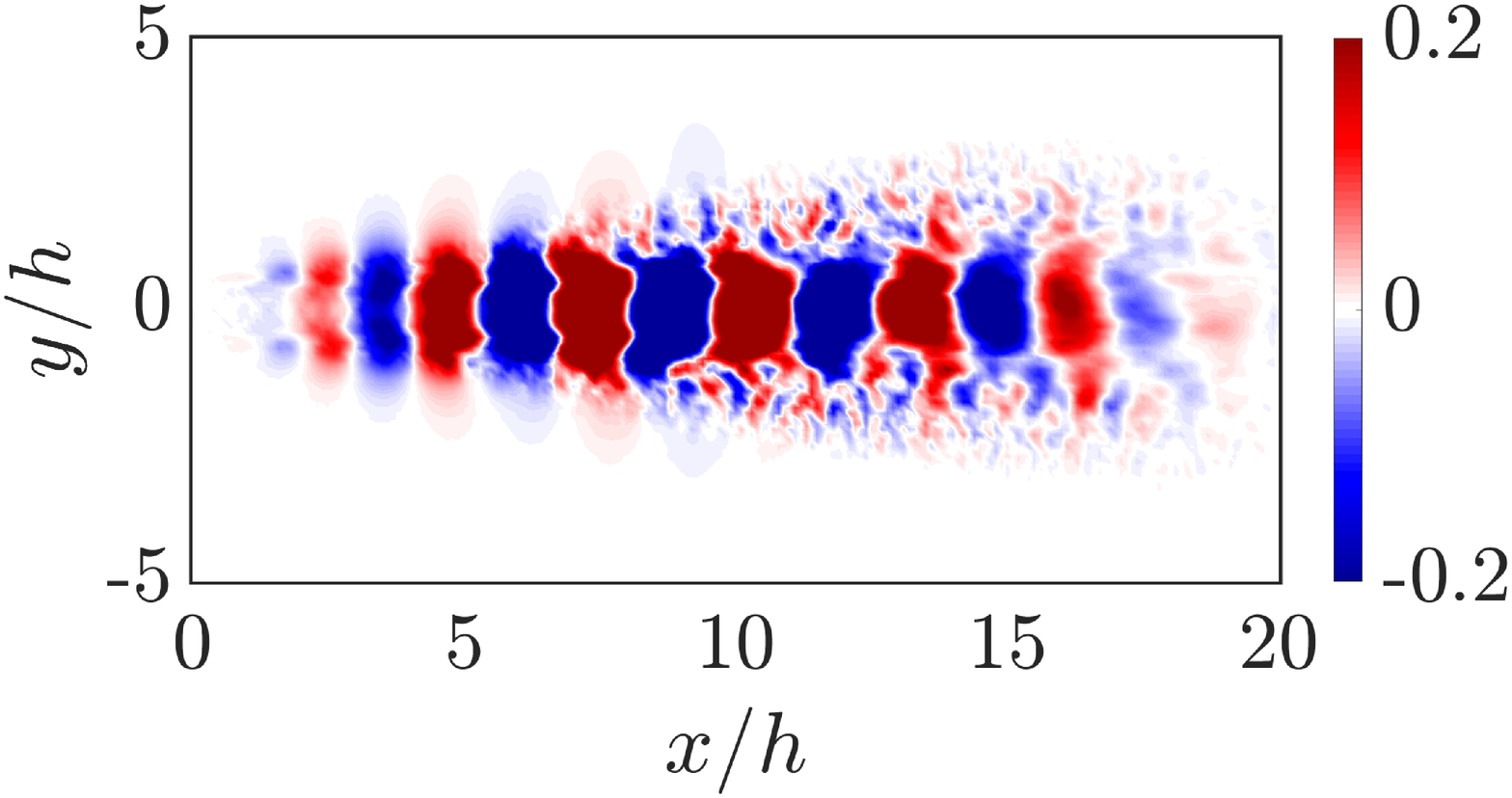} \\
    (c) & (d) \\
  \end{tabular} 
  \caption{Decomposition of the ensemble averaged leading SPOD modes for the transverse velocity fluctuations into the (a,c) upstream- and (b,d) downstream-propagating components. (a,b) Jet 1; (c,d) Jet 2.}
\label{fig:v_spod_decomposition}
\end{figure}

Figure~\ref{fig:v_spod_decomposition} visualizes the resulting isolated wave components for the transverse velocity fluctuations $v'$. Note that the upstream- and downstream-propagating components show out-of-phase coupling between the two jets at the screech frequency. The upstream-propagating modes include structures confined within the jet and outside of it, which resemble the mode first identified by~\citet{Tam1989}. The downstream-propagating modes predominantly correspond to the KH instability wavepackets, and they are used to represent the $k_{+}$ mode.

\begin{figure}
  \centering
  \begin{tabular}{c}
    \includegraphics[width=0.55\textwidth]{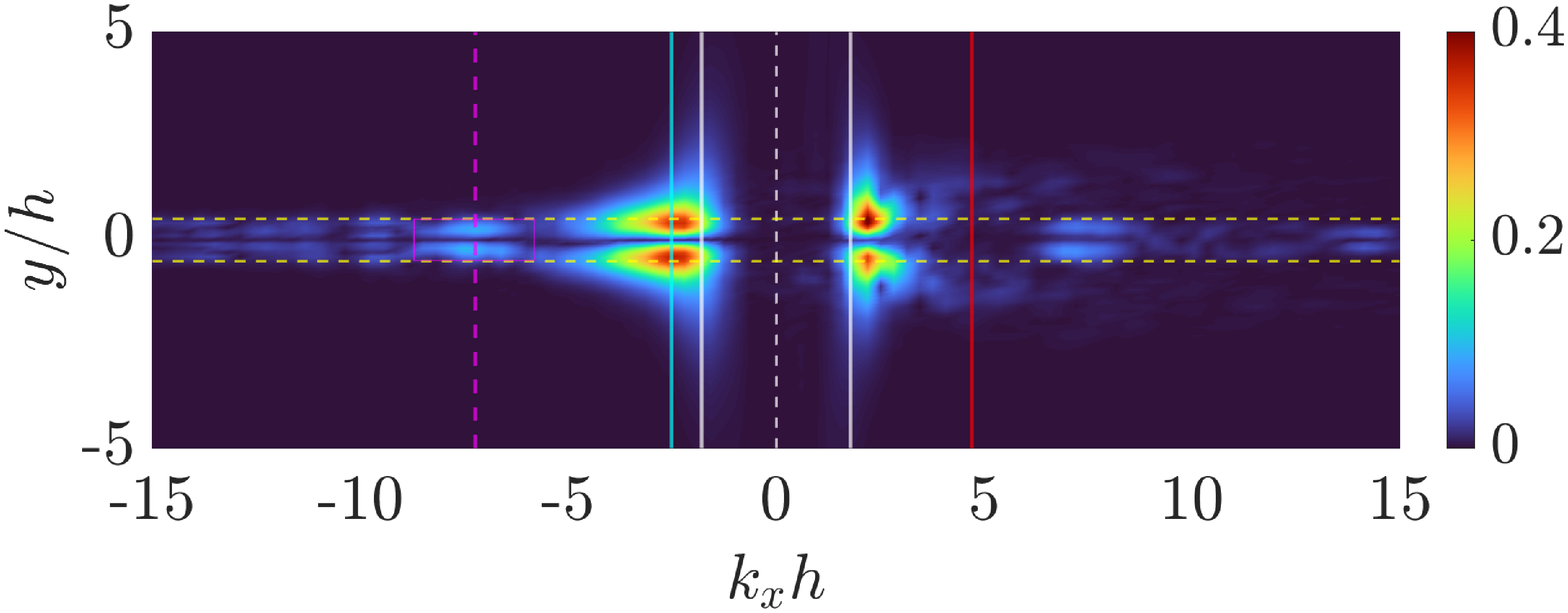} \\
    (a) \\
    \includegraphics[width=0.55\textwidth]{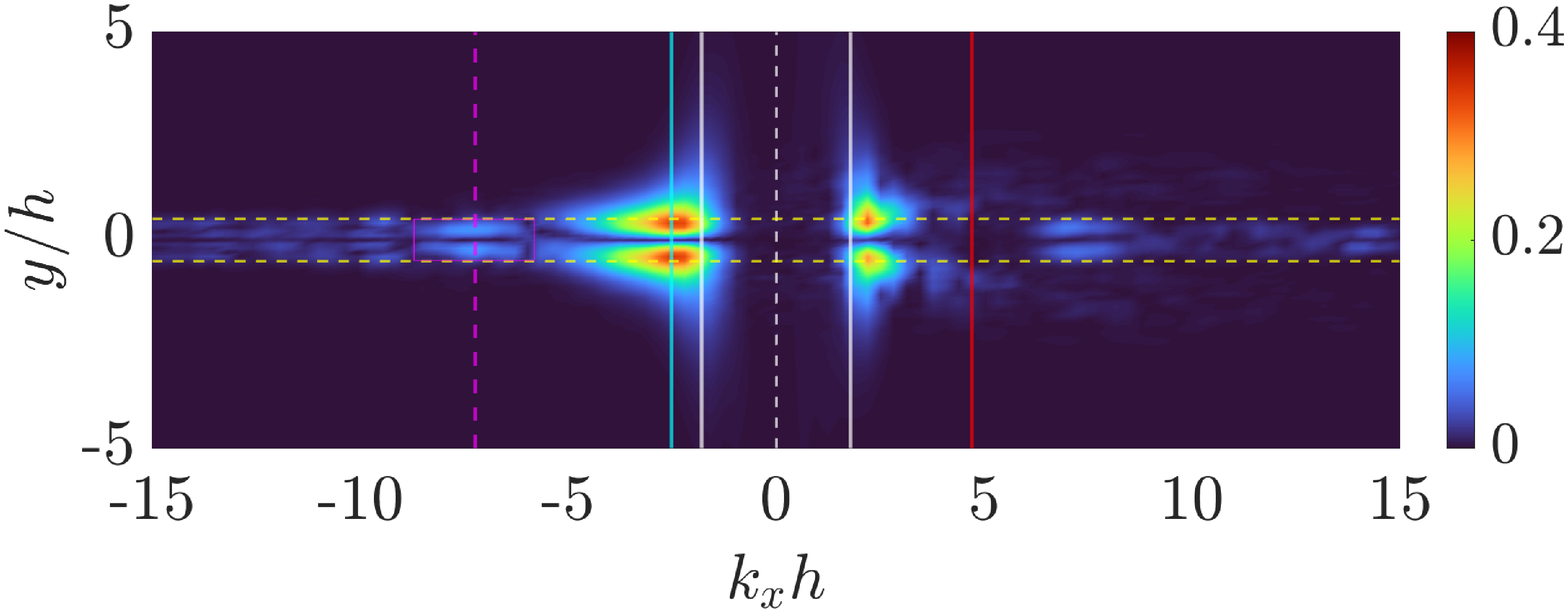} \\
    (b) \\
    \includegraphics[width=0.55\textwidth]{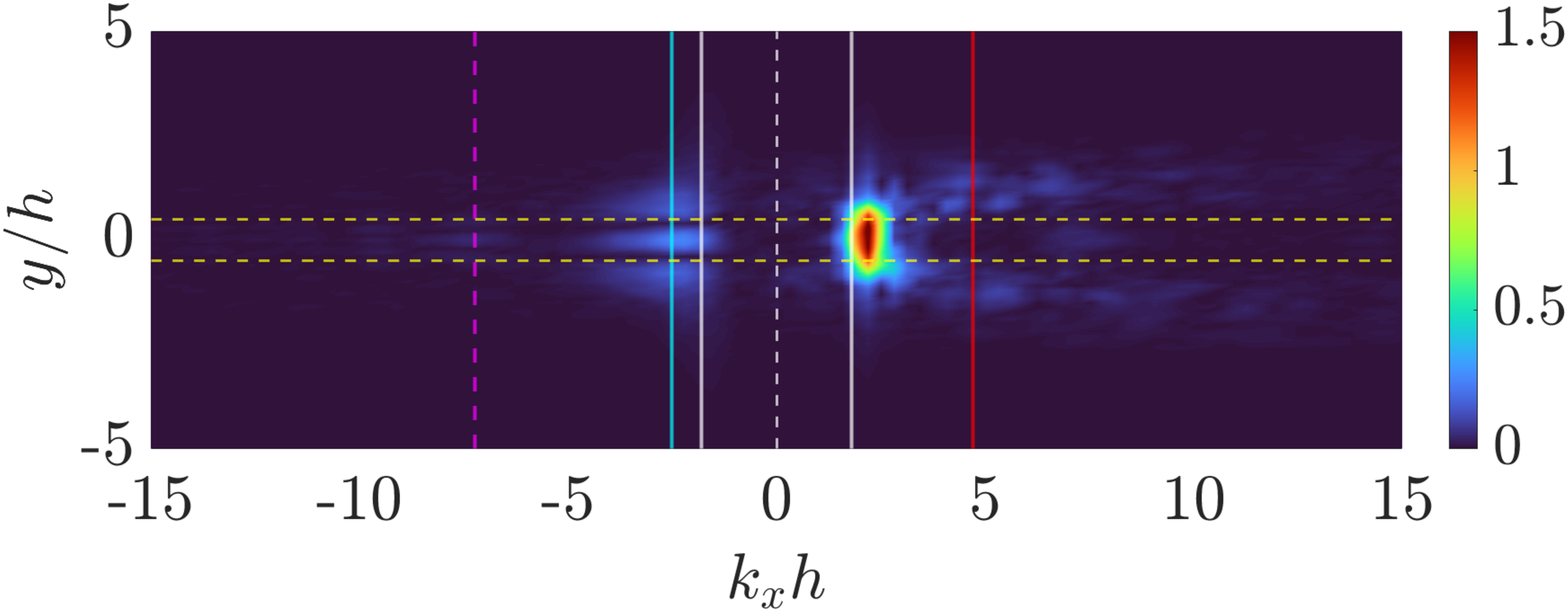} \\
    (c) \\
    \includegraphics[width=0.55\textwidth]{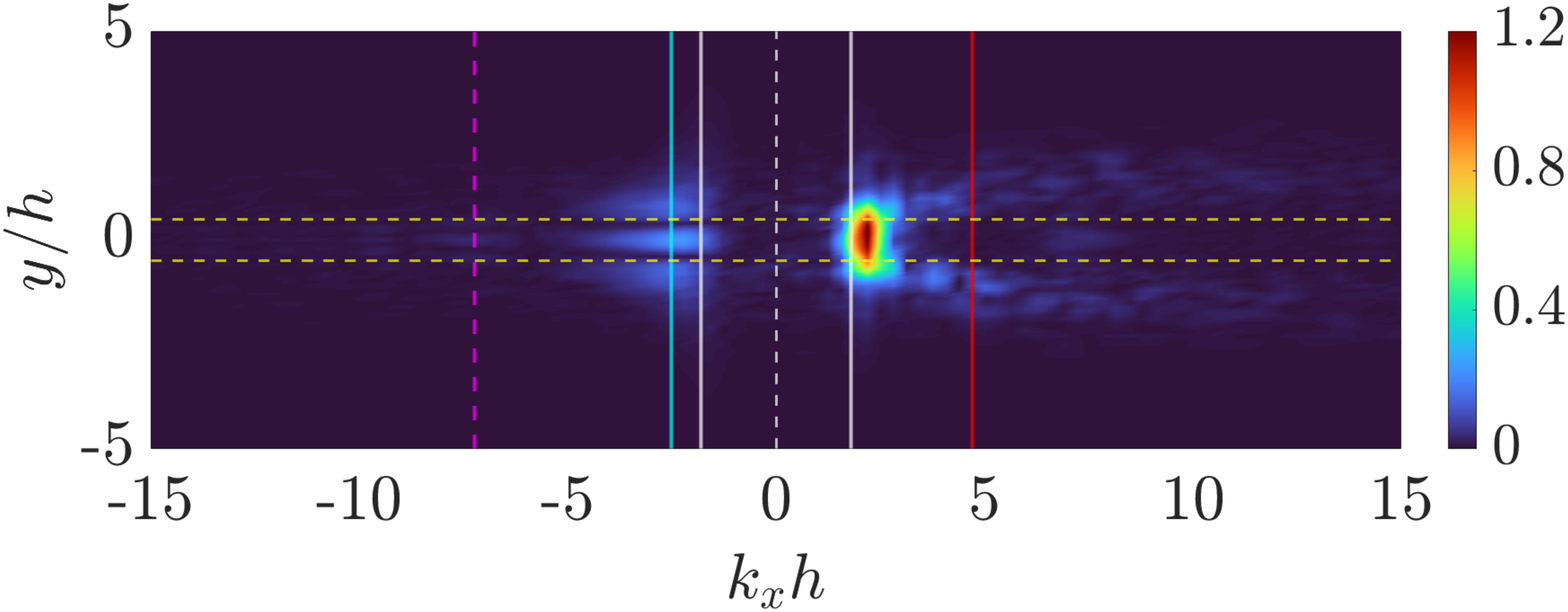} \\
    (d) \\
  \end{tabular}  
  \caption{Streamwise wavenumber spectra visualized by the modulus of the ensemble averaged leading SPOD mode shape for pressure fluctuations (a,b) and transverse velocity fluctuations (c,d). (a,c) Jet 1; (b,d) Jet 2. Cyan solid line, $k_{+,max}h - k_{s_1}h$; magenta dashed line, $k_{+,max}h - k_{s_2}h$; red solid line, $k_{s_1}h$; white solid lines, $\pm k_{c_{\infty}}h$; white dashed line, zero axis; yellow horizontal lines, $y/h$ = $\pm$0.5.}
\label{fig:wavenumber_spectra}
\end{figure}

The primary objective of this study is to explore the guided jet mode and the free-stream acoustic mode as potential closure mechanisms for the twin-jet screech coupling. This necessitates a demarcation between these two upstream-propagating modes, which are characterised by slightly different phase velocities. In this context we analyse the streamwise wavenumber spectra for more detailed insights. Figure~\ref{fig:wavenumber_spectra} displays the wavenumber spectra visualized by the modulus of the SPOD mode at the screech frequency. The contour plots unveil distinct properties of both the upstream- and downstream-propagating waves. Each of these waves comprises a wide range of Fourier modes, forming unique patterns. The downstream-propagating waves display a prominent band with high modulus, centered around $k_x h$ = 2.20. This aligns with a phase velocity of roughly 0.7$U_j$ and is closely related to the KH instability wavepackets. The upstream-propagating components, on the other hand, are characterised by a predominant band close to the wavenumber associated with the free-stream speed of sound, $c_\infty$, alongside several lobes with relatively low energy concentrations at much lower wavenumbers. For $k_x h < 0$, the peak moduli are observed at $k_x h$ = -2.51 for the fluctuating pressure and $k_x h$ = -2.20 for the fluctuating transverse velocity, respectively, which are slightly lower than the wavenumber corresponding to the free-stream speed of sound, $k_{c_\infty}h$ = $\pm$1.80 (white solid lines). The difference in peak wavenumbers between the two flow variables aligns with the uncertainty stemming from the constrained computational domain used in this study ($\Delta k_x h$ = $2\pi/20$), hence it is not considered particularly significant. The dominant energy blob exhibits significant support in the transverse direction within the jet, extending even beyond the jet shear layers. Meanwhile, the secondary lobes (denoted by magenta boxes in figure~\ref{fig:wavenumber_spectra}(a,b)) appear fully confined within the jet.

\begin{figure}
  \centering
  \begin{tabular}{cc}
    \includegraphics[width=0.45\textwidth]{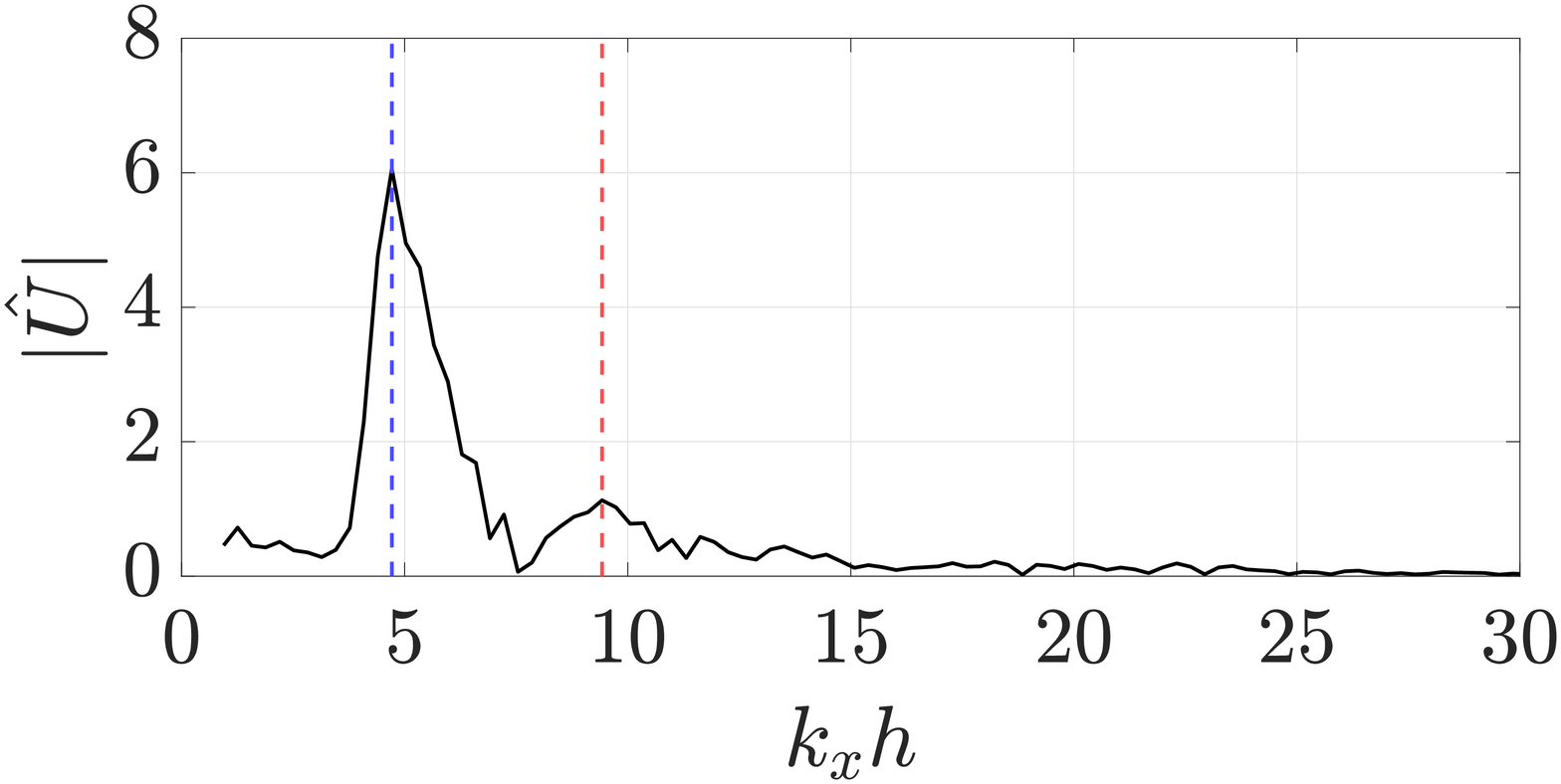} &
    \includegraphics[width=0.45\textwidth]{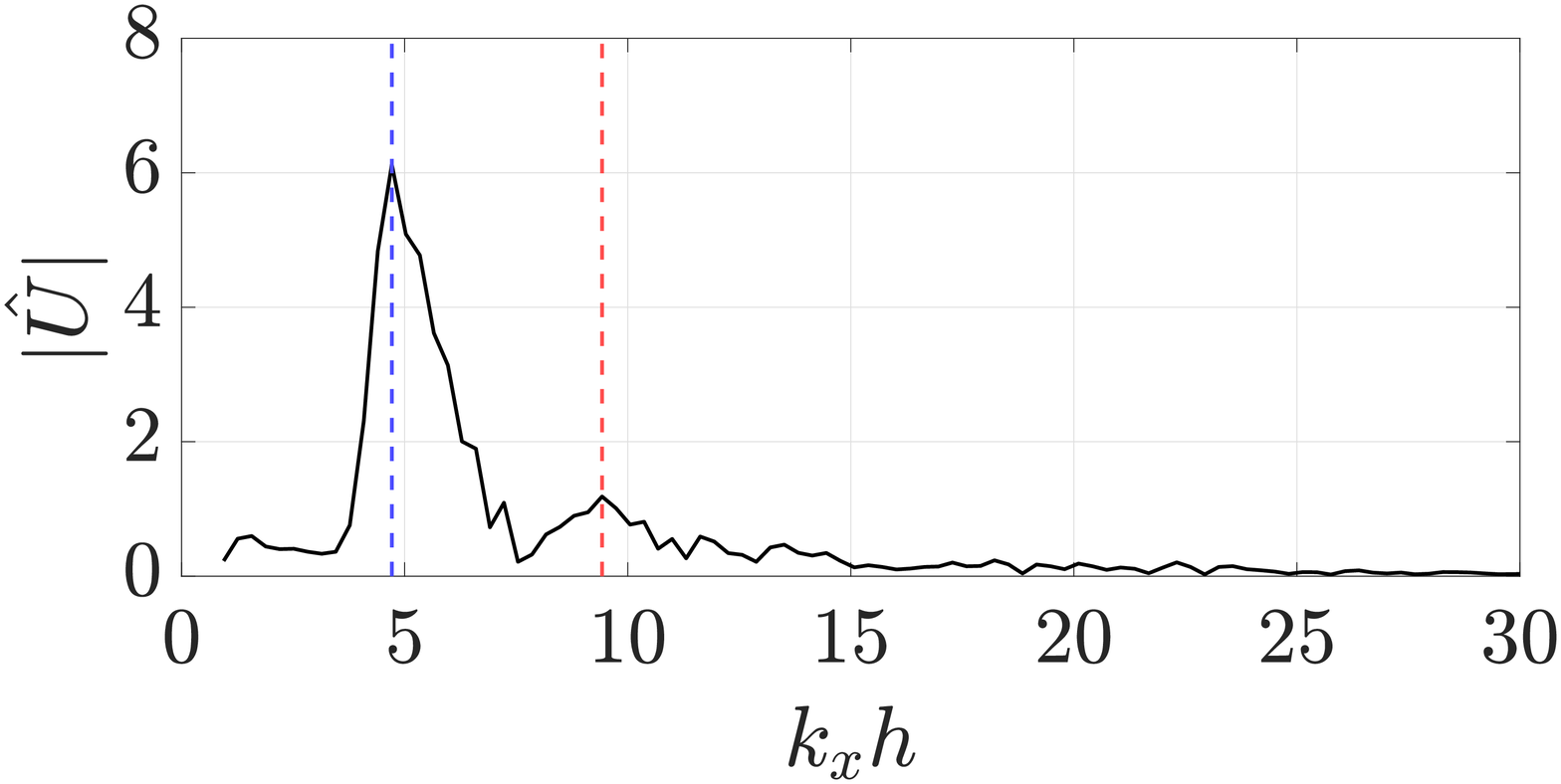} \\
    (a) & (b) \\
  \end{tabular}  
  \caption{Streamwise Fourier transform of the mean streamwise velocities measured along each jet centerline: (a) Jet 1; (b) Jet 2. Blue and red solid lines represent the peak shock wavenumber $k_{s_1}$ and the suboptimal shock wavenumber $k_{s_2}$, respectively.}
\label{fig:Uhat_centerline}
\end{figure}

Table~\ref{tab:wavenumbers} summarizes the peak wavenumbers of the upstream- and downstream-propagating waves, along with the wavenumbers associated with the shock-cell structures, and their differences for both jets. Here, $k_{s_1}h$ and $k_{s_2}h$ represent the primary and the second harmonic shock-cell peaks, respectively. These values are found by taking a streamwise Fourier transform of the mean centerline streamwise velocities, as shown in figure~\ref{fig:Uhat_centerline}. Notably, the peak wavenumber of the dominant energy blob of the upstream-propagating waves, denoted as $k_{-,max}h$, closely matches the difference between the positive peak wavenumber, $k_{+,max}h$, and the primary shock-cell wavenumber, $k_{s_1}h$. Conversely, the disparity $k_{+,max}h - k_{s_2}h$ aligns with the secondary lobes within the wavenumber spectra. Both of these modes are energised by the triadic interactions between the KH waves and the shock-cell structures. However, the energy distribution appears differently for each mode in the transverse direction. As reported in~\cite{Edgington-Mitchell2022}, the mode at $k_{+,max}h - k_{s_1}h$ is interpreted as the guided jet mode, while the waves at $k_{+,max}h - k_{s_2}h$ are identified as a duct-like mode.

\begin{table}
  \centering
  \begin{tabular}{ccccccccc}
  Jet & SPOD mode & $k_{c_{\infty}}h$ & $k_{-,max}h$ & $k_{+,max}h$ & $k_{s_1}h$ & $k_{s_2}h$ & $k_{+,max}h - k_{s_1}h$ & $k_{+,max}h - k_{s_2}h$ \\
  \hline
  1 & $p'$-SPOD & -1.80 & -2.51 & 2.20 & 4.71 & 9.42 & -2.51 & -7.22 \\
  1 & $v'$-SPOD & -1.80 & -2.20 & 2.20 & 4.71 & 9.42 & -2.51 & -7.22 \\   
  2 & $p'$-SPOD & -1.80 & -2.51 & 2.20 & 4.71 & 9.42 & -2.51 & -7.22 \\
  2 & $v'$-SPOD & -1.80 & -2.20 & 2.20 & 4.71 & 9.42 & -2.51 & -7.22 \\   
  \end{tabular}
  \caption{The peak wavenumbers of the upstream- and downstream-propagating waves, along with the wavenumbers associated with the shock-cell structures and their respective differences at the screech frequency.}
  \label{tab:wavenumbers}
\end{table}

Finally, the separation of the guided jet mode and the free-stream acoustic mode is achieved by employing bandpass filters in the wavenumber domain. Given that the interaction between the KH mode and the shock-cells is responsible for exciting upstream-propagating waves, the choice of bandwidth is determined based on the width of the high-energy KH blobs in the positive wavenumber domain. The KH energy band is specified by setting a threshold value of 10$\%$ of the maximum modulus for $k_xh > 0$, thereby establishing the low and high wavenumber boundaries as $k_x h$ = $[1.38,3.58]$. These values correspond to convection velocities of 0.43--1.12$U_j$, which are typical for the KH wavepackets in supersonic turbulent jets. This range also aligns with the variations in convection velocity in the shock-cell and along the jet shear layers, as reported in the authors' prior publication~\citep{Jeun2022}. These variations occur over the streamwise location of $x/h$ = [2.5,12], a region where a peak source location for screech is commonly identified in existing literature~\citep{Mercier2017}. Consequently, the allowable wavenumber range for the upstream-propagating modes is limited to $k_x h$ = [-3.33,-1.13], by subtracting $k_{s_1}h$ from the wavenumber interval corresponding to the KH band. Within this range, any modes exhibiting supersonic phase velocities are chosen to construct the free-stream acoustic mode. The remaining subsonic modes are utilized to recover the guided jet mode. 

\begin{figure}
  \centering
  \begin{tabular}{cc}
    \includegraphics[width=0.45\textwidth]{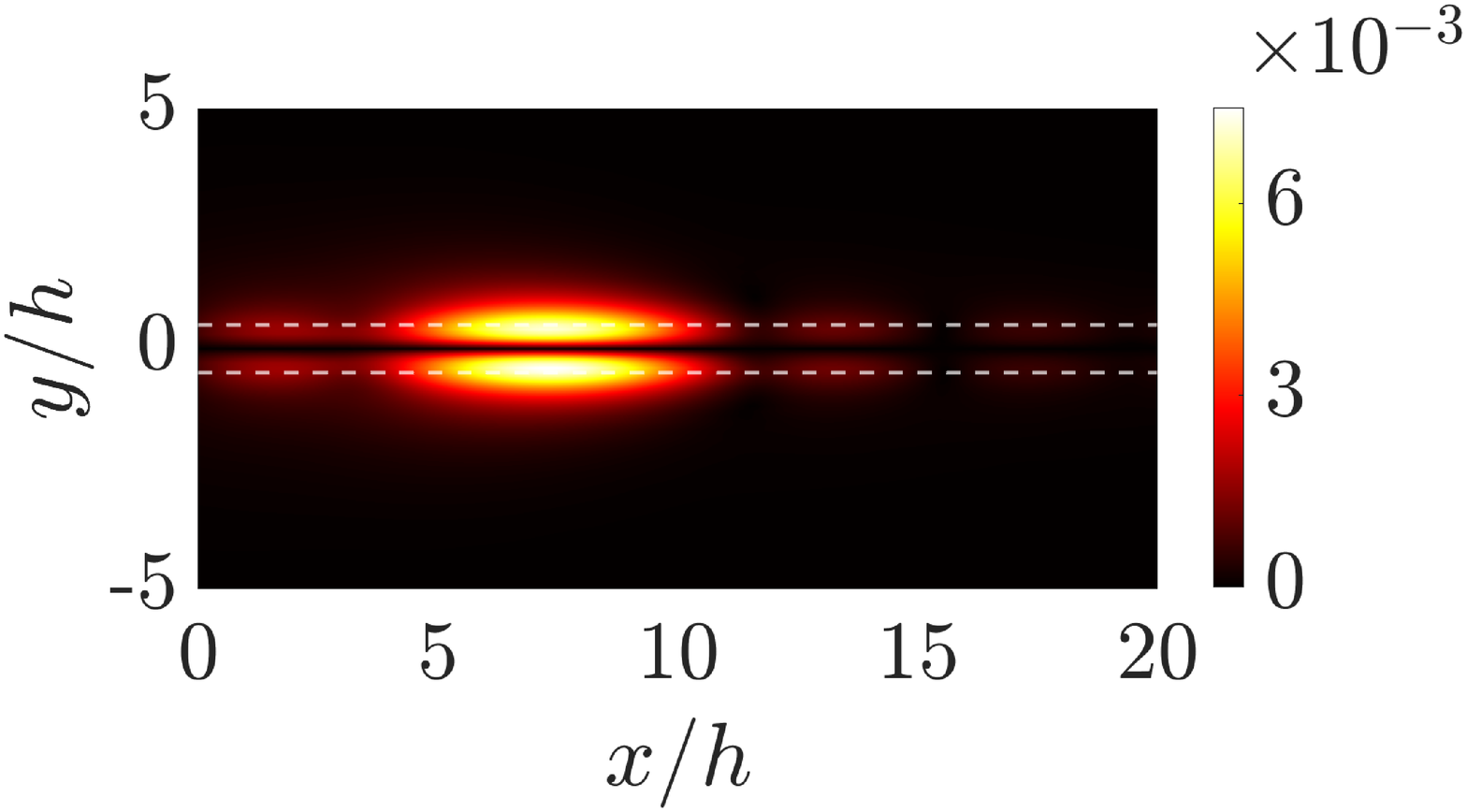} &
    \includegraphics[width=0.45\textwidth]{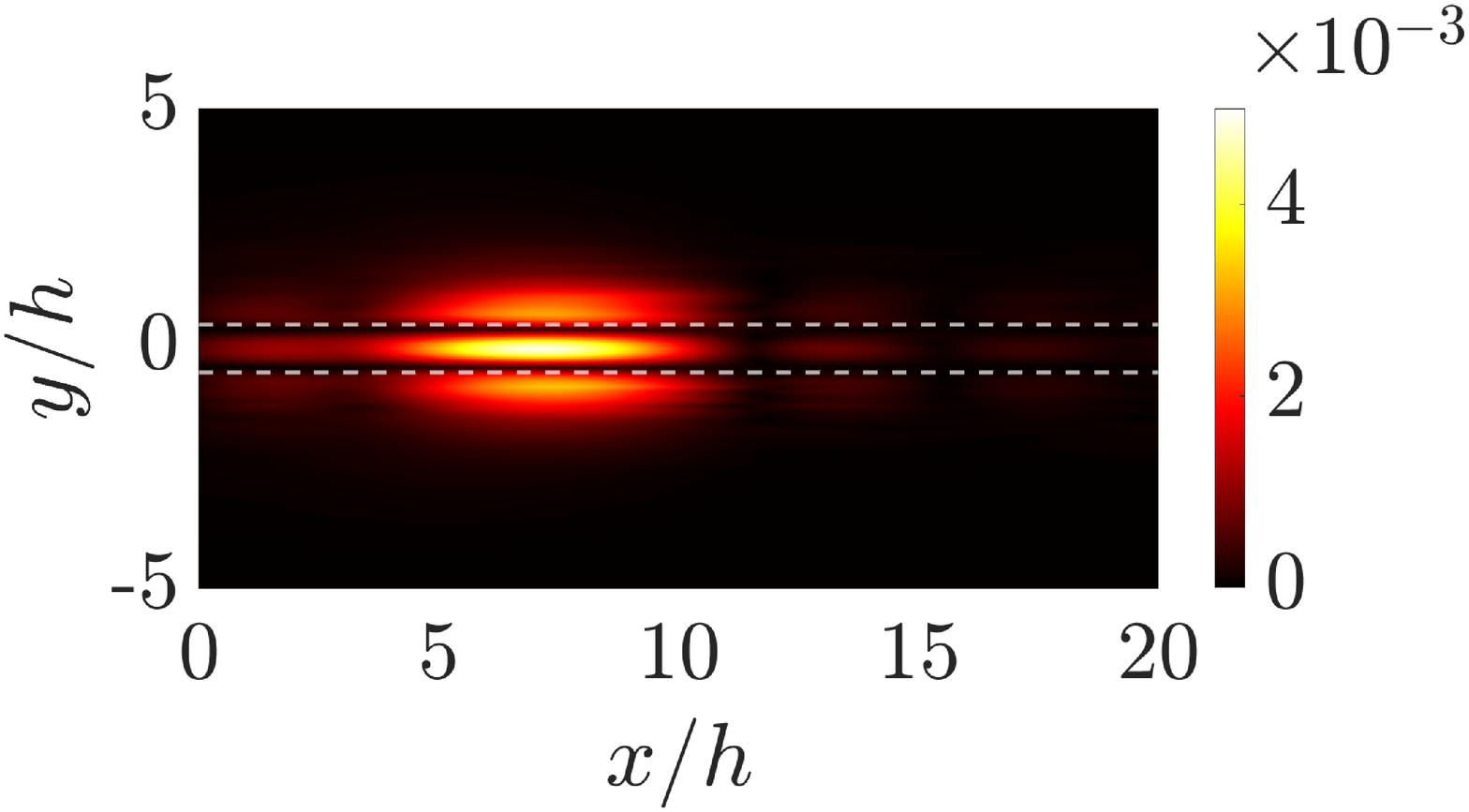} \\
    (a) & (b) \\
    \includegraphics[width=0.45\textwidth]{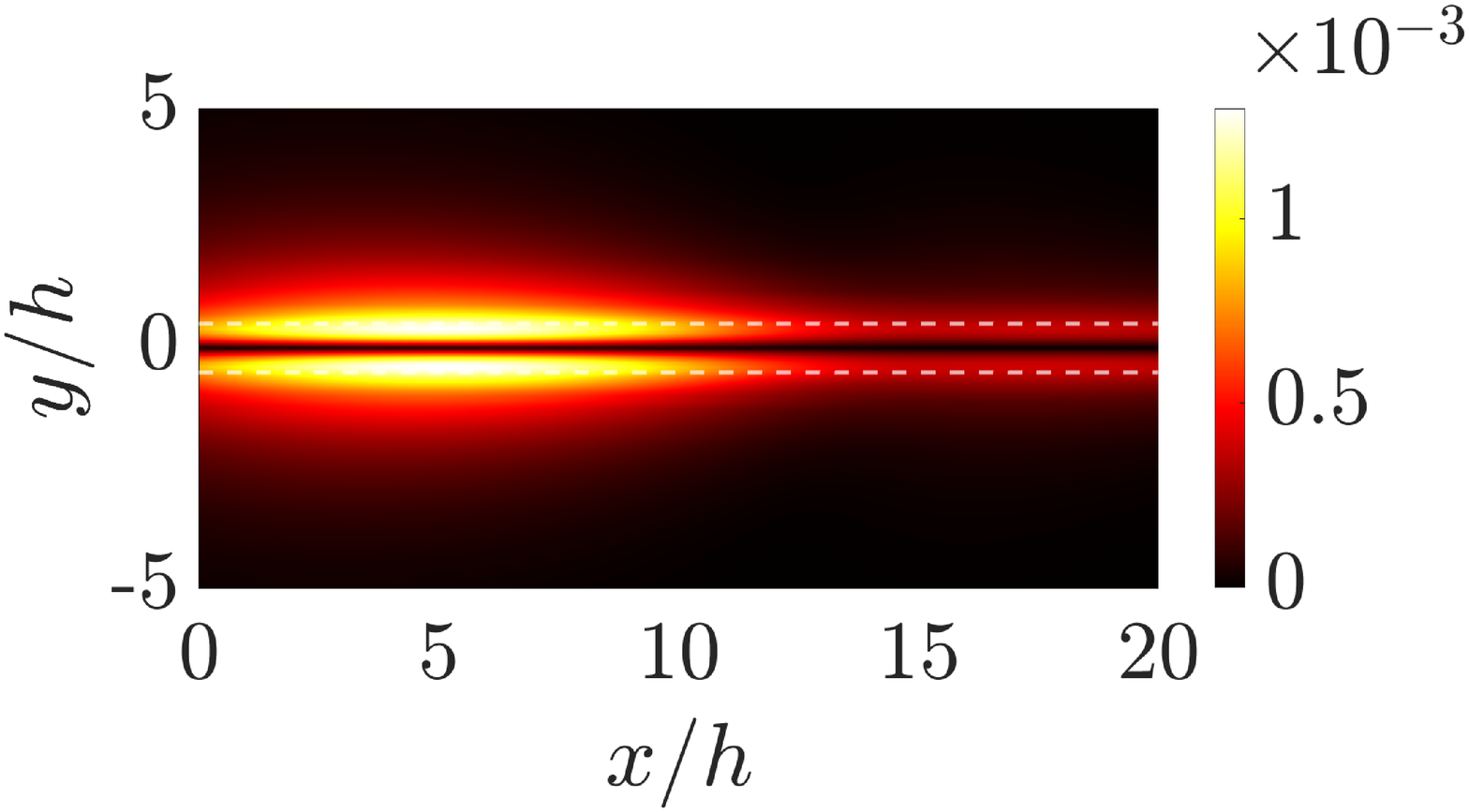} &
    \includegraphics[width=0.45\textwidth]{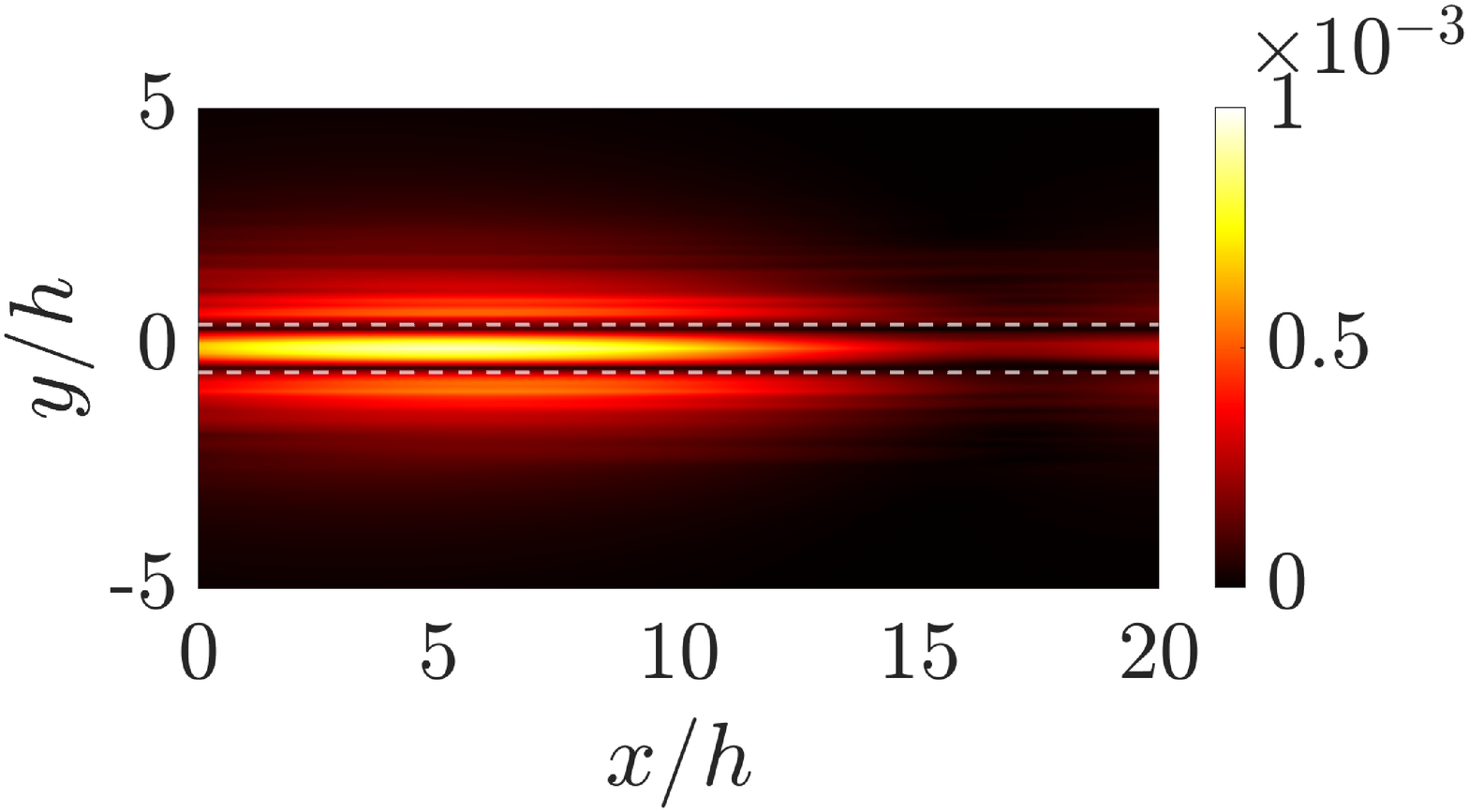} \\
    (c) & (d) \\
  \end{tabular}  
  \caption{Comparisons of the guided jet mode (a,b) and the free-stream acoustic mode (c,d) visualized by the respective modulus: (a,c) pressure fluctuations; (b,d) transverse velocity fluctuations. White dashed lines indicate the liplines. Results are shown for Jet 1 only. For brevity, results for Jet 2 are omitted.}
\label{fig:gjm_vs_fam}
\end{figure}

Figure~\ref{fig:gjm_vs_fam} offers a comparison between the resulting upstream-propagating modes obtained from the $p'$- and $v'$-SPOD modes. The $k_{-}$ guided jet mode has support both in the jet and outside of the shear layers, bearing qualitatively similar flow structures to those previously identified by~\cite{Edgington-Mitchell2021}. For the $v'$ component modes, the modulus of this mode is maximum in the jet core and also significant slightly outside of the lipline. Compared to the $c_{-}$ mode, the $k_{-}$ mode is more localised with the peak at $x/h$ = 7.5, approximately corresponding to the location of the fifth-sixth shock cell. The $c_{-}$ free-stream acoustic mode displays a peak shifted more upstream to $x/h$ = 5, and its mode shape is more extended downstream. Note that differences in modal shapes between $p'$- and $v'$-SPOD modes imply that data evaluated along one constant $y/h$ line may have different uncertainties for these components in the cross-correlation analysis provided in the next section.

\section{Analysis of the twin-jet screech feedback loops}
\label{sec:feedback}
\subsection{Spatial cross-correlation analysis}
\label{subsec:xcorr_analysis}
By following~\citet{Wu2020,Wu2023}, for a zero-mean stationary signal $q(\boldsymbol{x},t)$ of any flow variable detected by two probes placed at different locations $\boldsymbol{x_{1}} = (x_1,y_1,z_1)$ and $\boldsymbol{x_{2}} = (x_2,y_2,z_2)$, one can write a relation
\begin{equation}
    q(\boldsymbol{x_{2}},t) = \alpha q(\boldsymbol{x_{1}},t-\tau) + n(t),
\end{equation}
where $\alpha$ measures the growth/decay in amplitude, $\tau$ is the time delay for a wave travelling from one location to another, and $n(t)$ is the random noise.

The cross-correlation function between the two signals is computed by
\begin{align}
    R_{12}(\tau') &= E\left[q(\boldsymbol{x_{1}}, t) q(\boldsymbol{x_{2}}, t+\tau') \right] \nonumber\\
    &= E\left[ q(\boldsymbol{x_{1}}, t) \left(\alpha q(\boldsymbol{x_{1}}, t+\tau'-\tau) + n(t)  \right)\right] \nonumber \\
    &= \alpha R_{11}(\tau' - \tau),
    \label{eq:xcorr-autocorr}
\end{align}
where the auto-correlation function of one signal is
\begin{equation}
     R_{11}(\tau') = E\left[q(\boldsymbol{x_{1}}, t) q(\boldsymbol{x_{1}}, t+\tau')\right].
\end{equation}

The relationship between the cross-spectral density function $S_{12}(f)$ and the cross-correlation function $R_{12}(t)$ associated with $q$ is expressed by the Wiener-Khinchin theorem such that
\begin{equation}
     R_{12}(\tau')= \int_{-\infty}^{\infty} S_{12}(f) e^{\mathrm{i} 2\pi f \tau'} \mathrm{d}f.
\end{equation}
Similarly, 
\begin{equation}
     R_{11}(\tau')= \int_{-\infty}^{\infty} S_{11}(f) e^{\mathrm{i} 2\upi f \tau'} \mathrm{d}f,
\end{equation}
where $S_{11}$ is the auto-spectral density function of the signal $q(\boldsymbol{x_{1}},t)$. In this work, at the screech frequency $f_{sc}$ we decide to use the leading SPOD mode  $\hat{\phi}_1$ to represent the signal $q$ as
\begin{equation}
    q(\boldsymbol{x},t) = \hat{\phi}_1 (\boldsymbol{x},f_{sc})e^{ \mathrm{i} 2\upi f_{sc} t}.
\end{equation}
Now, $q$ is harmonic in time, and the cross-spectral density function becomes
\begin{eqnarray}
    S_{12}(f) \ &=& \ 
    \begin{cases}
    0 & \text{if} \ f = f_{sc} \\
    \frac{1}{T_{sc}}\hat{q}^*(\boldsymbol{x_{1}},f_{sc})\hat{q}(\boldsymbol{x_{2}},f_{sc}) & \text{otherwise} \\
    \end{cases},
\end{eqnarray}
where $T_{sc}$ is the screech period, $\hat{q}(f)$ is the Fourier transform of $q(t)$, and the superscript $*$ denotes the complex conjugate. Hence, the correlation functions are reduced to
\begin{equation}
    R_{12}(\tau') =  \hat{q}_1^{*}(f_{sc})\hat{q}_2(f_{sc}) e^{\mathrm{i} 2\upi f_{sc} \tau'}
    \label{eq:xcorr}
\end{equation}
and 
\begin{equation}
    R_{11}(\tau'-\tau) = \hat{q}_1^{*}(f_{sc})\hat{q}_1(f_{sc}) e^{\mathrm{i} 2\upi f_{sc} (\tau'-\tau)},
    \label{eq:autocorr}
\end{equation}
after dropping the entities for the probe locations in $q$. Instead, the subscripts denote the corresponding probes.

Finally, substituting \eqref{eq:xcorr} and \eqref{eq:autocorr} into \eqref{eq:xcorr-autocorr} and solving for $\tau$ and $\alpha$ gives
\begin{equation}
    \tau = -\frac{\mathrm{arg}\zeta}{2\upi f}, \quad
    \alpha = |\zeta|, \quad
    \zeta = \frac{\hat{q}^*_{1}\hat{q}^{}_{2}}{\hat{q}^*_{1}\hat{q}^{}_{1}}.
\end{equation}

Screech is an aeroacoustic resonance phenomenon that can be established when a constructive phase relationship is satisfied between the disturbances associated with the feedback loop. Assuming maximum receptivity at the nozzle exit, we seek to identify a downstream streamwise location $x$ where upstream waves originating from such a point arrive at the nozzle exit with an appropriate phase criterion. We trace disturbances by changing their streamwise location along a constant $y$ and $z$, considering it serves as the primary direction of energy propagation. Mathematically, for the self-excitation path of each jet, such points can be expressed as
\begin{equation}
    x \quad \mathrm{s.t.} \quad [\tau_{k_+}(x) - \tau_{-}(x)]/T_{sc} = \tau_{t}/T_{sc} = N, 
\label{eq:points_of_return}
\end{equation}
where $\tau_{-}$ is the negative time delay, which is either $\tau_{k_{-}}$ or $\tau_{c_{-}}$ depending on the choice of the guided jet mode or the external mode as a closure mechanism, $\tau_t$ means the total time delay involved in the feedback loop, and $N$ is a positive integer. 

Each self-excited screech feedback can be influenced by disturbances originating from its twin (cross-excitation path). With respect to a given (screech) source location in Jet 1 ($x_1$), disturbances originating from eligible points of return in Jet 2 ($x_2$) must satisfy a certain phase relationship at the nozzle exit. Note that this cross-excitation is achieved by waves propagating purely external to the jets. Depending on the coupling mode of the twin jets, $x_2$ can be written as
\begin{eqnarray}
    x_2 \quad \mathrm{s.t.} \quad \tau_{t,2 \to 1}/T_{sc} \
    &=& [\tau_{k_{+,1}}(x_1) - \tau_{c_{-,2 \to 1}}(x_2)] \big/ T_{sc} \nonumber \\
    &=& 
    \begin{cases}
    N & \text{(in-phase coupling)} \\
    N + \frac{1}{2} & \text{(out-of-phase coupling)} \\
    \end{cases}.
    \label{eq:cross_excitation_phase_criteria}
\end{eqnarray}
Here, $x_1$ and $x_2$ are not necessarily the same. Analogous relationship holds for the cross-excitation by Jet 1 onto Jet 2.

From the perspective of locating points of return, the present analysis may be viewed as an extension of Powell's phased array model~\citep{Powell1953} by merely relaxing the assumption of equidistant sources. It is important to note that our approach does not preclude the view of distributed sources~\citep{Tam1982}, which is supported by emerging evidence~\citep{Nogueira2021,Edgington-Mitchell2022,Stavropoulos2022}. Instead, our approach provides quantification of the phase relationship of acoustic sources that are spread over multiple shock spacings in jet turbulence, with reference to a certain receptivity location.

\subsection{Closure mechanism for screech coupling}
\label{subsec:closure_mechanism}
As depicted in figure~\ref{fig:wavenumber_spectra}, the upstream-propagating guided jet and free-stream acoustic modes are characterised by their distinct spatial support in the transverse direction. The $k_{-}$ mode is predominantly energetic within the core of the jet and experiences rapid decay far beyond the shear layers. In contrast, the $c_{-}$ waves exhibit support at even farther $y$ locations. Hence, for the spatial cross-correlation analysis, both the guided jet mode $k_{-}$ and the KH mode $k_{+}$ are taken within the jet plume, whereas the free-stream acoustic mode $c_{-}$ is traced far away from the jets. Similarly, the influence onto one jet by the other $c_{-,2 \rightarrow 1}$ or $c_{-,1 \rightarrow 2}$ is also extracted from regions well outside of the jet plume.

The cross-correlation analysis can be sensitive to the choice of the band-pass filters used to extract the $k_{-}$ and $c_{-}$ modes, as well as the selection of the flow variable for the base SPOD mode. Additionally, results might vary as the transverse locations of the probes change. Nevertheless, the overall trend remains consistent across various permissible combinations of these parameters. To examine the sensitivity to the flow variable, the analysis is repeated for the fluctuating pressure and the fluctuating transverse velocity components of ensemble-averaged SPOD modes. The $k_{+}$ and $k_{-}$ modes are extracted respectively along $y/h$ = 0.4 for the fluctuating pressure and $y/h$ = 0 for the fluctuating transverse velocity, where each peak modulus in the negative wavenumber domain is observed. Considering that the guided jet mode travels directly upstream, the reference point is set at the same transverse location at the nozzle exit as that for the corresponding mode. In both cases, the free-stream acoustic modes for the self-excitation $c_{-}$ and for the cross-excitation $c_{-,1 \rightarrow 2}$ or $c_{-,2 \rightarrow 1}$ are traced along $y/h$ = 5. The reference location for all free-stream acoustic modes is maintained at $(x/h,y/h)$ = $(0,0.5)$. Hence, there is a slight difference in $y$ between the chosen references in each case, but we anticipate this would not have a significant impact on the results.

\begin{figure}
  \centering
  \begin{tabular}{ccc}
    \includegraphics[width=0.31\textwidth]{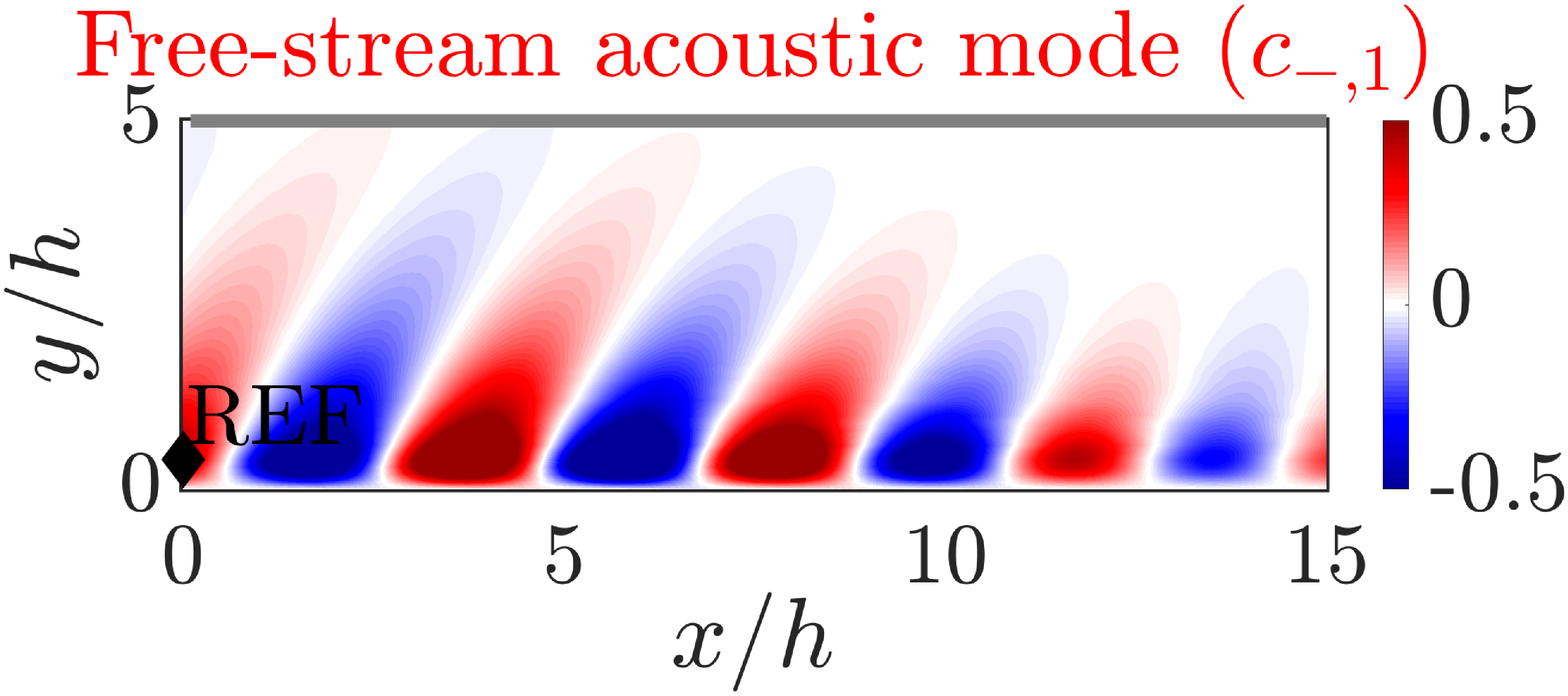} &
    \includegraphics[width=0.31\textwidth]{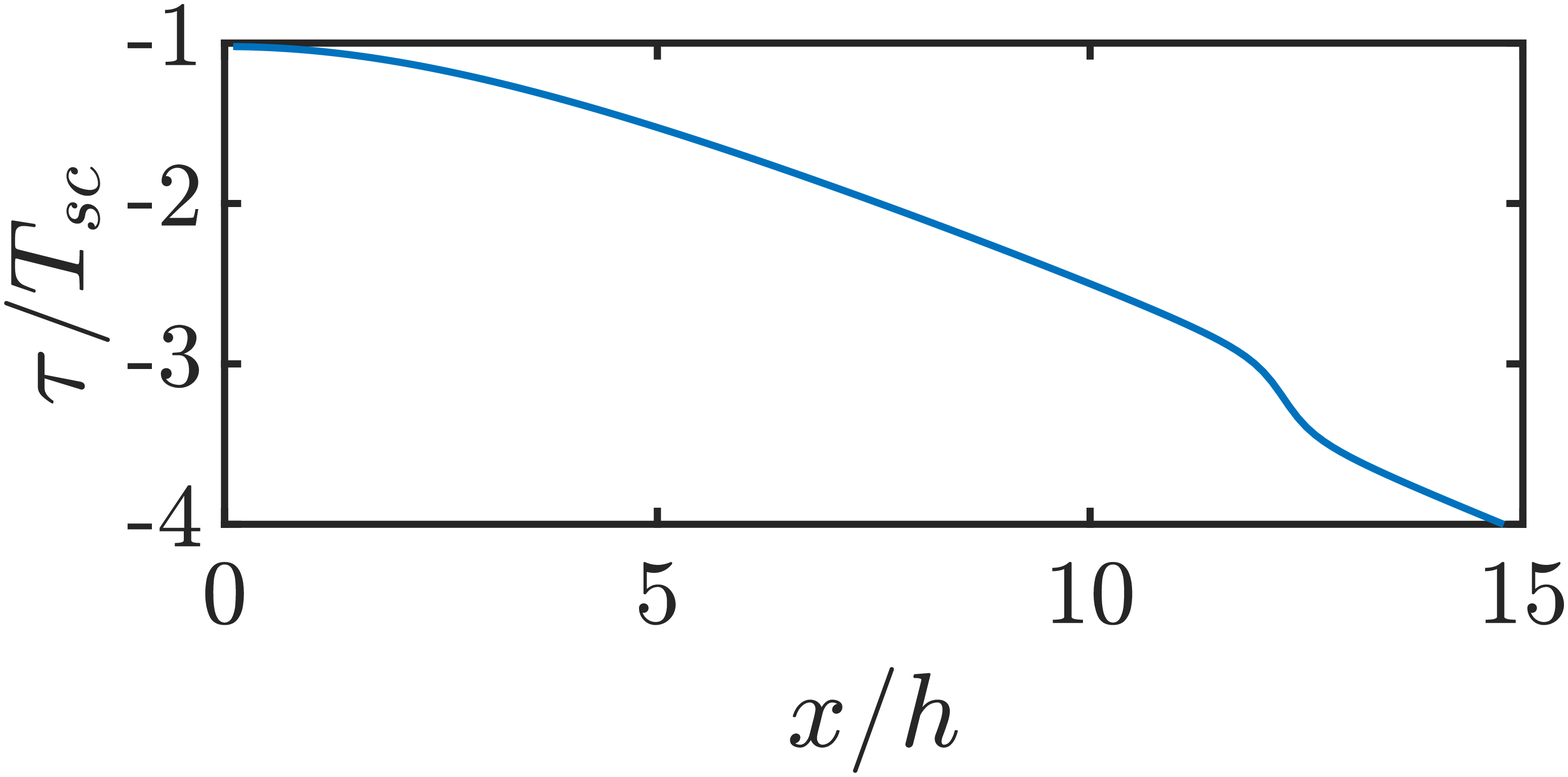} &
    \includegraphics[width=0.31\textwidth]{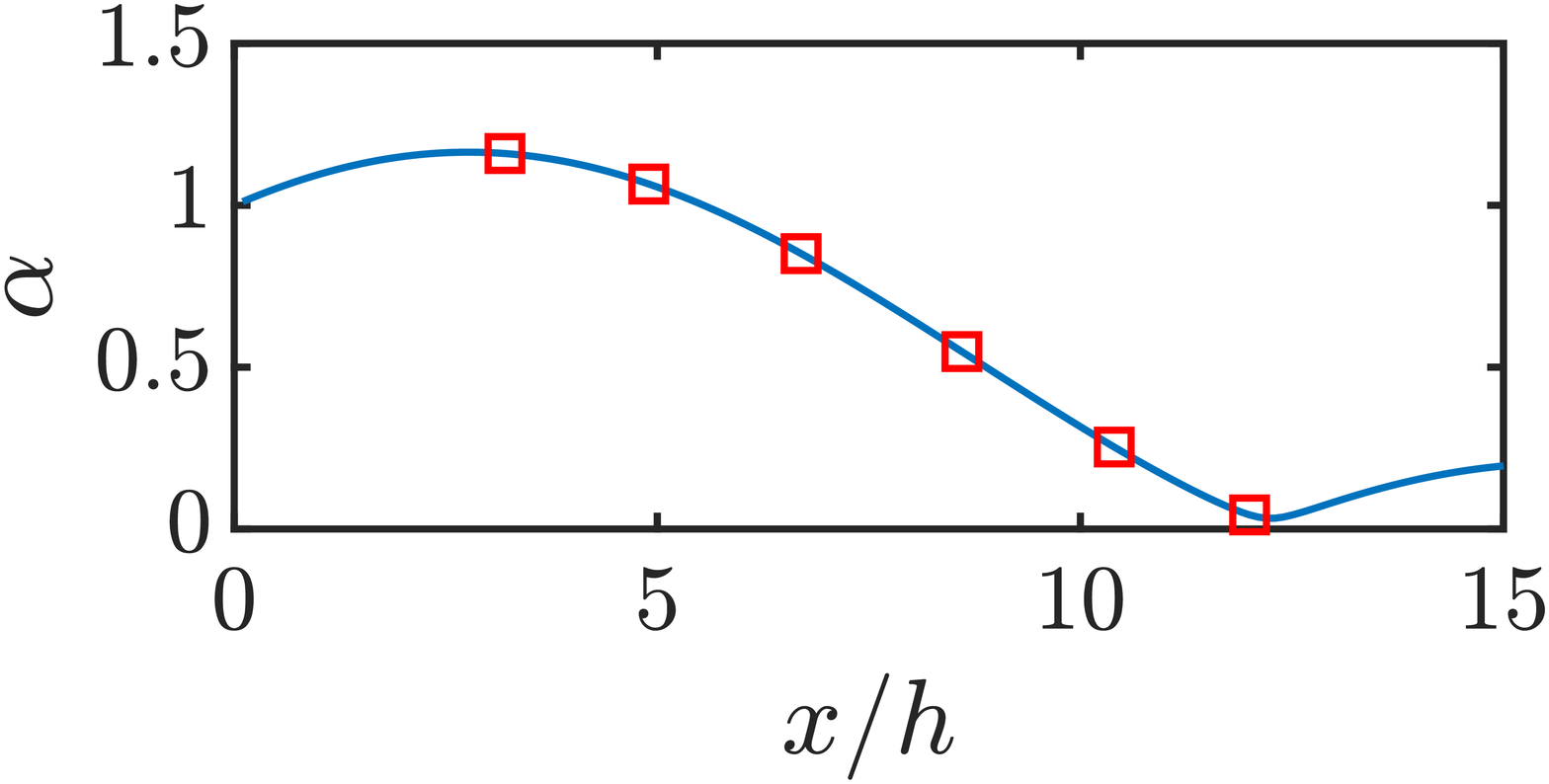} \\
    (a) & (b) & (c) \\
    \includegraphics[width=0.31\textwidth]{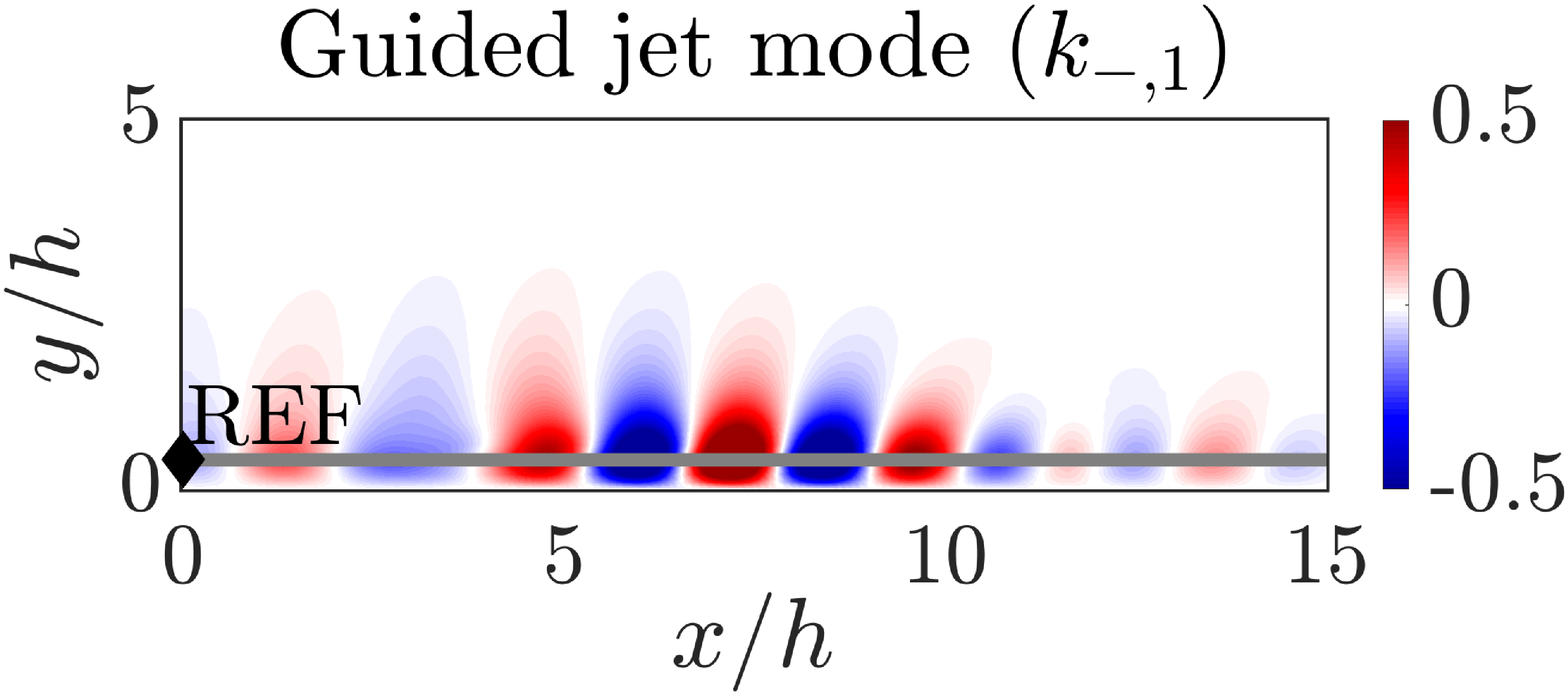} &
    \includegraphics[width=0.31\textwidth]{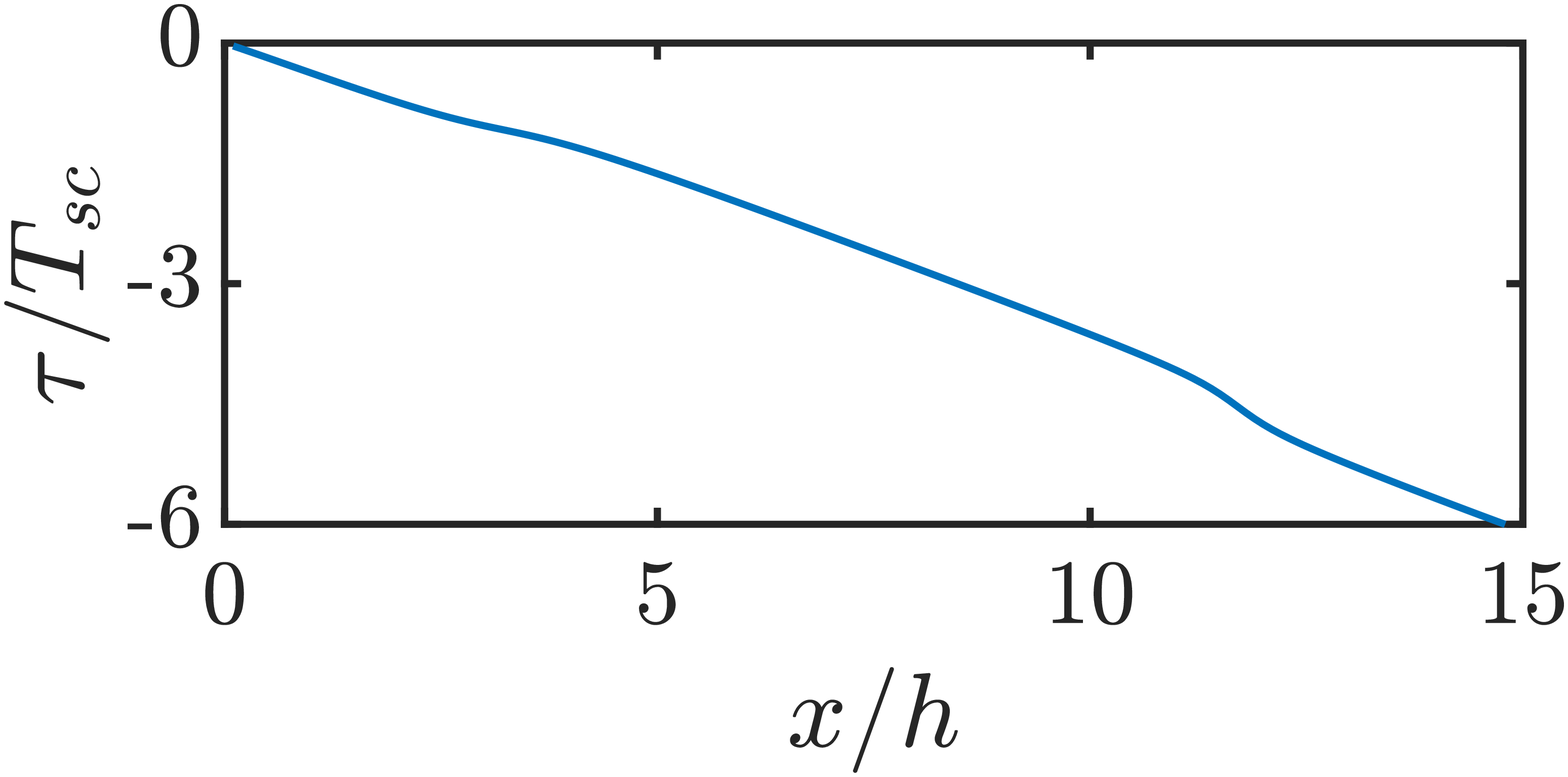} &
    \includegraphics[width=0.31\textwidth]{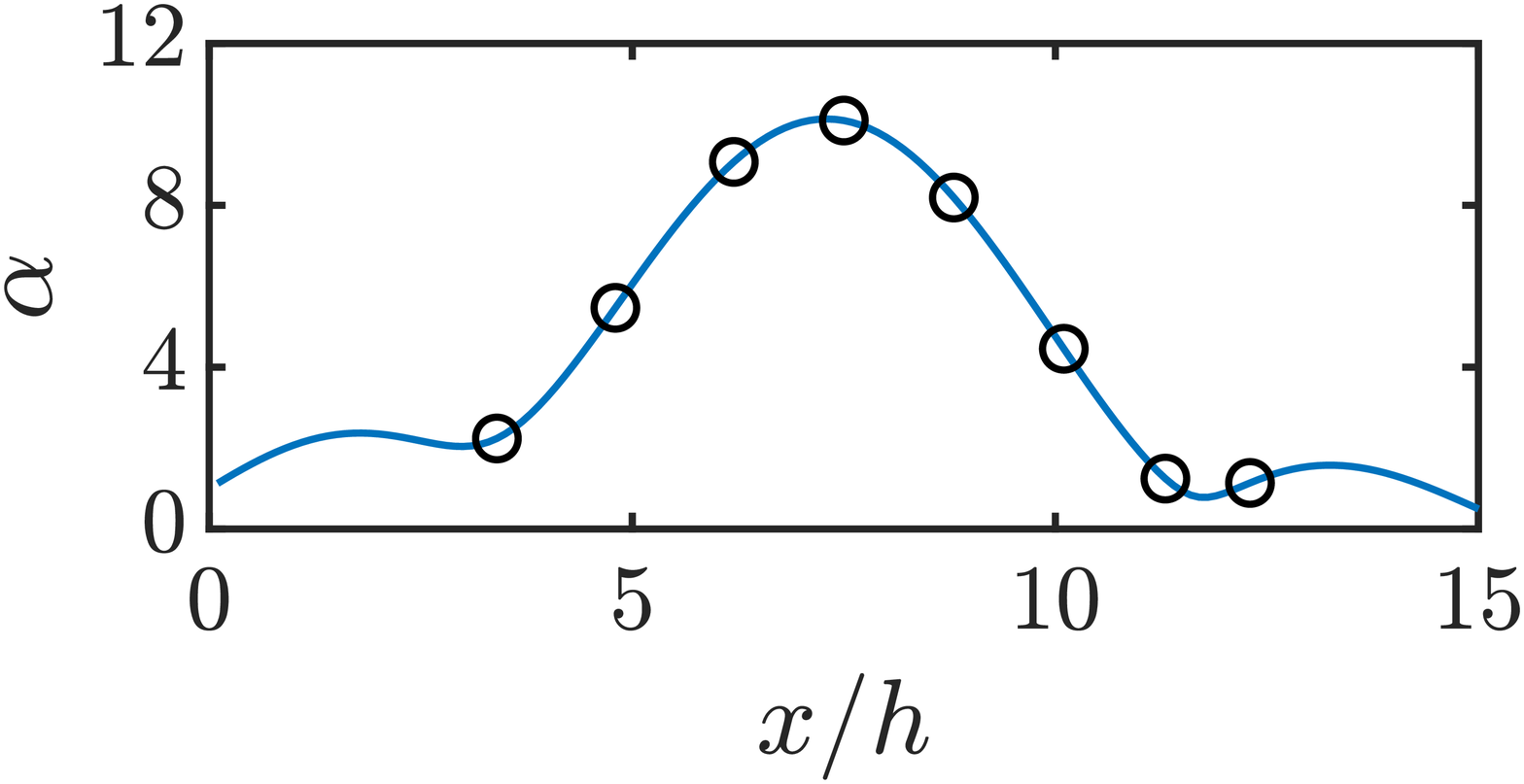} \\
    (d) & (e) & (f) \\
    \includegraphics[width=0.31\textwidth]{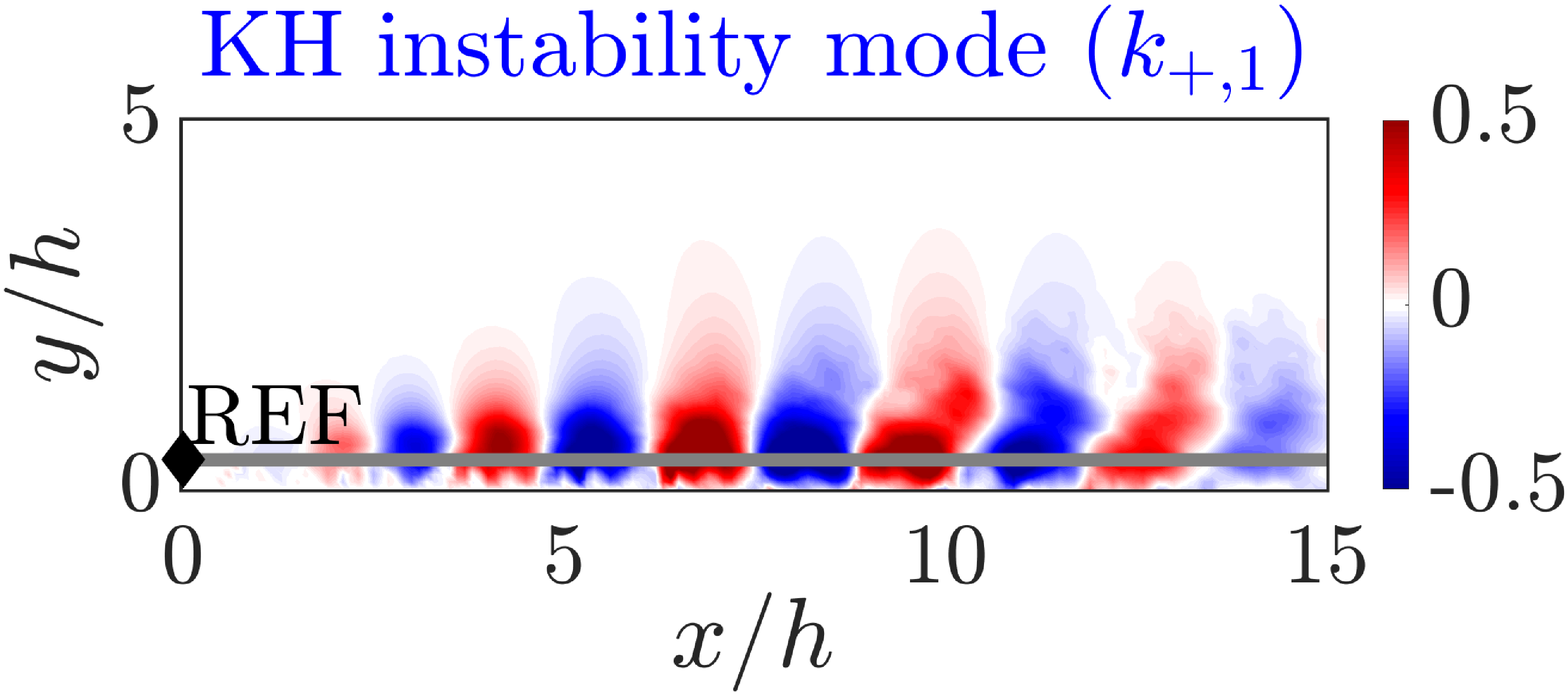} &
    \includegraphics[width=0.31\textwidth]{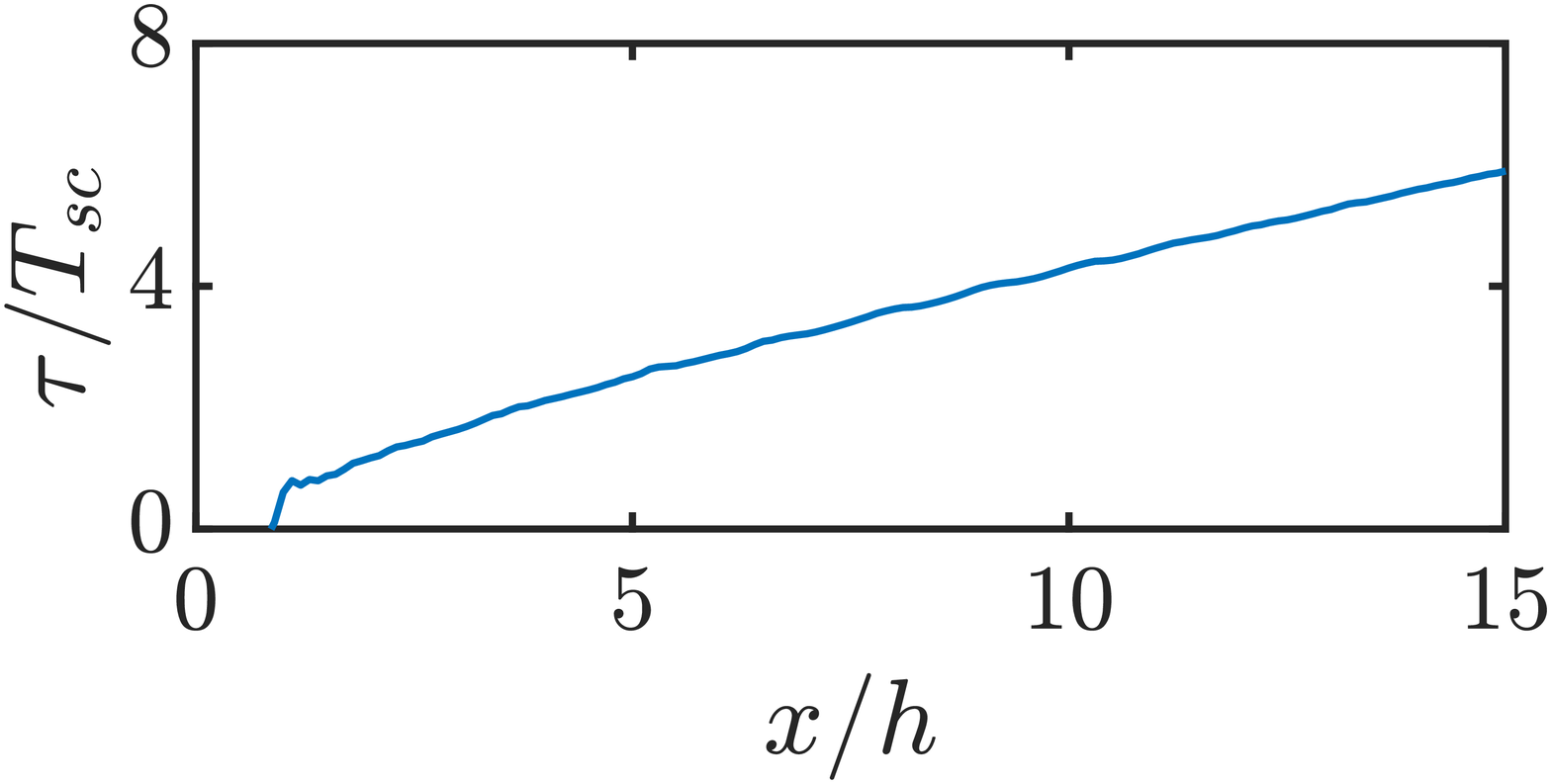} &
    \includegraphics[width=0.31\textwidth]{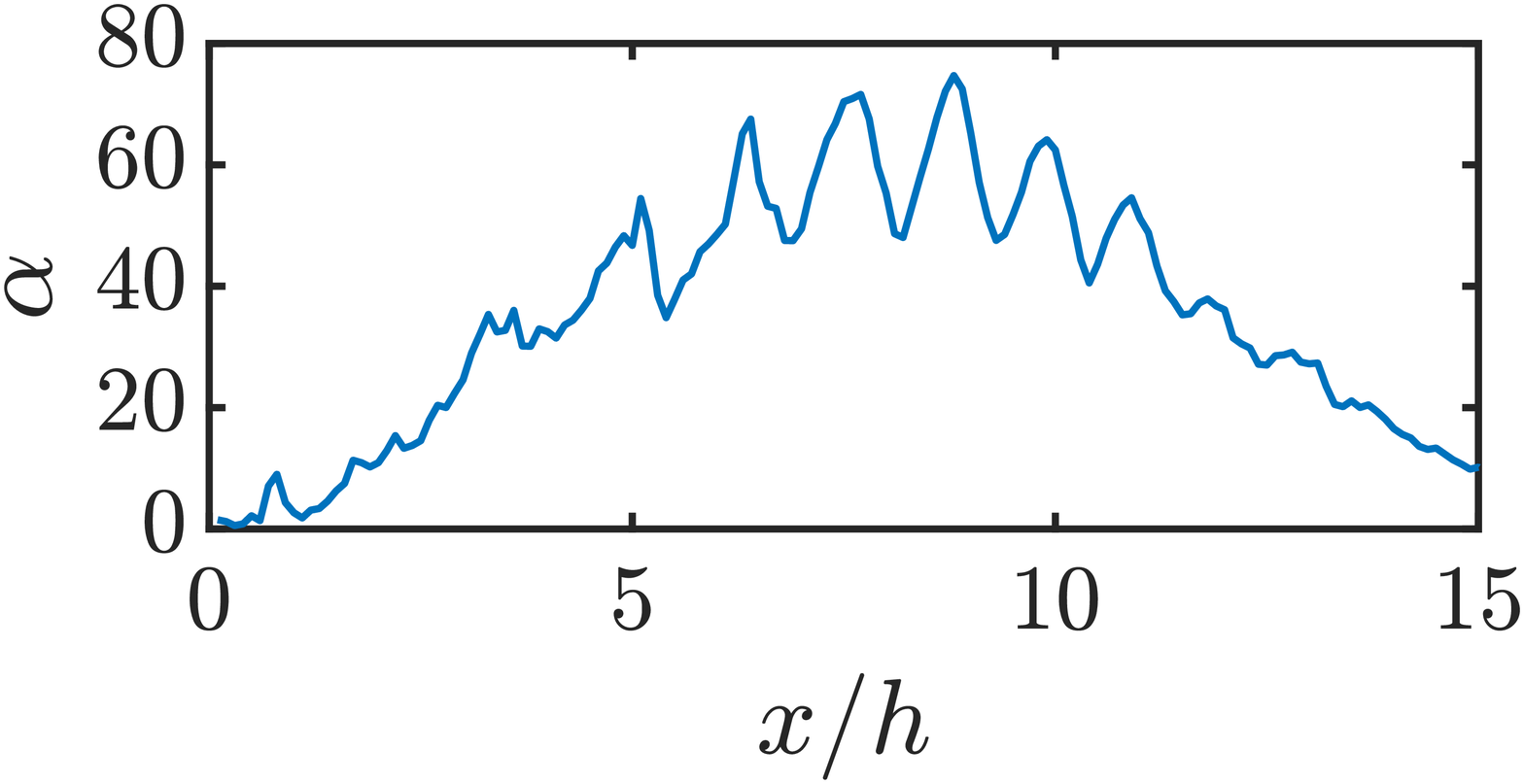} \\
    (g) & (h) & (i) \\
    \includegraphics[width=0.31\textwidth]{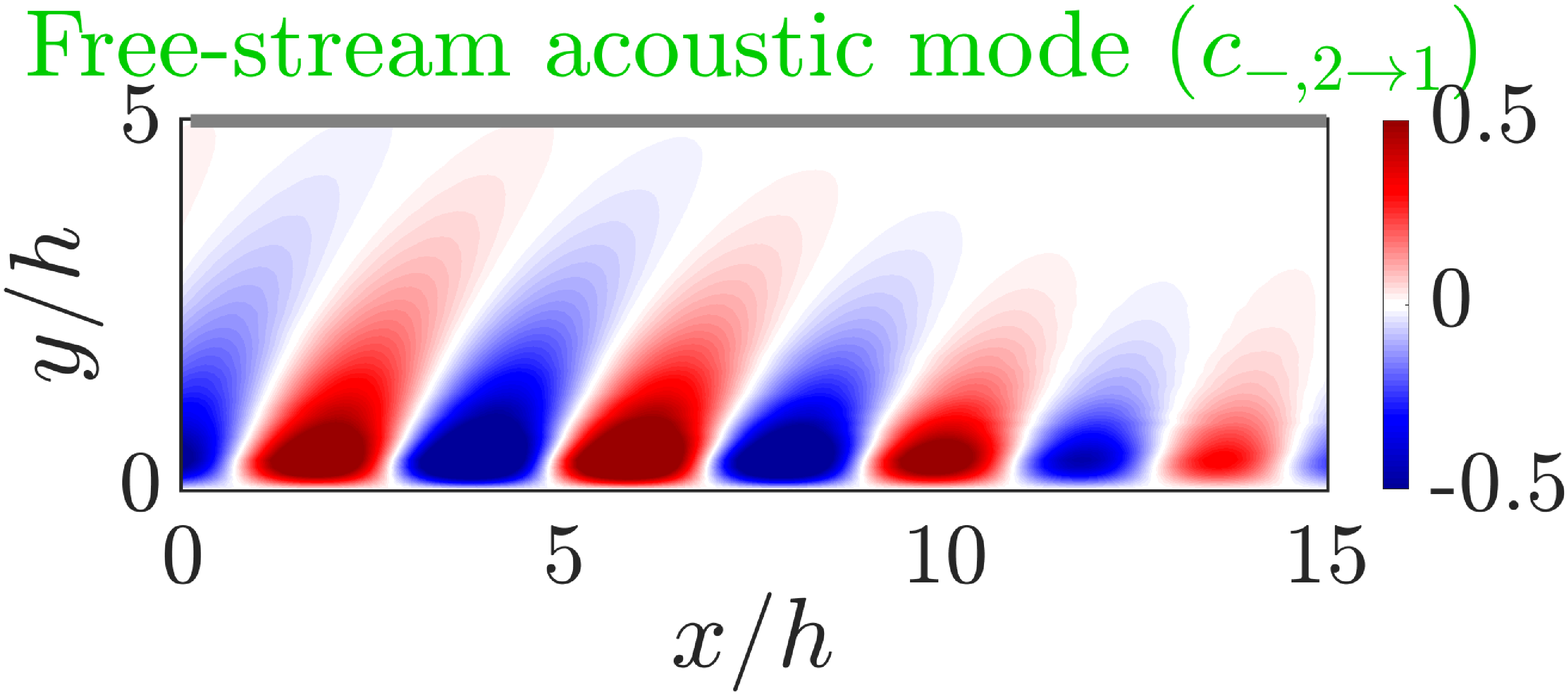} &
    \includegraphics[width=0.31\textwidth]{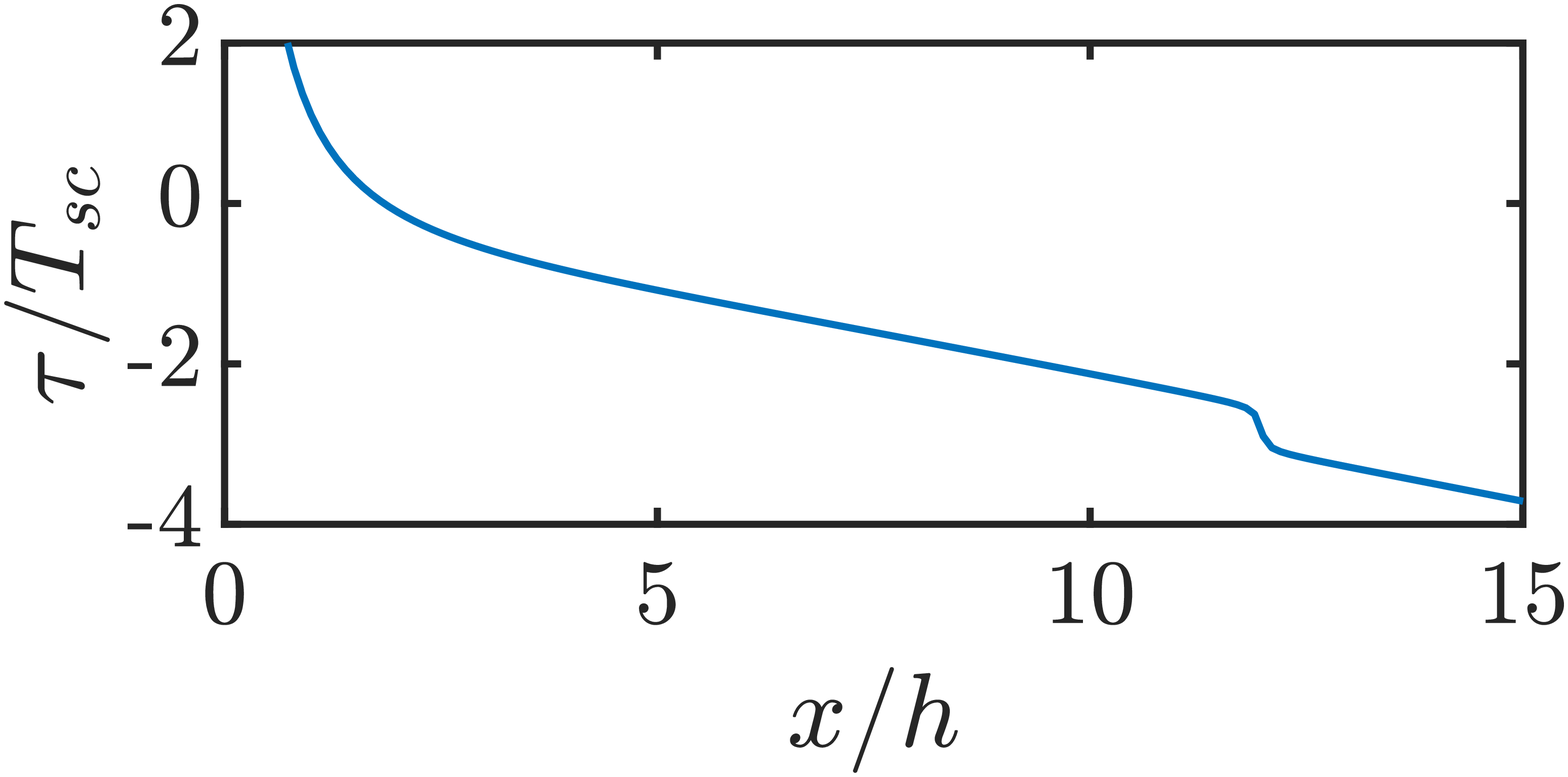} &
    \includegraphics[width=0.31\textwidth]{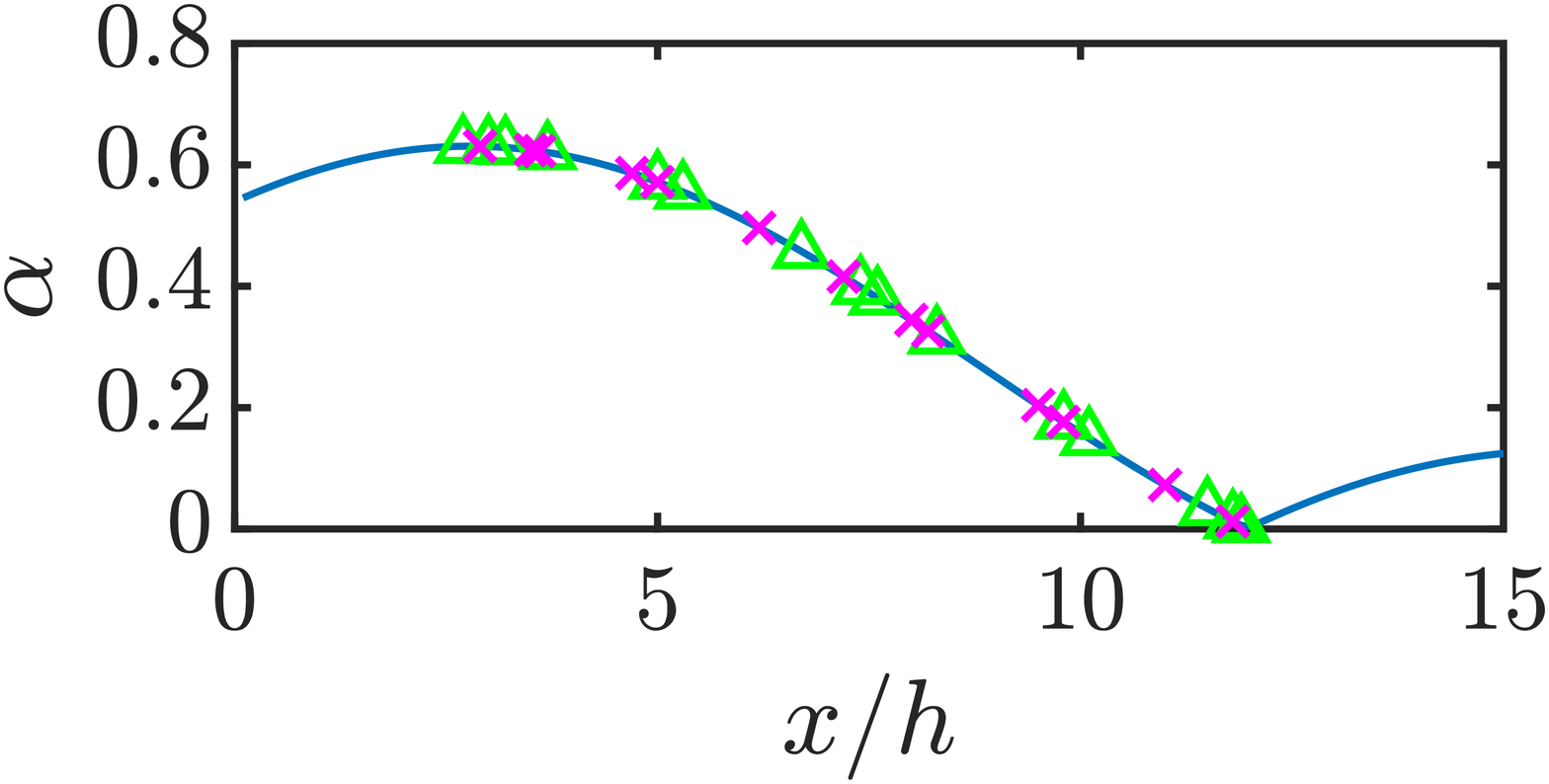} \\
    (j) & (k) & (l) \\
  \end{tabular} \\
  \caption{Spatial cross-correlation analysis for the self-excitation by Jet 1 itself and cross-excitation by Jet 2 onto Jet 1: (a-c) free-stream acoustic mode, $c_{-,1}$; (d-f) guided jet mode, $k_{-,1}$; (g-i) KH mode, $k_{+,1}$; (j-l) free-stream acoustic mode by Jet 2, $c_{-,2\rightarrow 1}$. (a,d,g,j) probe location is denoted by the grey solid line with respect to the reference point marked by the black diamond; (b,e,h,k) time lag with respect to the reference point; and (c,f,i,l) relative amplitude variation overlaid with the identified points of return represented by symbols. {$\color{red}\square$}, Closure SA; $\color{black}\bigcirc$, Closure SG; $\color{magenta}\times$, Closure CA; $\color{green}\triangle$, Closure CG.}
\label{fig:xcorr_analysis}
\end{figure}

Figure~\ref{fig:xcorr_analysis} shows the time delay and the relative amplitude variations of the four different modes associated with the screech coupling with reference to the receptivity location in Jet 1. From top to bottom, results are obtained by the $c_{-,1}$, $k_{-,1}$, $k_{+,1}$, and $c_{-,2 \rightarrow 1}$ modes of the fluctuating pressure. Results for the fluctuating transverse velocity can be found similarly and are omitted here for simplicity. Cross-excitation is considered with respect to the self-excitation feedback closed either by the free-stream mode $c_{-,2 \rightarrow 1 | c_{-,1}}$ or by the guided jet mode $c_{-,2 \rightarrow 1 | k_{-,1}}$. For each mode, harmonic signals are tracked along the grey solid line as shown in the left columns. Screech feedback loop of Jet 2 reinforced by Jet 1 can likewise be obtained but omitted for simplicity. 

As shown in the middle columns the total time delay of each signal changes almost linearly as the probe location moves downstream. The slope represents the phase velocity of it, indicating that the upstream-propagating modes have negative phase velocities. The free-stream acoustic mode has a supersonic phase velocity, while the guided jet mode exhibits a phase velocity of approximately $u_{p,k_{-}}/U_j \approx 0.71$. Considering the broad spectrum of the KH energy blob results in a phase velocity that is much slower than $c_{\infty}$ for the guided jet mode, but this value is 
in reasonable agreement with that recently reported by linear stability analysis~\citep{Edgington-Mitchell2023}. The downstream-propagating KH wave has a positive phase velocity of $u_{p,k_{+}}/U_j \approx$ 0.78, which is close to the typical convection velocity of large-scale eddies in turbulent jets. Here, free-stream acoustic waves additionally take account of the travel distance in the cross-stream directions. Their time delay shows a rapid variation with distance near the nozzle exit due to scattering of sound at the nozzle. 

The relative amplitude variations are shown in the right columns of figure~\ref{fig:xcorr_analysis}. For the free-stream acoustic modes, they vary as $\sim1/r$ where $r$ is the distance between the probe location and the reference point, as expected for sound propagation. For each feedback path, the eligible points of return are overlaid from $x/h$ = 2.5 to 12.5, to highlight parts of the jets with strong acoustic sources for screech.  Points for the {\it{self-excitation}} closed by the free-stream {\it{acoustic}} mode are represented by red squares on the $c_{-}$ mode (Closure SA), and those for the {\it{self-excitation}} closed by the {\it{guided}} jet mode are marked by black circles on the $k_{-}$ mode (Closure SG). Concerning the {\it{cross-excitation}} path, on top of the variations of the amplitudes of the $c_{-,2 \rightarrow 1}$ mode, eligible points of return identified with respect to the self-excitation screech feedback closed by the $c_{-,1}$ mode (Closure CA) and the $k_{-,1}$ mode (Closure CG) are represented by magenta crosses and green triangles, respectively. These cases are summarized in table~\ref{tab:closure_summary}. At this frequency, the KH mode exhibits very large amplitudes around 5 < $x/h$ < 12. It should be also noted that the relative amplitude of the $k_{-,1}$ mode is appreciably larger than that of the $c_{-}$ mode. It increases as the probe location moves downstream, peaks around $x/h \approx$ 7.5 which corresponds to the fifth-sixth shock cells, and then decays rapidly farther downstream. Lastly, for the $c_{-,2 \rightarrow 1}$ mode, the CA and CG closure cases show comparable maximum amplitudes. 

\begin{table}
  \centering
  \begin{tabular}{ccc}
  Cases & Labels & Symbols \\
  \hline
  Self-excitation closed by the free-stream acoustic mode & SA & {\color{red}$\square$} \\ 
  Self-excitation closed by the guided jet mode & SG & {\color{black}$\bigcirc$} \\ 
  Cross-excitation with respect to SA & CA & {\color{magenta}$\times$} \\ 
  Cross-excitation with respect to SG & CG & {\color{green}$\bigtriangleup$} \\ 
  \end{tabular}
  \caption{Summary of the possible closure scenarios.}
  \label{tab:closure_summary}
\end{table}

The fact that the guided jet mode shows substantially larger amplitudes compared to those of the free-stream mode indicates that it may be more effective in closing the screech feedback. The heightened amplitude of the guided jet mode may appear to simply imply a stronger correlation with the signal at the nozzle exit, rather than serving as causal evidence of its preferential closure mechanism. However, it is worth noting that the guided jet mode experiences a more rapid decrease as it travels towards the nozzle. This mode is generated somewhere downstream of the nozzle exit and within the potential core, pumping more energy to feed the resonance loop compared to the free-stream acoustic mode. To further examine the dominance of the guided jet mode, similar analyses can be repeated at immediately neighboring, non-resonating frequencies. As indicated in table~\ref{tab:amplitudes_p}, the upstream-propagating guided jet modes can still be extracted at these frequencies, but their amplitudes are lower compared to those measured at the screech frequency. While the strength of the KH instability decreases, the relative importance of the free-stream mode becomes greater at these frequencies. Analysis of the fluctuating transverse velocity components lead to analogous results as shown in table~\ref{tab:amplitudes_u1}.

\begin{table}
    \centering
    \begin{tabular}{cccccc}
         $St$ & $max(\alpha_{c_{-,1}})$ & $max(\alpha_{k_{-,1}})$ & $max(\alpha_{k_{+,1}})$ & $max(\alpha_{c_{-,2 \rightarrow 1 | c_{-,1}}})$ & $max(\alpha_{c_{-,2 \rightarrow 1 | k_{-,1}}})$  \\
         0.367 & 2.25 & 8.05 & 48.56 & 2.30 & 2.39 \\
         0.373 & 1.16 & 10.10 & 74.71 & 0.63 & 0.63 \\
         0.380 & 1.13 & 3.13 & 65.11 & 1.07 & 0.96 \\
    \end{tabular}
    \caption{Maximum amplitude of each component of the screech feedback loop measured at the screech frequency ($St$ = 0.373) and the two neighboring non-resonant frequencies. Results are obtained using the fluctuating pressure components.}
    \label{tab:amplitudes_p}
\end{table}

\begin{table}
    \centering
    \begin{tabular}{cccccc}
         $St$ & $max(\alpha_{c_{-,1}})$ & $max(\alpha_{k_{-,1}})$ & $max(\alpha_{k_{+,1}})$ & $max(\alpha_{c_{-,2 \rightarrow 1 | c_{-,1}}})$ & $max(\alpha_{c_{-,2 \rightarrow 1 | k_{-,1}}})$  \\
         0.367 & 2.68 & 6.90 & 33.11 & 1.32 & 2.68 \\
         0.373 & 1.11 & 9.18 & 141.44 & 0.58 & 0.58 \\
         0.380 & 2.08 & 4.68 & 49.71 & 1.18 & 1.20 \\
    \end{tabular}
    \caption{Maximum amplitude of each component of the screech feedback loop measured at the screech frequency ($St$ = 0.373) and the two neighboring non-resonant frequencies. Results are obtained using the fluctuating transverse velocity components.}
    \label{tab:amplitudes_u1}
\end{table}

\begin{figure}
\centering
  \begin{tabular}{cc}
    \includegraphics[width=0.45\textwidth]{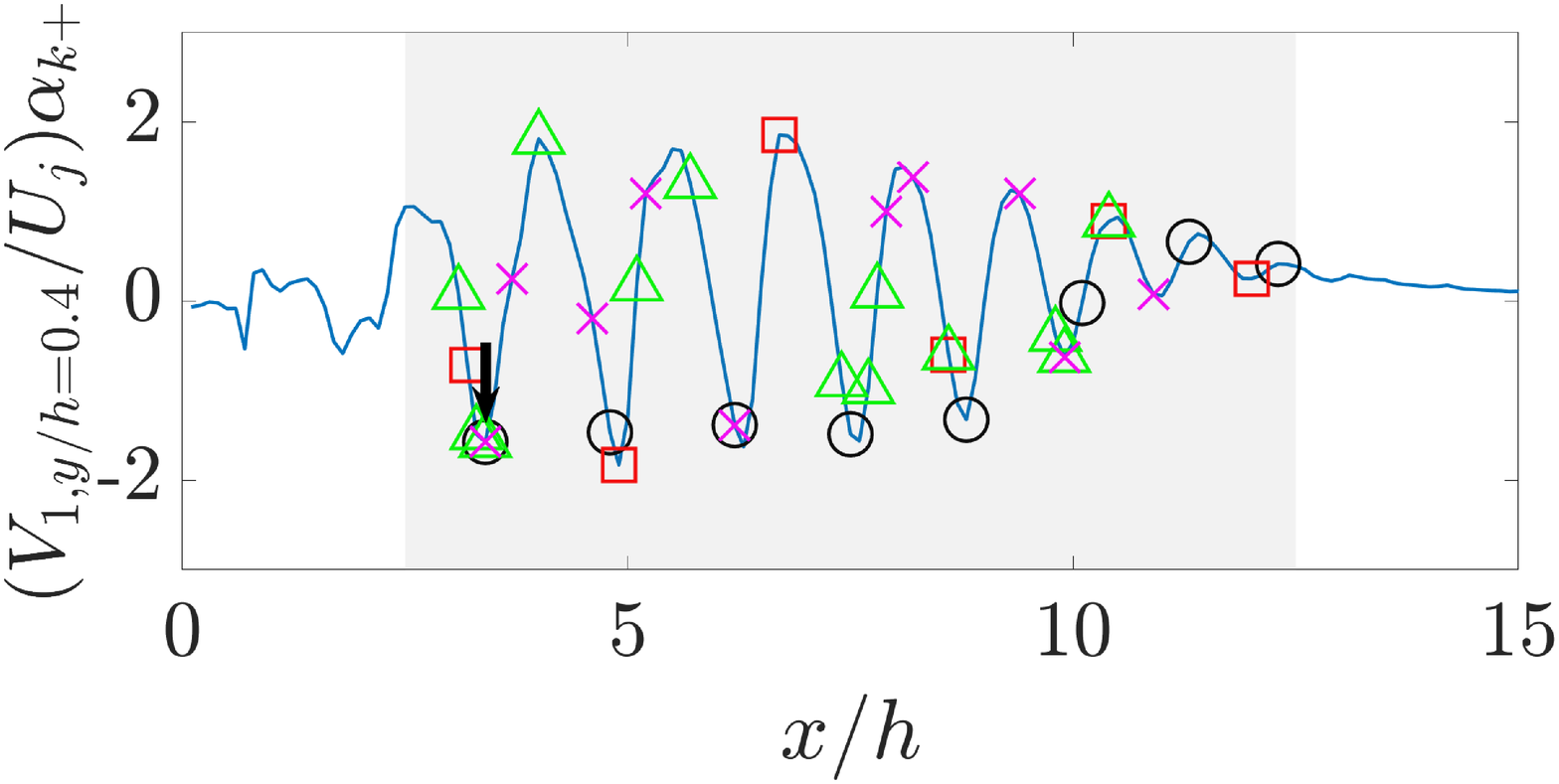} &
    \includegraphics[width=0.45\textwidth]{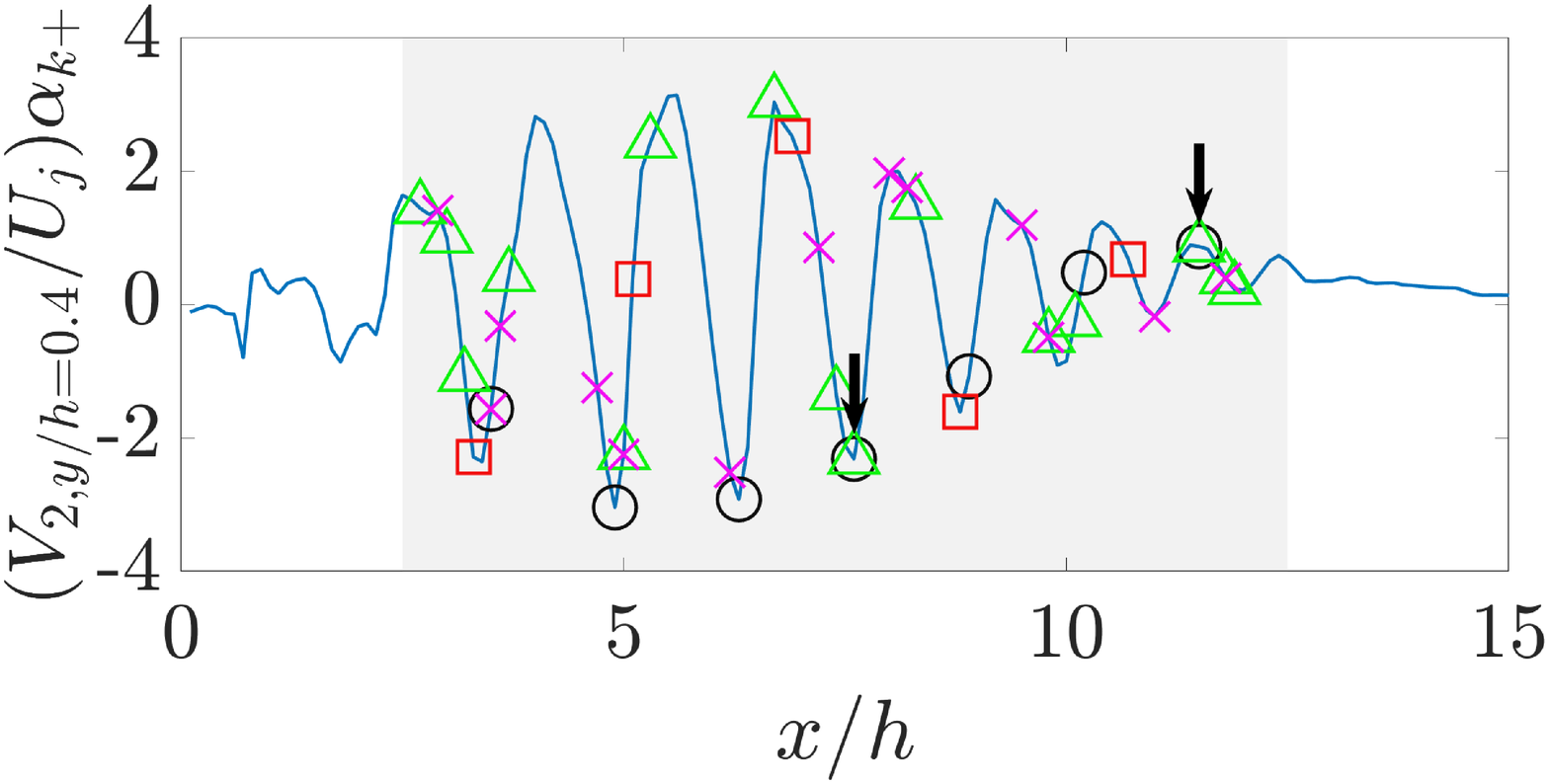} \\
    (a) & (b) \\
  \end{tabular}
\caption{Eligible points of return overlaid on top of $(V_{y/h=0.4}/U_j)\alpha_{k_{+}}$, obtained using the fluctuating pressure components: (a) Jet 1; (b) Jet 2. {$\color{red}\square$}, Closure SA; $\color{black}\bigcirc$, Closure SG; $\color{magenta}\times$, Closure CA; $\color{green}\triangle$, Closure CG. Downward arrows ($\boldsymbol{\downarrow}$) count the synchronisation of points of return for the Closure CG and SG, while the synchronisation of the Closure CA and SA is missing.}
  \label{fig:shock_KH_alpha}
\end{figure}

Considering that upstream-propagating waves are driven by interaction of the shock structure and the KH waves, we investigate whether the identified points of return can be related to the locations where such interactions are strong. The strength of shock/instability wave interactions is quantified by the product of the normalised mean transverse velocity $V/U_j$ and the relative amplitude of the KH mode $\alpha_{k_{+}}$. Each set of the identified points of return are also displayed on top of it, as shown in figure~\ref{fig:shock_KH_alpha}. 

Figure~\ref{fig:shock_KH_alpha} shows that both the free-stream acoustic mode and the guided jet mode include several eligible points of return. However, for each jet, the guided jet mode (black circles) contains more candidates than the free-stream acoustic mode (red squares), and they are mostly located at the troughs of $(V/U_j)\alpha_{k_{+}}$ curves (or conversely, the peaks of $(V/U_j)\alpha_{k_{+}}$ curves in terms of magnitude). Also plotted are the eligible points of return from which the free-stream acoustic waves arrive at the other jet's nozzle lip with an appropriate phase difference (out-of-phase) to reinforce the self-excited screech feedback of that jet. Such points are found with respect to the acoustic waves extracted from both the free-stream acoustic mode (magenta crosses) and the guided jet mode (green triangles). Between the two scenarios, synchronisation with the acoustic source locations for the corresponding self-excitation screech feedback occurs mostly in the case of the guided jet mode (black downward arrows). This implies that feedback waves for the cross-excitation are produced from locations where the guided jet mode (for the self-excitation) is excited, thereby confirming the dominant role of this mode in completing the rectangular twin-jet screech. 

Despite the twin geometry, the points of return for the cross-excitation are observed to be not perfectly identical for the two jets. While perfect convergence of the LES is not guaranteed, the ensemble averaged SPOD modes exhibit sufficient convergence, as mentioned earlier, and thus may not be solely responsible for this small discrepancy. Our filtering scheme for separating the guided jet mode and the free-stream acoustic mode is sensitive and may somewhat amplify discrepancies when identifying points of return for the cross-excitation. Nevertheless, the points of return for the self-excitation feedback are still found at nearly identical locations for both jets.

\subsection{At non-resonating frequencies}
\label{subsec:closure_mechanis_offpeak}
\begin{figure}
\centering
  \begin{tabular}{cc}
    \includegraphics[width=0.45\textwidth]{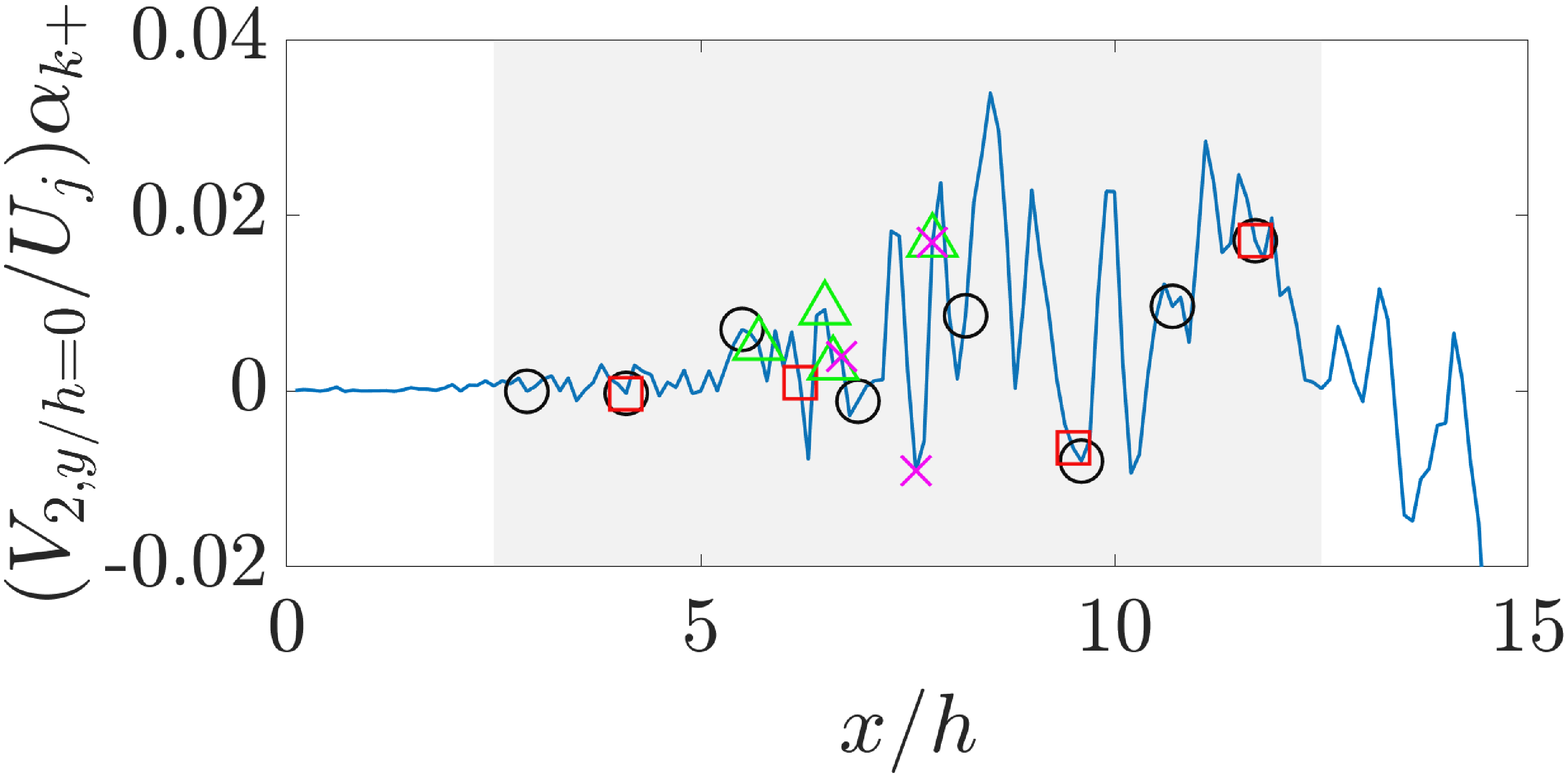} &
    \includegraphics[width=0.45\textwidth]{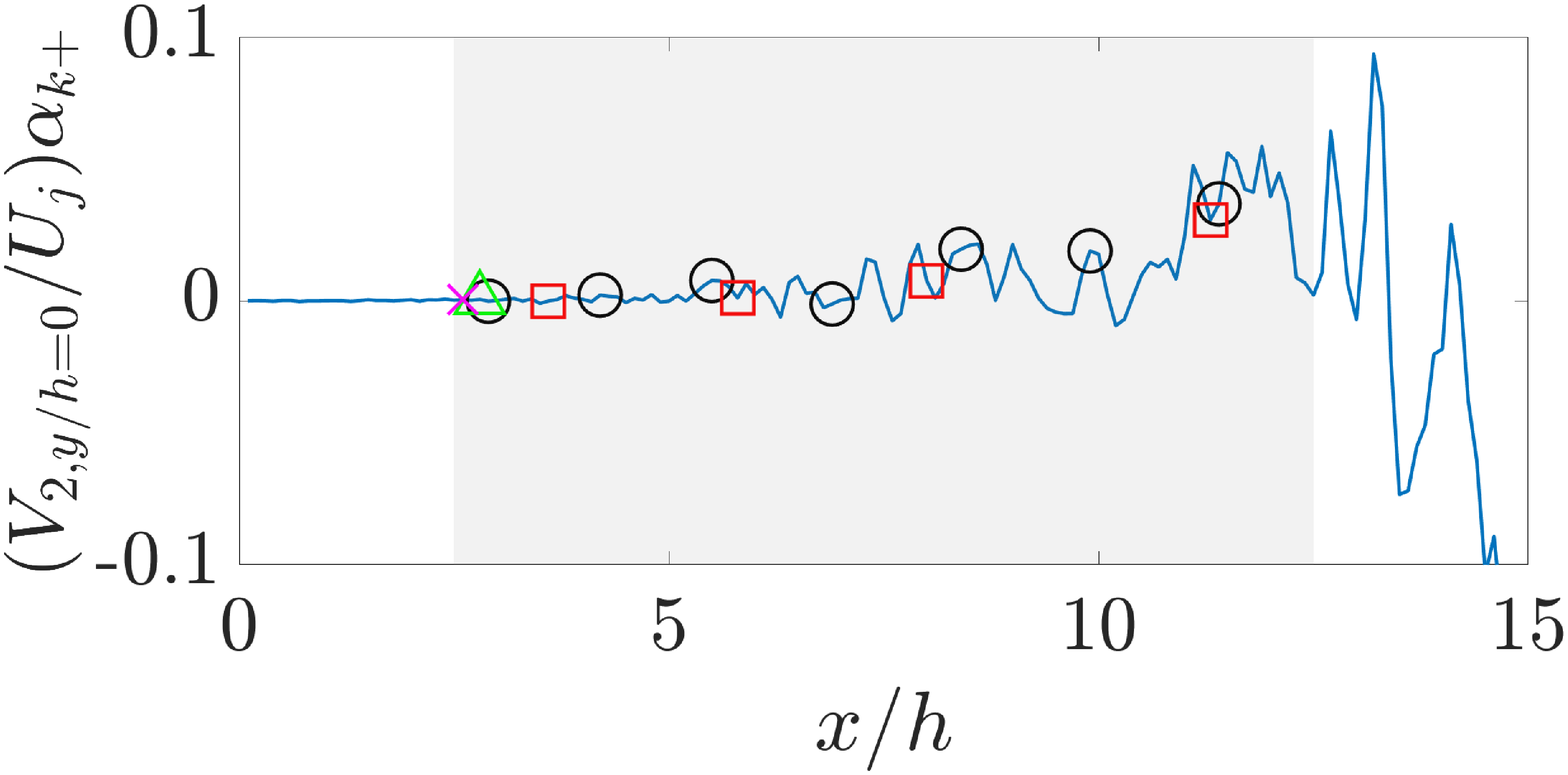} \\
    (a) & (b) \\
  \end{tabular}
\caption{Eligible points of return overlaid on top of $(V/U_j)\alpha_{k_{+}}$ at the two immediate neighboring non-resonant frequencies. Results are shown for the fluctuating transverse velocity components. (a) $St$ = 0.367 and (b) $St$ = 0.383. Symbols: {$\color{red}\square$}, Closure SA; $\color{black}\bigcirc$, Closure SG; $\color{magenta}\times$, Closure CA; $\color{green}\triangle$, Closure CG}.
\label{fig:shock_KH_alpha_off_peaks}
\end{figure}

\begin{figure}
  \centering
  \begin{tabular}{cc}
    \includegraphics[width=0.45\textwidth]{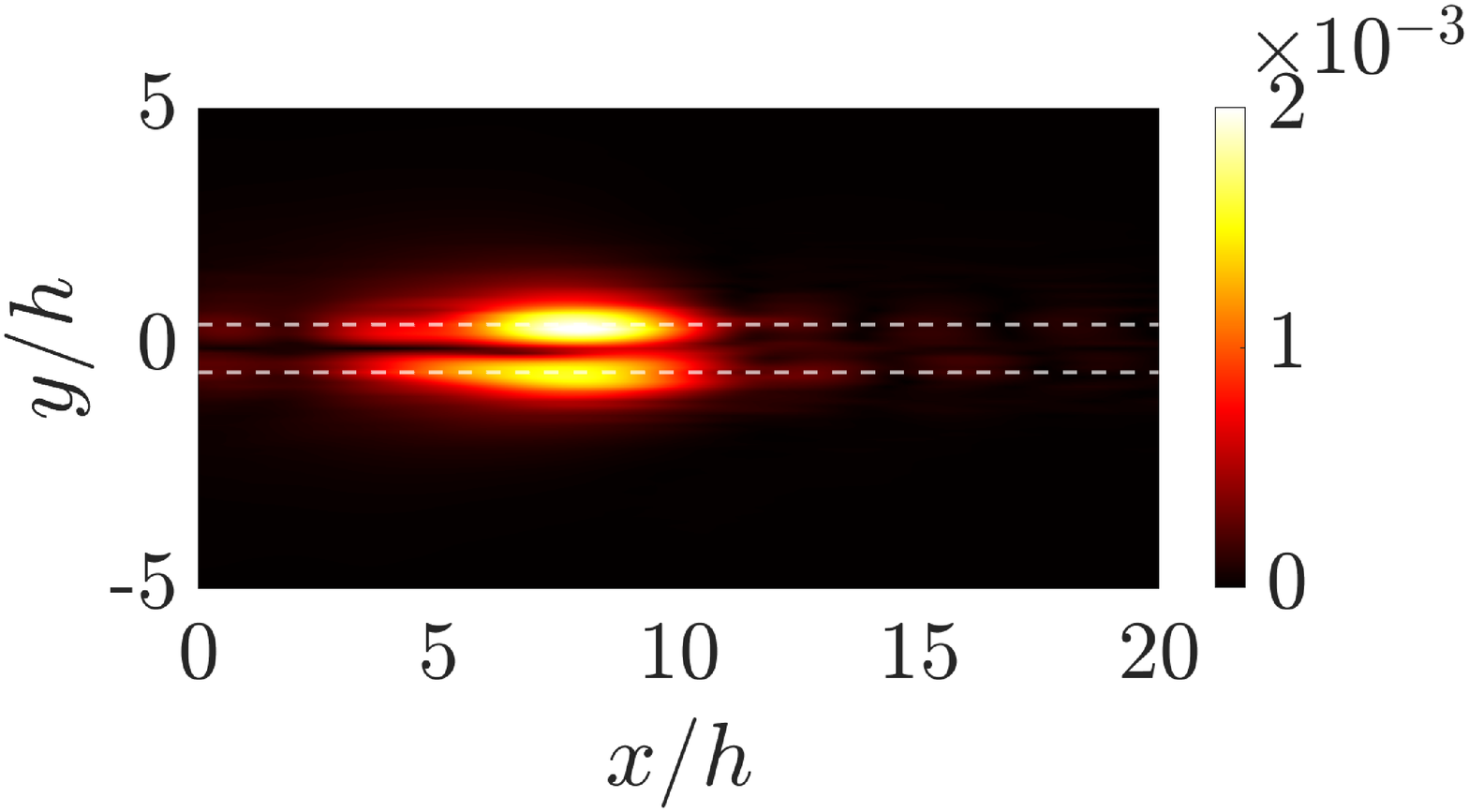} &
    \includegraphics[width=0.45\textwidth]{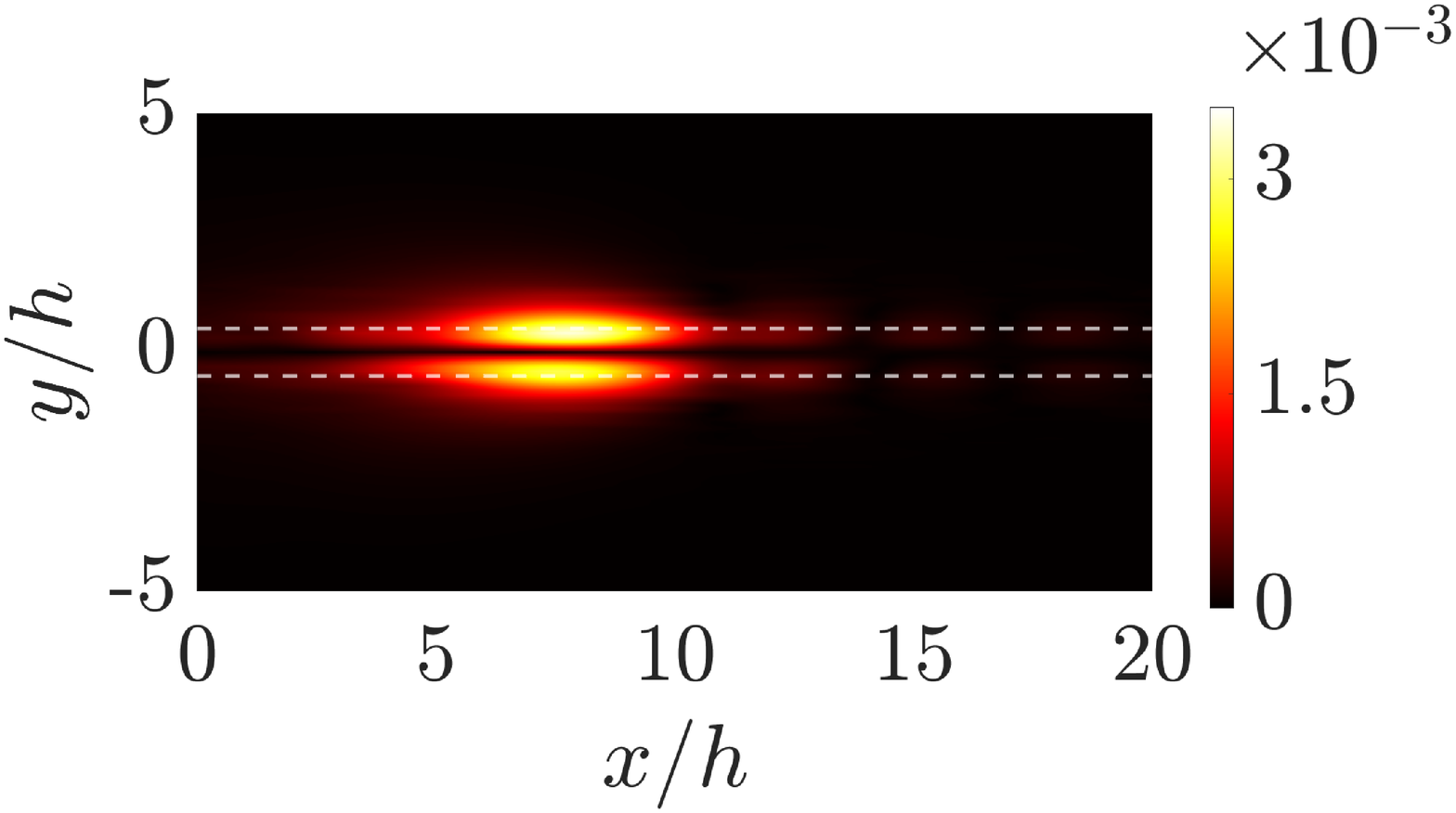} \\
    (a) & (b) \\
    \includegraphics[width=0.45\textwidth]{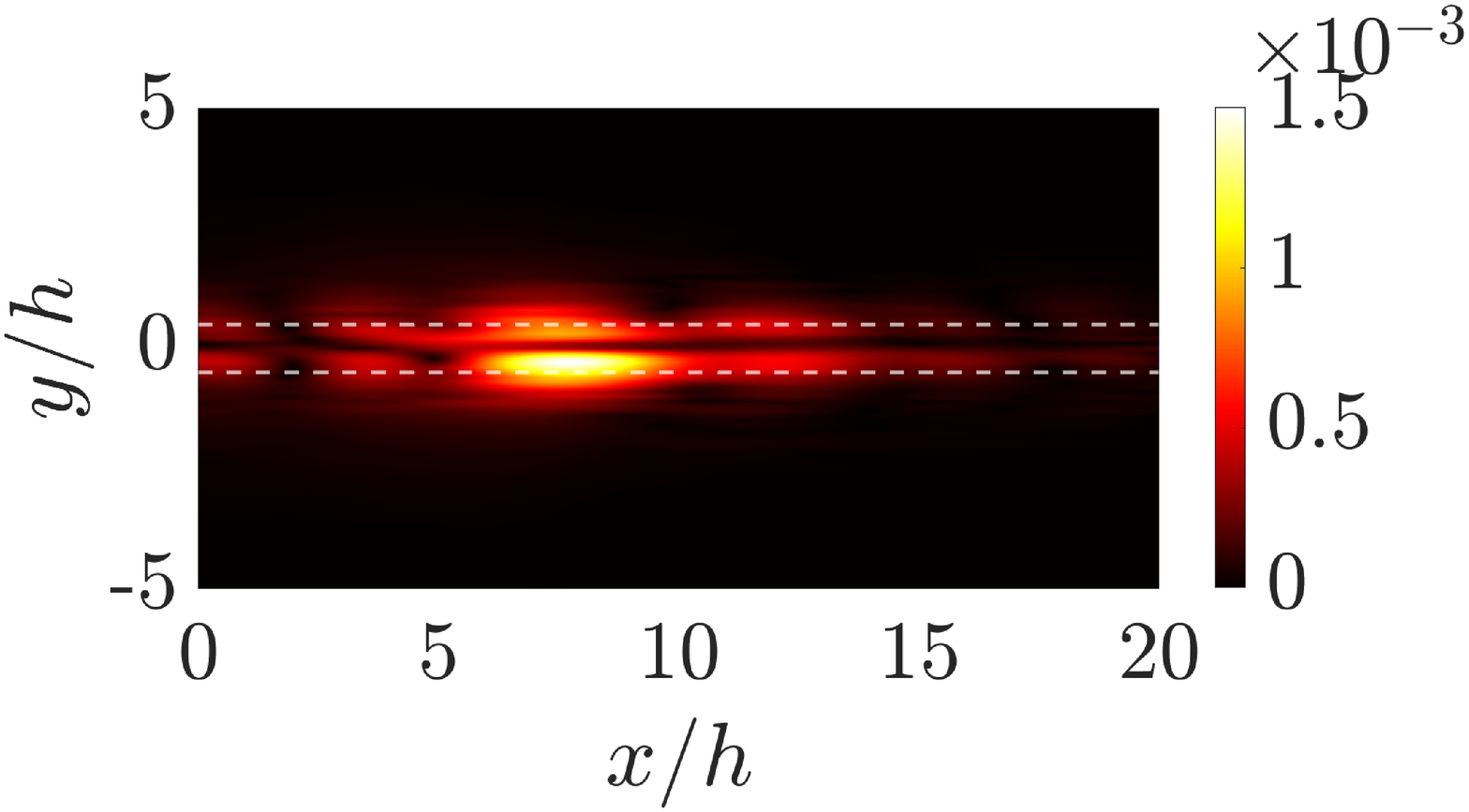} &
    \includegraphics[width=0.45\textwidth]{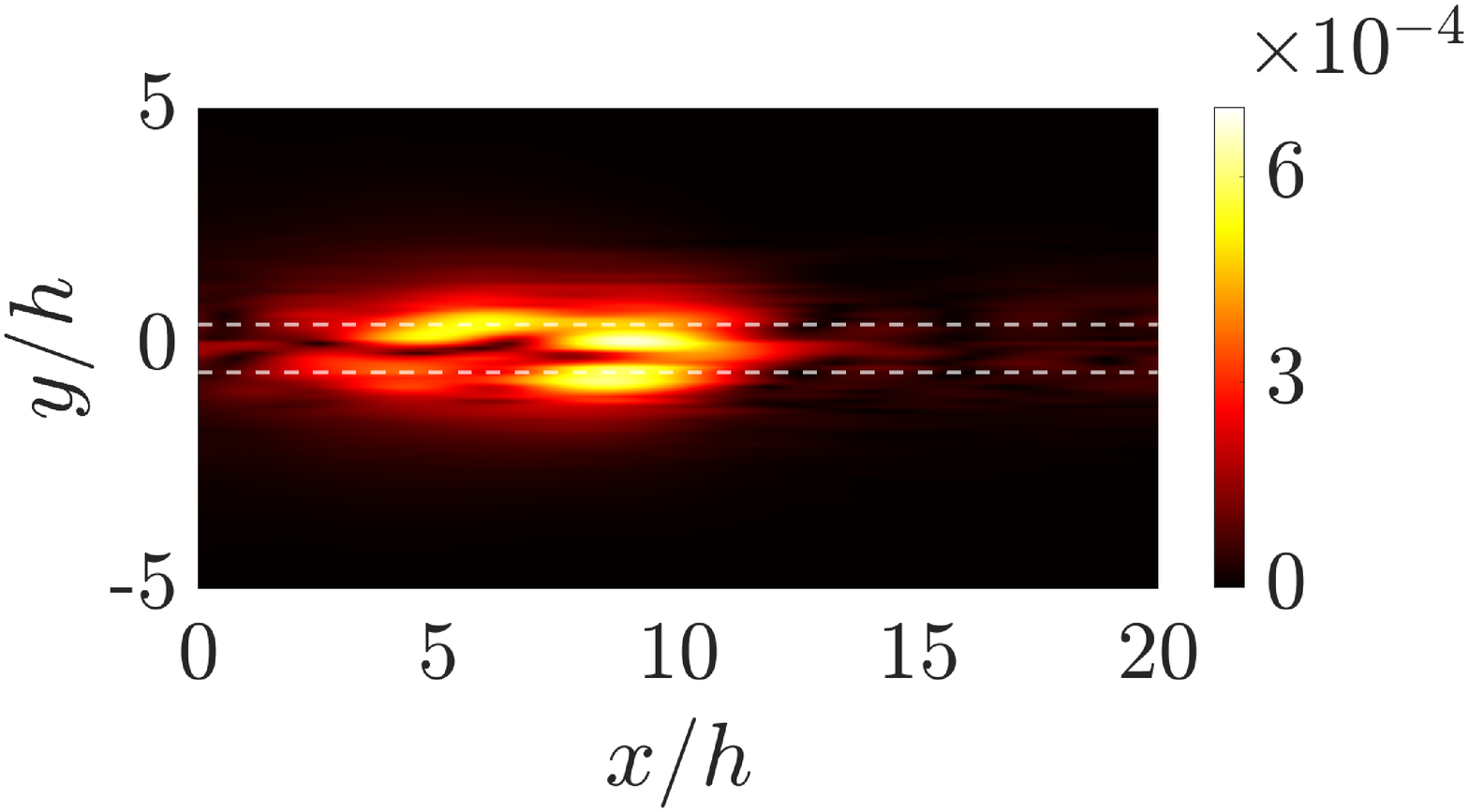} \\
    (c) & (d) \\
  \end{tabular}  
  \caption{The guided jet modes identified at $St$ = 0.367 (a,b) and $St$ = 0.383 (c,d): (a,c) Jet 1; (b,d) Jet 2. Modes are visualized by the modulus of the fluctuating pressure components. White dashed lines indicate the liplines. Note that these modes are much weaker than those found at the screech frequency.}
\label{fig:gjm_offpeak}
\end{figure}

\begin{figure}
  \centering
  \begin{tabular}{c}
    \includegraphics[width=0.55\textwidth]{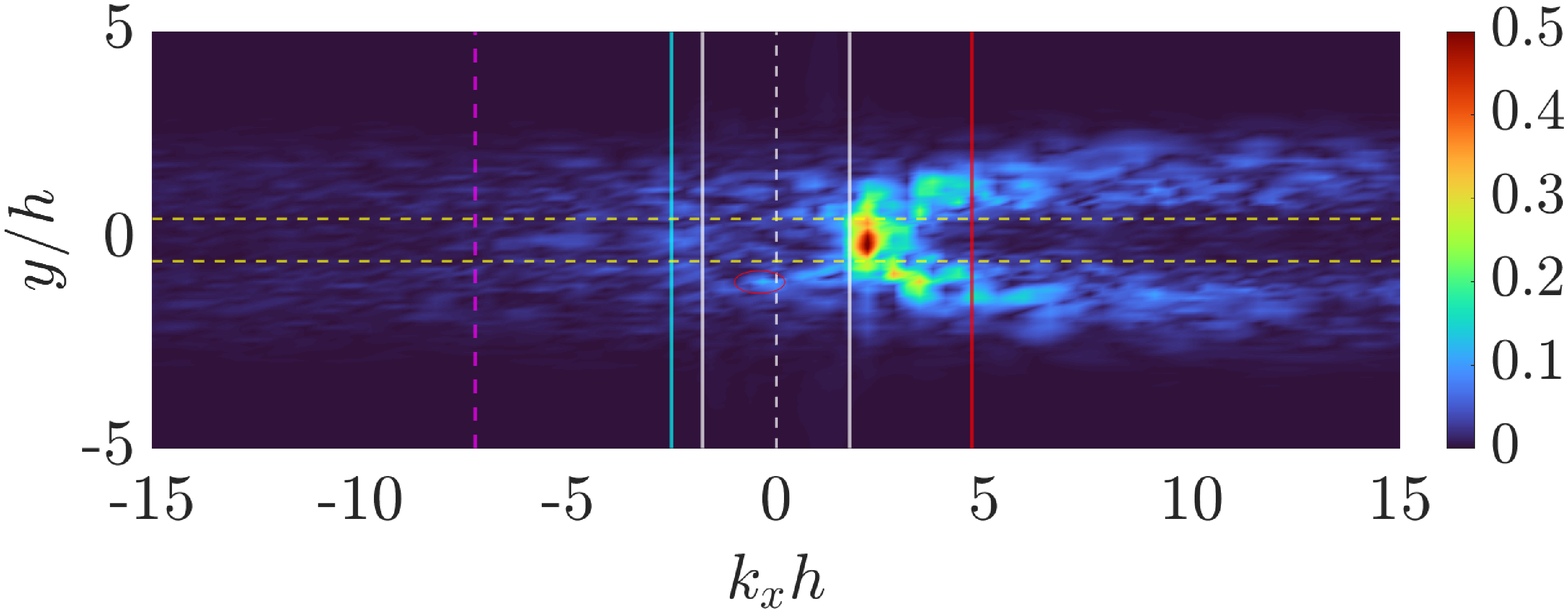} \\
    (a) \\
    \includegraphics[width=0.55\textwidth]{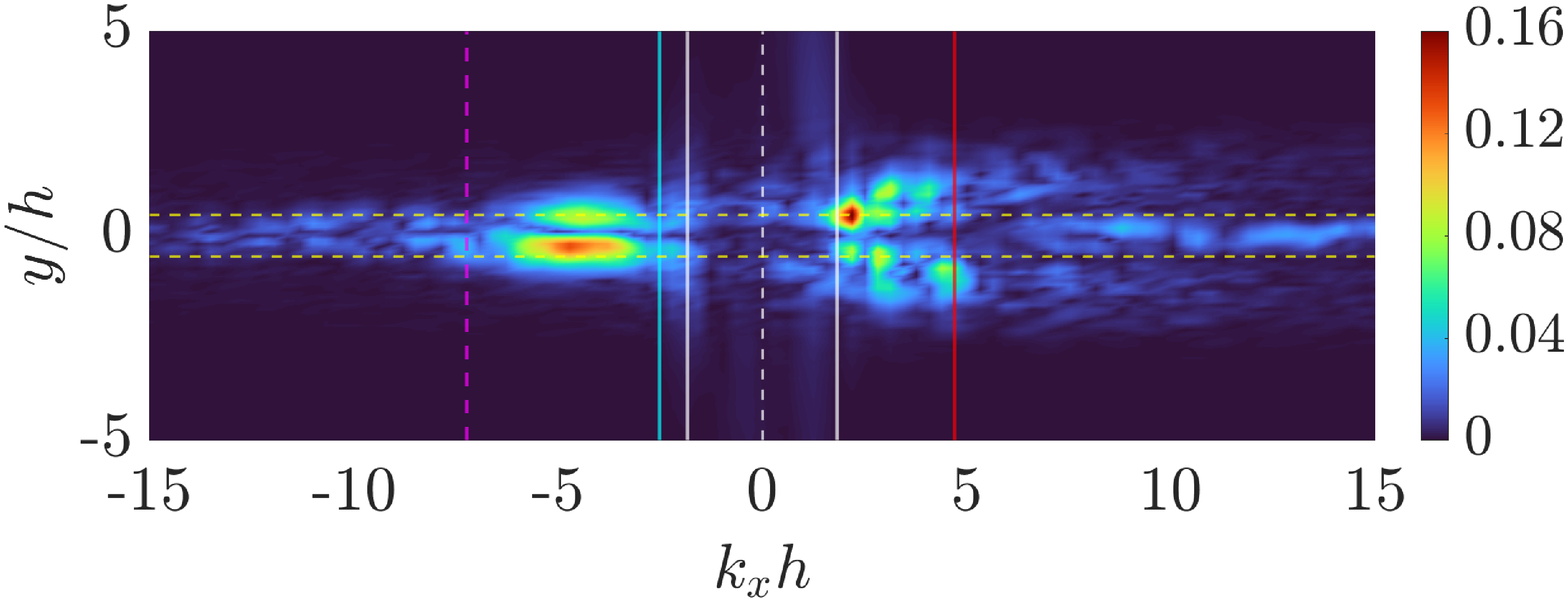} \\
    (b) \\
  \end{tabular}  
  \caption{Streamwise wavenumber spectra visualized by the modulus of the ensemble averaged leading SPOD mode shape: (a) fluctuating transverse velocity component at $St$ = 0.367 and (b) fluctuating pressure component at $St$ = 0.383. Cyan solid line, $k_{+,max} - k_{s_1}$;  magenta dashed line, $k_{+,max} - k_{s_2}$; red solid line, $k_{s_1}$; white solid lines, $\pm k_{c_{\infty}}$; white dashed line, zero axis; yellow horizontal lines, $y/h$ = $\pm$0.5. Note that the peak modulus is found to be much lower than the value observed at the screech frequency.}
\label{fig:wavenumber_spectra_offpeak}
\end{figure}

At the off-peak frequencies, such synchronisation with respect to the guided jet mode rarely happens as shown in figure~\ref{fig:shock_KH_alpha_off_peaks}. In this case, weakened guided jet mode hinders each jet's self-excitation as shown in tables~\ref{tab:amplitudes_p} and~\ref{tab:amplitudes_u1}, and the free-stream acoustic waves from its twin fails to reinforce the coupling between the two jets. In fact, even though the guided jet modes are recovered via the same procedure applied at the screech frequency, loss of previously observed properties raises questions about its validity. While the guided jet modes are organized with perfect symmetry about $y/h$ = 0 as well as between the twin jets at the screech frequency, figure~\ref{fig:gjm_offpeak} shows that those at the non-resonating frequencies do not exhibit such symmetries. As depicted in figure~\ref{fig:wavenumber_spectra_offpeak}, the wavenumber spectra at these frequencies look significantly dispersed. The wavenumber corresponding to the peak modulus in the negative wavenumber domain does not necessarily align with the difference between the wavenumbers associated with the maximum modulus of the KH blob and the shock-cell system. From these spectra, the peak negative wavenumber region (as highlighted by orange ellipse in figure~\ref{fig:wavenumber_spectra_offpeak}(a)) is often found at $k > k_{c_{\infty}}$, which violates the indicative characteristics of the guided jet mode. 

It is important to note that guided jet modes are known to be supported in a very narrow range~\citep{Towne2017}, and, to be accurate, their existence at neighboring frequencies should be examined using stability analysis. Even if the modes identified here may not be the true guided jet mode because they cannot be found at the non-resonating frequencies, their absence at these frequencies would still highlight its pivotal role in the screech coupling.

\section{Conclusions}
\label{sec:conclusion}
In this work the effectiveness of the free-stream acoustic mode or the guided jet mode as a closure mechanism for the rectangular twin-jet screech coupling is assessed. The jets studied herein produce intermittent screech tones resulting from a competition between the out-of-phase and the in-phase coupling modes. To consider the wave components active in the screech coupling that is perfectly synchronised to the out-of-phase mode, an ensemble average of leading SPOD modes is obtained from several segments of the LES data, which correspond to periods marked by invariant phase differences. The streamwise wavenumber spectra of the resulting ensemble averaged mode reveal that both the upstream- and downstream-propagating modes consist of a wide range of the Fourier modes. The separation of the guided jet mode and the free-stream acoustic mode is then achieved by retaining a series of wavenumber modes with appropriate phase velocity. The determination of wavenumbers associated with upstream propagation is based on the fact that the these modes are energised by the interaction of the KH waves and the shock-cell system. The KH blob in the positive wavenumber domain is bounded by setting a threshold value at 10\% of the maximum modulus in the wavenumber spectra. Thereafter, the difference between the wavenumbers corresponding to the KH blob and the peak wavenumber of the shock-cell is used to design a bandpass filter that effectively restricts the upstream-propagating modes. Within the realm of eligible wavenumber modes, any modes with supersonic phase velocity are used to retrieve the free-stream acoustic mode. The remaining modes are treated as the guided jet mode.

Via the spatial cross-correlation, the phase and amplitude variations of each mode with respect to the receptivity location are computed. Several eligible points of return for each path are identified, where the upstream-propagating waves from such points complete the screech feedback loop satisfying the appropriate phase criteria. The present analysis shows that the guided jet mode yields significantly larger amplitudes and admits more number of eligible points of return, which mostly coincide with the peaks of the shock/KH instability waves interaction. Free-stream acoustic waves from such points propagate to the other jet's receptivity location with a 180$^\circ$ phase difference, reinforcing its self-excited screech feedback loop. At the immediate off-peak frequencies, these observations regarding the gain and phasing are not discernible. The reliability of the guided jet mode at these frequencies is questionable. In fact, the existence of the guided jet mode should be examined more accurately using linear stability analysis~\citep{Towne2017}. Nonetheless, even if the guided jet mode is not realised at non-resonating frequencies, the absence of it at these frequencies would serve as compelling evidence to support that the guided jet mode plays a crucial role in the screech coupling. In summary, the upstream-propagating guided jet mode seems to work as a preferred closure mechanism for the rectangular twin jets as it does for singles jets.

\backsection[Acknowledgements]{The authors acknowledge Cascade Technologies for granting us the access to their numerical software. We would like to thank Prof. Ephraim Gutmark and Aatresh Karnam at the University of Cincinnati for generously sharing their laboratory measurements and extended discussions. We also acknowledge helpful comments by anonymous referees which led to improved analysis and demarcation between the guided jet mode and the upstream-propagating acoustic waves.}

\backsection[Funding]{This work was supported by the Office of Naval Research under Grant No. N00014-18-1-2391. Computational resources for LES were provided by the Extreme Science and Engineering Discovery Environment (XSEDE).}

\backsection[Declaration of interests]{The authors report no conflict of interest.}



\bibliographystyle{jfm}
\bibliography{references_twinjet}

\end{document}